\documentclass[twocolumn]{aastex61}
\usepackage{xspace} 
\usepackage{graphicx} 
\usepackage{natbib}
\usepackage{amsmath}
\newcommand{\nraoblurb}{The National Radio Astronomy Observatory is
a facility of the National Science Foundation operated under cooperative
agreement by Associated Universities, Inc.}

\newcommand{\lsim}{\ensuremath{\,\lesssim\,}\xspace}
\newcommand{\gl}{\ensuremath{\ell}\xspace}

\newcommand{\vlsr}{\ensuremath{V_{\rm LSR}}\xspace}

\newcommand{\kms}{\ensuremath{\,{\rm km\,s^{-1}}}\xspace}

\newcommand{\kpc}{\ensuremath{\,{\rm kpc}}\xspace}

\newcommand{\hi}{H\,{\sc i}}
\newcommand{\hii}{H\,{\sc ii}}

\newcommand{\co} {\ensuremath{^{\rm 12}{\rm CO}}\xspace}

\newcommand{\az}{\theta_{\rm Az}}
\newcommand{\azerr}{\sigma_{\rm Az}}
\revised{\today, accepted to ApJ -- tvw}

\shorttitle{Kinematic Distances}
\shortauthors{Wenger et al.}

\begin{document}

\title{KINEMATIC DISTANCES: A MONTE CARLO METHOD}

\author{Trey V. Wenger} 
\affiliation{Astronomy Department, University of Virginia, P.O. Box
  400325, Charlottesville, VA 22904-4325, USA.}
\affiliation{National Radio Astronomy Observatory, 520 Edgemont Road,
  Charlottesville, VA 22903, USA.}
\email{tvw2pu@virginia.edu}

\author{Dana S. Balser}
\affiliation{National Radio Astronomy Observatory, 520 Edgemont Road,
  Charlottesville, VA 22903, USA.}

\author{L. D. Anderson}
\affiliation{Department of Physics and Astronomy, West Virginia
  University, Morgantown, WV 26505, USA.}
\affiliation{Center for Gravitational Waves and Cosmology, West
Virginia University, Morgantown, Chestnut Ridge Research Building, 
Morgantown, WV 26505, USA.}
\affiliation{Adjunct Astronomer at the Green Bank Observatory, 
P.O. Box 2, Green Bank, WV 24944, USA.}

\author{T. M. Bania}
\affiliation{Institute for Astrophysical Research, Astronomy 
  Department, Boston University, 725 Commonwealth Ave., Boston, MA 
  02215, USA.}

\begin{abstract}
  Distances to high mass star forming regions (HMSFRs) in the Milky
  Way are a crucial constraint on the structure of the Galaxy.  Only
  kinematic distances are available for a majority of the HMSFRs in
  the Milky Way. Here we compare the kinematic and parallax distances
  of 75 Galactic HMSFRs to assess the accuracy of kinematic
  distances. We derive the kinematic distances using three different
  methods: the traditional method using the \citet{brand1993} rotation
  curve (Method A), the traditional method using the \citet{reid2014}
  rotation curve and updated Solar motion parameters (Method B), and a
  Monte Carlo technique (Method C). Methods B and C produce kinematic
  distances closest to the parallax distances, with median differences
  of \(13\%\) (\(0.43\kpc\)) and \(17\%\) (\(0.42\kpc\)),
  respectively. Except in the vicinity of the tangent point, the
  kinematic distance uncertainties derived by Method C are smaller
  than those of Methods A and B. In a large region of the Galaxy, the
  Method C kinematic distances constrain both the distances and the
  Galactocentric positions of HMSFRs more accurately than parallax
  distances. Beyond the tangent point along \(\ell=30^\circ\), for
  example, the Method C kinematic distance uncertainties reach a
  minimum of \(10\%\) of the parallax distance uncertainty at a
  distance of \(14\kpc\). We develop a prescription for deriving and
  applying the Method C kinematic distances and distance
  uncertainties. The code to generate the Method C kinematic distances
  is publicly available and may be utilized through an on-line tool.
\end{abstract}


\section{Introduction}

Revealing the morphological and chemical structure of the Milky Way
requires knowing the locations of objects on a Galaxy-wide scale. In
the Solar neighborhood, the distances to stars can be accurately
derived by measuring their parallax. Far from the Solar neighborhood,
distances to stars may be determined using spectrophotometric
techniques \citep[e.g.,][]{moises2011} and red clump stars
\citep[e.g.,][]{bovy2014}. Distances to gas clouds can be gotten from
both Very Long Baseline Interferometry (VLBI) parallax measurements of
molecular maser emission from high mass star forming regions (HMSFRs)
\citep[e.g.,][]{reid2014} as well as kinematic distance determinations
\citep[e.g.,][]{anderson2012}.

Kinematic distances are derived by measuring the local standard of
rest (LSR) velocity, \vlsr, of an object and assuming a model of
Galactic rotation. If the object is on a circular orbit following this
Galactic rotation model (GRM), then the LSR velocity of the object
uniquely identifies the object's Galactocentric radius, \(R\). Beyond
the Solar orbit, this technique also uniquely determines the object's
Galactocentric azimuth, \(\theta\), and distance from the Sun, \(d\).
Within the Solar orbit, kinematic distances suffer from the kinematic
distance ambiguity (KDA). Here, a single LSR velocity may correspond
to two distances: a ``near'' and ``far'' kinematic distance. We must
use additional information to identify the kinematic distance
ambiguity resolution (KDAR). The kinematic method is commonly used to
determine the distances to HMSFRs in the study of Galactic structure.
Recently, for example, \citet{balser2015} used \hii\ region kinematic
distances to probe the metallicity distribution across the Galactic
disk.

The Green Bank Telescope \hii\ Region Discovery Survey (GBT HRDS) and
its successors discovered more than \({\sim}1000\) new Galactic \hii\
regions by measuring their centimeter wavelength radio recombination
line (RRL) emission \citep{bania2010,bania2012,anderson2015}. \hii\
regions are the zones of ionized gas surrounding recently-formed
high-mass (OB-type) stars. They are the archetypical tracer of
Galactic spiral structure. \citet{anderson2012} derived the kinematic
distances to 149 \hii\ regions in the original GBT HRDS, and today the
\textit{WISE} Catalog of Galactic \hii\ Regions \citep{anderson2014}
lists \({\sim}1500\) \hii\ region kinematic distances.

Errors in kinematic distances are caused by both inaccurate GRMs and
incorrect KDARs. The rotation of the Milky Way is affected by
non-circular streaming motions induced by the Galactic bar and spiral
arms \citep[e.g.,][]{burton1971,gomez2006,moises2011}. These
deviations from circular motion will affect the accuracy of GRMs. A
variety of techniques have been used to resolve the KDA for Galactic
\hii\ regions, for example, \hi\ emission/absorption experiments
\citep{kuchar1994,kolpak2003,anderson2009a,anderson2012,urquhart2012,brown2014},
\hi\ self-absorption experiments \citep{roman-duval2009,urquhart2012},
and H\(_2\)CO absorption experiments
\citep{araya2002,watson2003,sewilo2004}. If these KDAR techniques are
inaccurate, the derived kinematic distances will be as well.

Very Long Baseline Interferometric (VLBI) trigonometric parallax
measurements of molecular masers are an independent and accurate way
to measure the distances to HMSFRs. Over the past decade, the Bar and
Spiral Structure Legacy Survey
(BeSSeL)\footnote{\url{http://bessel.vlbi-astrometry.org/}}, the
Japanese VLBI Exploration of Radio Astrometry
(VERA)\footnote{\url{http://veraserver.mtk.nao.ac.jp/}}, and the
European VLBI Network (EVN)\footnote{\url{http://www.evlbi.org/}}
projects have accumulated a sample of more than 100 VLBI parallaxes
and proper motions for masers associated with HMSFRs
\citep{reid2014}. These trigonometrically-derived distances do not
suffer from the same problems as kinematic distances. With a typical
parallax uncertainty of \({\sim}20\,\mu\text{as}\), these parallax
distances are accurate to about 10\% at distances of 5\kpc
\citep{reid2014rev}.

Although parallaxes are the ``gold standard'' distances for HMSFRs,
they are difficult and time-consuming to measure.  To constrain the
parallax and proper motion of four HMSFRs, including W51 Main/South,
\citet{sato2010} used the National Radio Astronomy Observatory (NRAO)
Jansky Very Large array to locate background extragalactic position
reference objects together with the NRAO Very Long Baseline Array
(VLBA) for the accurate astrometry.  The VLBA observations totaled
\({\sim}28\) hours spread over \({\sim}12\) months. Such observations
are impractical to make for all \({\sim}4000\) \hii\ regions in the
\textit{WISE} Catalog. Furthermore, the majority of the \hii\ regions
in the \textit{WISE} Catalog will not have detectable maser emission.

\startlongtable
\begin{deluxetable*}{lcccD@{ \(\pm\!\)}DD@{ \(\pm\!\)}Dr}
\tablewidth{0pt}
\tabletypesize{\normalsize}
\tablecaption{HMSFR Sample\label{tab:sample}}
\tablehead{
\colhead{Name} & \colhead{Alias} & \colhead{RA (J2000)} & \colhead{Decl. (J2000)} & \multicolumn{4}{c}{Parallax} & \multicolumn{4}{c}{\(\vlsr\)} & \colhead{Refs.} \\
\colhead{} & \colhead{} & \colhead{(hh:mm:ss)} & \colhead{(dd:mm:ss)} & \multicolumn{4}{c}{(mas)} & \multicolumn{4}{c}{(km s\(^{-1}\))} & \colhead{}
}
\decimals
\startdata
G015.03\(-\)00.67 & M 17            & 18:20:24.81 & \(-\)16:11:35.3 & 0.505 & 0.033 & 22 & 3 & 10 \\
G016.58\(-\)00.05 &  & 18:21:09.08 & \(-\)14:31:48.8 & 0.279 & 0.023 & 60 & 5 & 4 \\
G023.00\(-\)00.41 &  & 18:34:40.20 & \(-\)09:00:37.0 & 0.218 & 0.017 & 80 & 3 & 11 \\
G023.44\(-\)00.18 &  & 18:34:39.19 & \(-\)08:31:25.4 & 0.170 & 0.032 & 97 & 3 & 11 \\
G023.65\(-\)00.12 &  & 18:34:51.59 & \(-\)08:18:21.4 & 0.313 & 0.039 & 83 & 3 & 12 \\
G023.70\(-\)00.19 &  & 18:35:12.36 & \(-\)08:17:39.5 & 0.161 & 0.024 & 73 & 5 & 7 \\
G025.70+00.04 &  & 18:38:03.14 & \(-\)06:24:15.5 & 0.098 & 0.029 & 93 & 5 & 4 \\
G027.36\(-\)00.16 &  & 18:41:51.06 & \(-\)05:01:43.4 & 0.125 & 0.042 & 92 & 3 & 10 \\
G028.86+00.06 &  & 18:43:46.22 & \(-\)03:35:29.6 & 0.135 & 0.018 & 100 & 10 & 4 \\
G029.86\(-\)00.04 &  & 18:45:59.57 & \(-\)02:45:06.7 & 0.161 & 0.020 & 100 & 3 & 6 \\
G029.95\(-\)00.01 & W 43S           & 18:46:03.74 & \(-\)02:39:22.3 & 0.190 & 0.019 & 98 & 3 & 6 \\
G031.28+00.06 &  & 18:48:12.39 & \(-\)01:26:30.7 & 0.234 & 0.039 & 109 & 3 & 6 \\
G031.58+00.07 & W 43Main        & 18:48:41.68 & \(-\)01:09:59.0 & 0.204 & 0.030 & 96 & 5 & 6 \\
G032.04+00.05 &  & 18:49:36.58 & \(-\)00:45:46.9 & 0.193 & 0.008 & 97 & 5 & 4 \\
G033.64\(-\)00.22 &  & 18:53:32.56 & +00:31:39.1 & 0.153 & 0.017 & 60 & 3 & 1 \\
G034.39+00.22 &  & 18:53:18.77 & +01:24:08.8 & 0.643 & 0.049 & 57 & 5 & 13 \\
G035.02+00.34 &  & 18:54:00.67 & +02:01:19.2 & 0.430 & 0.040 & 52 & 5 & 2 \\
G035.19\(-\)00.74 &  & 18:58:13.05 & +01:40:35.7 & 0.456 & 0.045 & 30 & 7 & 14 \\
G035.20\(-\)01.73 &  & 19:01:45.54 & +01:13:32.5 & 0.306 & 0.045 & 42 & 3 & 14 \\
G037.43+01.51 &  & 18:54:14.35 & +04:41:41.7 & 0.532 & 0.021 & 41 & 3 & 2 \\
G043.16+00.01 & W 49N           & 19:10:13.41 & +09:06:12.8 & 0.090 & 0.007 & 10 & 5 & 15 \\
G043.79\(-\)00.12 & OH 43.8\(-\)0.1     & 19:11:53.99 & +09:35:50.3 & 0.166 & 0.005 & 44 & 10 & 2 \\
G043.89\(-\)00.78 &  & 19:14:26.39 & +09:22:36.5 & 0.121 & 0.020 & 54 & 5 & 2 \\
G045.07+00.13 &  & 19:13:22.04 & +10:50:53.3 & 0.125 & 0.005 & 59 & 5 & 2 \\
G045.45+00.05 &  & 19:14:21.27 & +11:09:15.9 & 0.119 & 0.017 & 55 & 7 & 2 \\
G048.60+00.02 &  & 19:20:31.18 & +13:55:25.2 & 0.093 & 0.005 & 18 & 5 & 15 \\
G049.19\(-\)00.33 &  & 19:22:57.77 & +14:16:10.0 & 0.189 & 0.007 & 67 & 5 & 2 \\
G049.48\(-\)00.36 & W 51 IRS2       & 19:23:39.82 & +14:31:05.0 & 0.195 & 0.071 & 56 & 3 & 16 \\
G049.48\(-\)00.38 & W 51M           & 19:23:43.87 & +14:30:29.5 & 0.185 & 0.010 & 58 & 4 & 17 \\
G052.10+01.04 & IRAS 19213+1723 & 19:23:37.32 & +17:29:10.5 & 0.251 & 0.060 & 42 & 5 & 18 \\
G059.78+00.06 &  & 19:43:11.25 & +23:44:03.3 & 0.463 & 0.020 & 25 & 3 & 16 \\
G069.54\(-\)00.97 & ON 1            & 20:10:09.07 & +31:31:36.0 & 0.406 & 0.013 & 12 & 5 & 19,20,21 \\
G074.03\(-\)01.71 &  & 20:25:07.11 & +34:49:57.6 & 0.629 & 0.017 & 5 & 5 & 21 \\
G075.29+01.32 &  & 20:16:16.01 & +37:35:45.8 & 0.108 & 0.005 & -58 & 5 & 22 \\
G075.76+00.33 &  & 20:21:41.09 & +37:25:29.3 & 0.285 & 0.022 & -9 & 9 & 21 \\
G075.78+00.34 & ON 2N           & 20:21:44.01 & +37:26:37.5 & 0.261 & 0.030 & 1 & 5 & 23 \\
G076.38\(-\)00.61 &  & 20:27:25.48 & +37:22:48.5 & 0.770 & 0.053 & -2 & 5 & 21 \\
G078.12+03.63 & IRAS 20126+4104 & 20:14:26.07 & +41:13:32.7 & 0.610 & 0.030 & -4 & 5 & 24 \\
G078.88+00.70 & AFGL 2591       & 20:29:24.82 & +40:11:19.6 & 0.300 & 0.024 & -6 & 7 & 25 \\
G079.73+00.99 & IRAS 20290+4052 & 20:30:50.67 & +41:02:27.5 & 0.737 & 0.062 & -3 & 5 & 25 \\
G079.87+01.17 &  & 20:30:29.14 & +41:15:53.6 & 0.620 & 0.027 & -5 & 10 & 21 \\
G080.79\(-\)01.92\tablenotemark{a} & NML Cyg         & 20:46:25.54 & +40:06:59.4 & 0.620 & 0.047 & -3 & 3 & 26 \\
G080.86+00.38 & DR 20           & 20:37:00.96 & +41:34:55.7 & 0.687 & 0.038 & -3 & 5 & 25 \\
G081.75+00.59 & DR 21           & 20:39:01.99 & +42:24:59.3 & 0.666 & 0.035 & -3 & 3 & 25 \\
G081.87+00.78 & W 75N           & 20:38:36.43 & +42:37:34.8 & 0.772 & 0.042 & 7 & 3 & 25 \\
G090.21+02.32 &  & 21:02:22.70 & +50:03:08.3 & 1.483 & 0.038 & -3 & 5 & 21 \\
G092.67+03.07 &  & 21:09:21.73 & +52:22:37.1 & 0.613 & 0.020 & -5 & 10 & 21 \\
G094.60\(-\)01.79 & AFGL 2789       & 21:39:58.27 & +50:14:21.0 & 0.280 & 0.030 & -46 & 5 & 18,28 \\
G095.29\(-\)00.93 &  & 21:39:40.51 & +51:20:32.8 & 0.205 & 0.015 & -38 & 5 & 28 \\
G097.53+03.18 &  & 21:32:12.43 & +55:53:49.7 & 0.133 & 0.017 & -73 & 5 & 27 \\
G100.37\(-\)03.57 &  & 22:16:10.37 & +52:21:34.1 & 0.291 & 0.010 & -37 & 10 & 28 \\
G105.41+09.87 &  & 21:43:06.48 & +66:06:55.3 & 1.129 & 0.063 & -10 & 5 & 21 \\
G107.29+05.63 & IRAS 22198+6336 & 22:21:26.73 & +63:51:37.9 & 1.288 & 0.107 & -11 & 5 & 29 \\
G108.18+05.51 & L 1206          & 22:28:51.41 & +64:13:41.3 & 1.289 & 0.153 & -11 & 3 & 19 \\
G108.20+00.58 &  & 22:49:31.48 & +59:55:42.0 & 0.229 & 0.028 & -49 & 5 & 28 \\
G108.47\(-\)02.81 &  & 23:02:32.08 & +56:57:51.4 & 0.309 & 0.010 & -54 & 5 & 28 \\
G108.59+00.49 &  & 22:52:38.30 & +60:00:52.0 & 0.398 & 0.031 & -52 & 5 & 28 \\
G109.87+02.11 & Cep A           & 22:56:18.10 & +62:01:49.5 & 1.430 & 0.080 & -7 & 5 & 30 \\
G111.23\(-\)01.23 &  & 23:17:20.79 & +59:28:47.0 & 0.288 & 0.044 & -53 & 10 & 28 \\
G111.25\(-\)00.76 &  & 23:16:10.36 & +59:55:28.5 & 0.294 & 0.016 & -43 & 5 & 28 \\
G111.54+00.77 & NGC 7538        & 23:13:45.36 & +61:28:10.6 & 0.378 & 0.017 & -57 & 5 & 30 \\
G121.29+00.65 & L 1287          & 00:36:47.35 & +63:29:02.2 & 1.077 & 0.039 & -23 & 5 & 19 \\
G122.01\(-\)07.08 & IRAS 00420+5530 & 00:44:58.40 & +55:46:47.6 & 0.460 & 0.020 & -50 & 5 & 31 \\
G123.06\(-\)06.30 & NGC 281         & 00:52:24.70 & +56:33:50.5 & 0.355 & 0.030 & -30 & 5 & 32 \\
G123.06\(-\)06.30 & NGC 281W        & 00:52:24.20 & +56:33:43.2 & 0.421 & 0.022 & -29 & 3 & 19 \\
G133.94+01.06 & W 3OH           & 02:27:03.82 & +61:52:25.2 & 0.512 & 0.010 & -47 & 3 & 33,34 \\
G134.62\(-\)02.19\tablenotemark{a} & S Per           & 02:22:51.71 & +58:35:11.4 & 0.413 & 0.017 & -39 & 5 & 35 \\
G135.27+02.79 & WB 89\(-\)437       & 02:43:28.57 & +62:57:08.4 & 0.167 & 0.011 & -72 & 3 & 36 \\
G209.00\(-\)19.38 & Orion Nebula    & 05:35:15.80 & \(-\)05:23:14.1 & 2.410 & 0.030 & 3 & 5 & 43,44,45 \\
G211.59+01.05 &  & 06:52:45.32 & +01:40:23.1 & 0.228 & 0.007 & 45 & 5 & 1 \\
G229.57+00.15 &  & 07:23:01.84 & \(-\)14:41:32.8 & 0.221 & 0.014 & 47 & 10 & 28 \\
G232.62+00.99 &  & 07:32:09.78 & \(-\)16:58:12.8 & 0.596 & 0.035 & 21 & 3 & 40 \\
G236.81+01.98 &  & 07:44:28.24 & \(-\)20:08:30.2 & 0.298 & 0.018 & 43 & 7 & 28 \\
G239.35\(-\)05.06\tablenotemark{a} & VY CMa          & 07:22:58.33 & \(-\)25:46:03.1 & 0.855 & 0.057 & 20 & 3 & 46,47 \\
G240.31+00.07 &  & 07:44:51.92 & \(-\)24:07:41.5 & 0.212 & 0.021 & 67 & 5 & 28 \\
\enddata
\tablenotetext{a}{Red supergiants}
\tablerefs{(1) BeSSeL Survey unpublished; (2) \citet{wu2014};
  (4) \citet{sato2014}; (6) \citet{zhang2014}; (7) \citet{sanna2014};
  (10) \citet{xu2011}; (11) \citet{brunthaler2009}; (12)
  (12) \citet{bartkiewicz2008}; (13) \citet{kurayama2011};
  (14) \citet{zhang2009}; (15) \citet{zhang2013};
  (16) \citet{xu2009}; (17) \citet{sato2010}; (18) \citet{oh2010};
  (19) \citet{rygl2010}; (20) \citet{nagayama2011};
  (21) \citet{xu2013}; (22) \citet{sanna2012}; (23) \citet{ando2011};
  (24) \citet{moscadelli2011}; (25) \citet{rygl2012};
  (26) \citet{zhang2012b}; (27) \citet{hachisuka2015};
  (28) \citet{choi2014}; (29) \citet{hirota2008};
  (30) \citet{moscadelli2009}; (31) \citet{moellenbrock2009};
  (32) \citet{sato2008}; (33) \citet{xu2006};
  (34) \citet{hachisuka2006}; (35) \citet{asaki2010};
  (36) \citet{hachisuka2009}; (40) \citet{reid2009a};
  (43) \citet{sandstrom2007}; (44) \citet{menten2007};
  (45) \citet{kim2008}; (46) \citet{choi2008};
  (47) \citet{zhang2012a}
}

\end{deluxetable*}

With such a large sample of HMSFR maser parallaxes, we can now compare
the parallax and kinematic distances and judge the accuracy of the
kinematic distance technique. \citet{reid2009b} performed a similar
study comparing the kinematic and parallax distances of 18
HMSFRs. They found that the kinematic distance method gives distances
much larger (up to a factor of 2) than the parallax distances for a
majority of their sample. After correcting the LSR velocities using
updated Solar motion parameters, however, the mean difference between
the kinematic and parallax distances became close to zero and only
half of their sample had kinematic distances larger than their
parallax distances. Here we expand upon the \citet{reid2009b} analysis
using a larger sample of HMSFRs.

\section{Sample Selection}

Our sample of HMSFRs comes from the maser parallax catalog in
\citet{reid2014} that contains parallaxes and proper motions for 103
HMSFRs and HMSFR proxies in the Milky Way. These data stem from
measurements made using the NRAO VLBA, the VERA project, and the
EVN. The \citet{reid2014} catalog contains the parallax, maser LSR
velocity, and their associated uncertainties for each HMSFR. This
provides the necessary information to derive both the parallax
distance and kinematic distance to each object.

Kinematic distances are unreliable in the direction of the Galactic
Center (GC; \(\gl = 0^\circ\)) and the Galactic Anti-center (GAC;
\(\gl = 180^\circ\)) due to velocity crowding: LSR velocities due to
circular motion tend towards zero in these directions. As in previous
studies using kinematic distances \citep[e.g.,][]{balser2015}, we
exclude all objects within \(15^\circ\) of the GC and \(20^\circ\) of
the GAC.

Our final sample contains 72 HMSFRs and 3 red supergiants (HMSFR
proxies). The positions, parallaxes, and LSR velocities (\(V_{\rm
  LSR}\)) from the \citet{reid2014} catalog are reproduced in
Table~\ref{tab:sample}.  According to \citet{reid2014}, the listed LSR
velocities are those of methanol masers when available, otherwise they
are the \co emission line velocities from associated giant molecular
clouds (GMCs). The LSR velocity uncertainties include both measurement
uncertainties as well as an added uncertainty relating the maser spot
motion to the bulk HMSFR motion. This added component ranges from
\(5\kms\) to \(20\kms\) \citep[see][]{reid2014}.

\section{Parallax Distances}

\begin{figure}[ht]
  \centering
  \includegraphics[width=\linewidth]{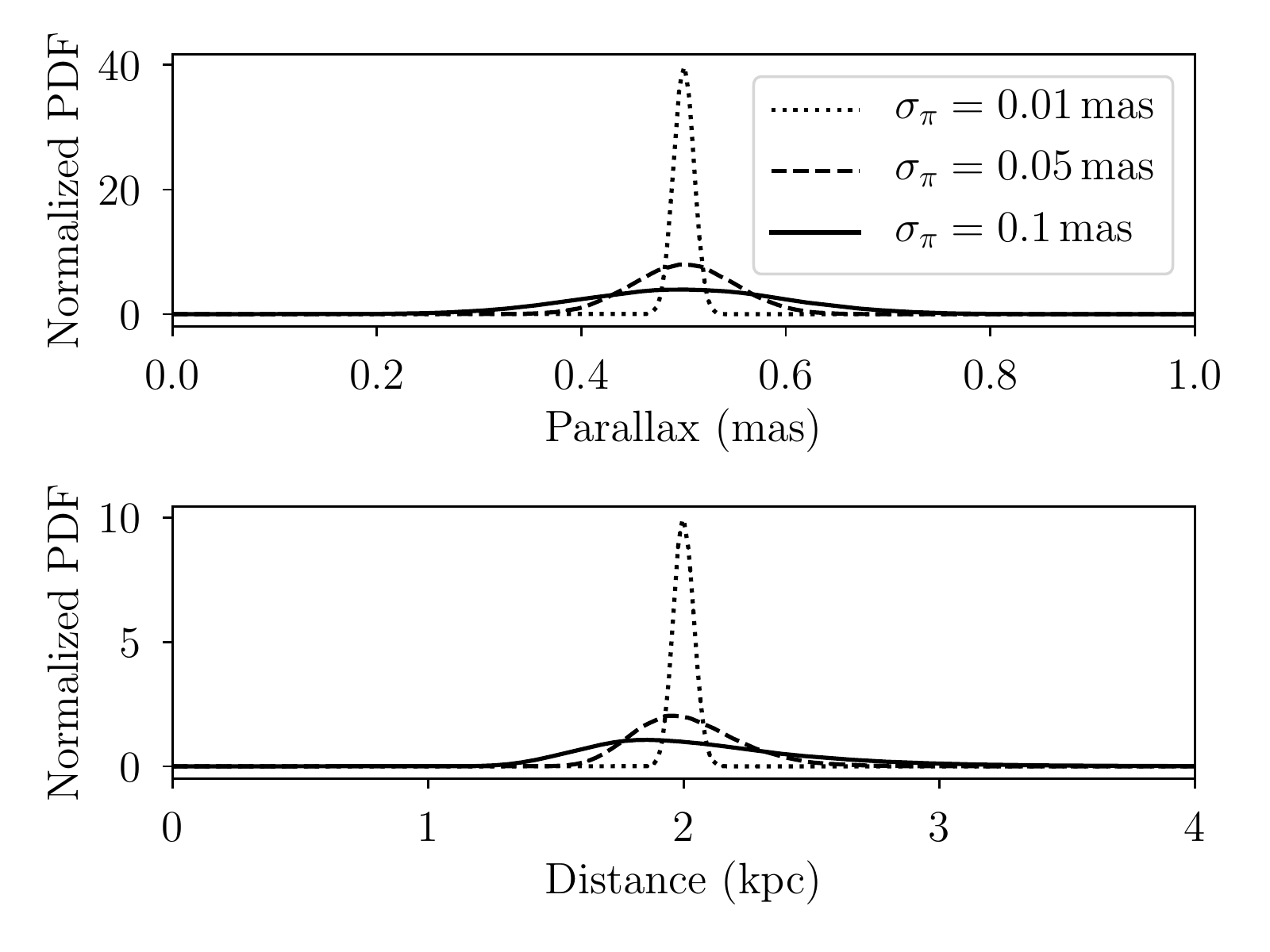}
  \caption{Normalized parallax probability distribution function (PDF;
    top) and parallax distance PDF (bottom). The parallax in this
    example is \(\pi = 0.5\,\text{mas}\) and the parallax uncertainty
    is \(\sigma_\pi = 0.01\,\text{mas}\) (dotted),
    \(0.05\,\text{mas}\) (dashed) and \(0.1\,\text{mas}\) (solid).
    The parallax distance PDF is determined by Monte Carlo re-sampling
    the Gaussian parallax PDF. For large relative parallax
    uncertainties, the parallax distance PDF is skewed.}
  \label{fig:parallax_pdf_example}
\end{figure}

The parallax distance is defined as
\begin{equation}
  D_P = \frac{1}{\pi} \label{eq:parallax}
\end{equation}
where the parallax distance, \(D_P\), has units of kpc when the
parallax, \(\pi\), has units of milli-arcseconds (mas). If the
parallax uncertainty, \(\sigma_\pi\), is small compared to the
parallax, i.e. \(\sigma_\pi/\pi \ll 1\), then the parallax distance
uncertainty, \(\sigma_P\), is determined by propagating the parallax
uncertainty through Equation~\ref{eq:parallax},
\begin{equation}
  \sigma_P = \frac{\sigma_\pi}{\pi^2}.
\end{equation}
If the fractional parallax uncertainty is large, however, the shape of
the parallax distance probability distribution function (PDF) is
skewed. Thus the peak (\(D_P\)) and the shape of the wings change and
the parallax distance uncertainty is non-symmetric around the peak
\citep[see][]{kovalevsky1998}. Figure~\ref{fig:parallax_pdf_example}
shows an example of the parallax distance PDF skew for different
parallax uncertainties.

\begin{figure*}[ht]
  \centering
  \includegraphics[width=0.49\linewidth]{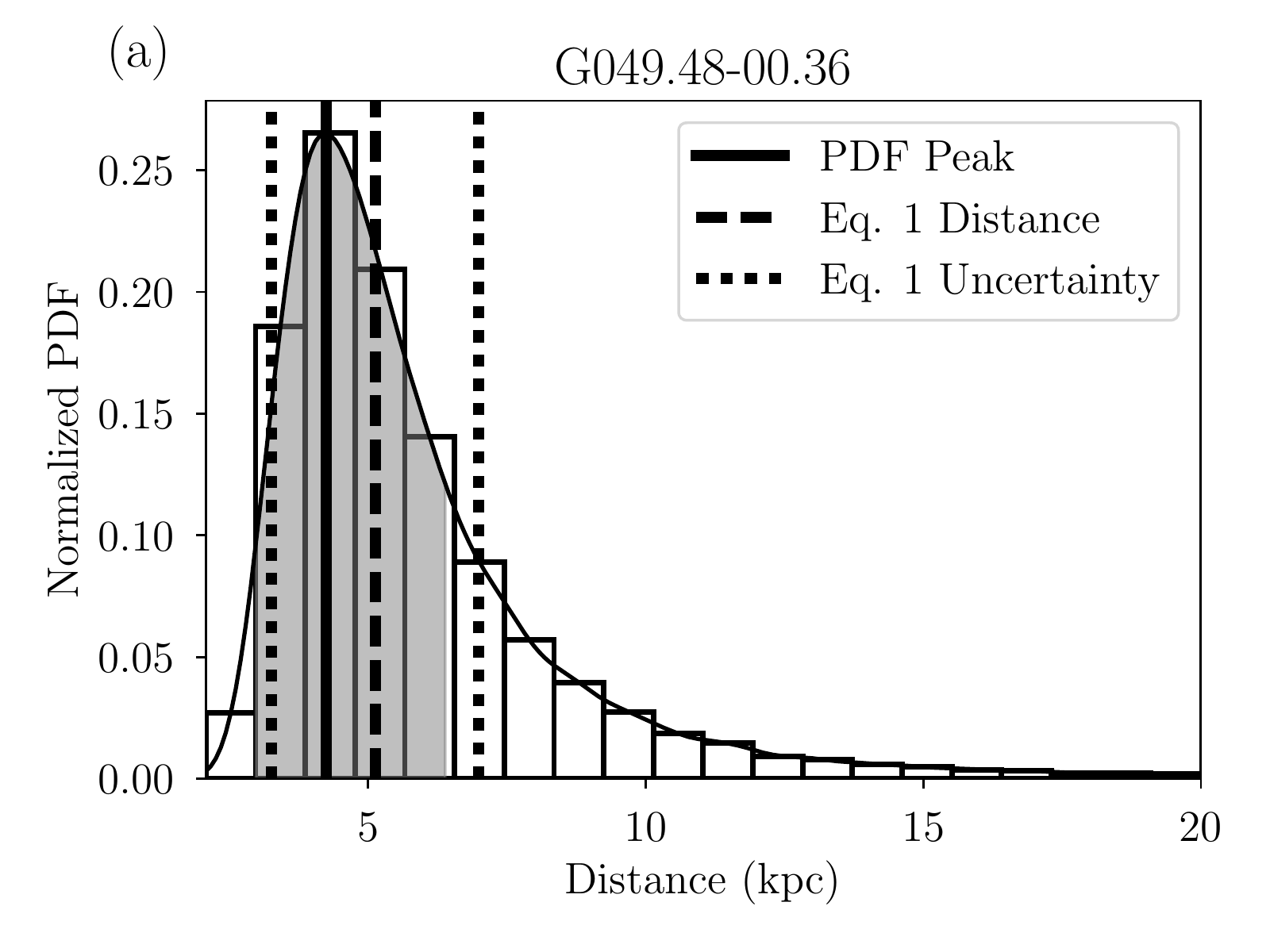}
  \includegraphics[width=0.49\linewidth]{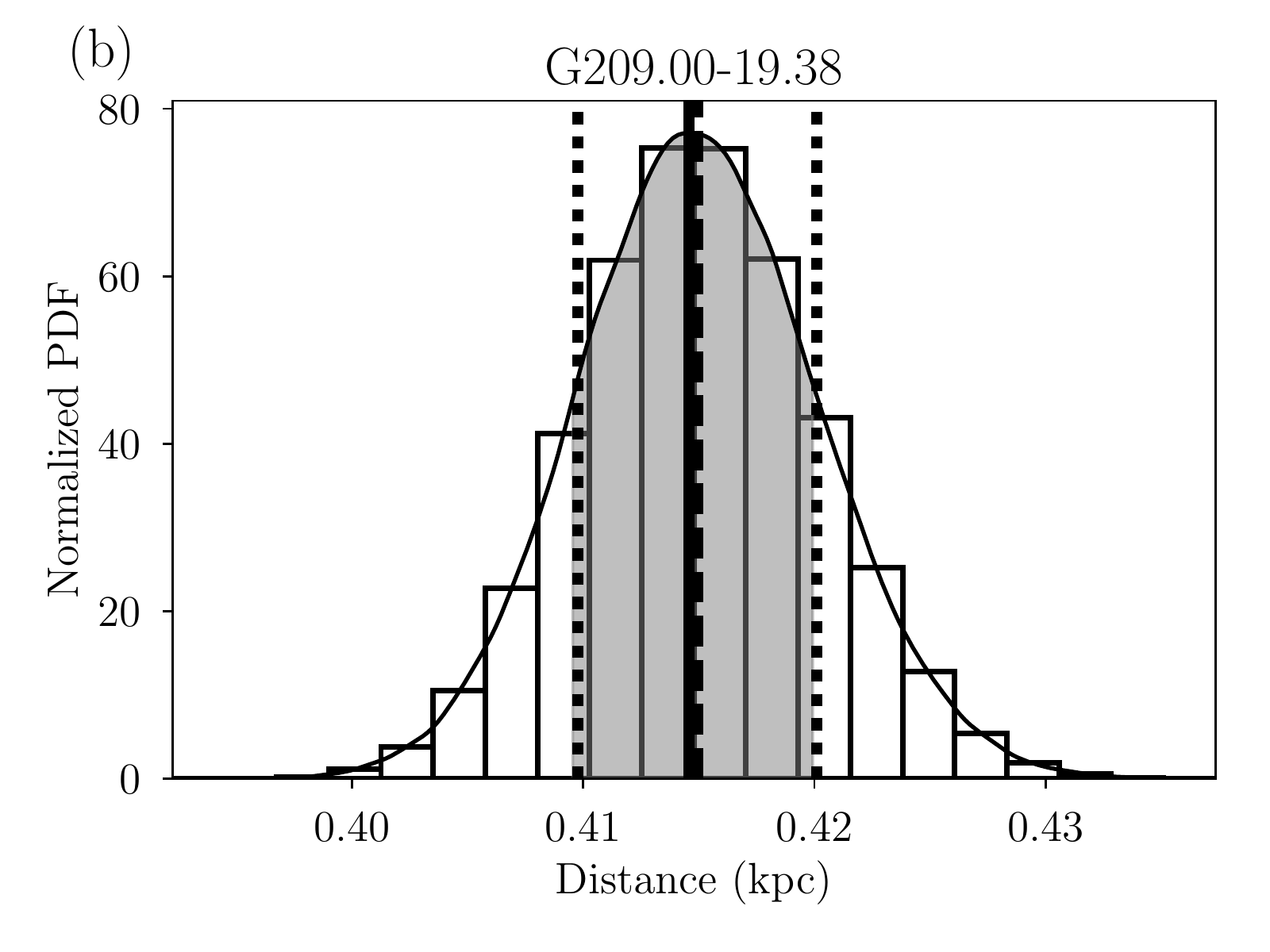} \\
  \includegraphics[width=0.49\linewidth]{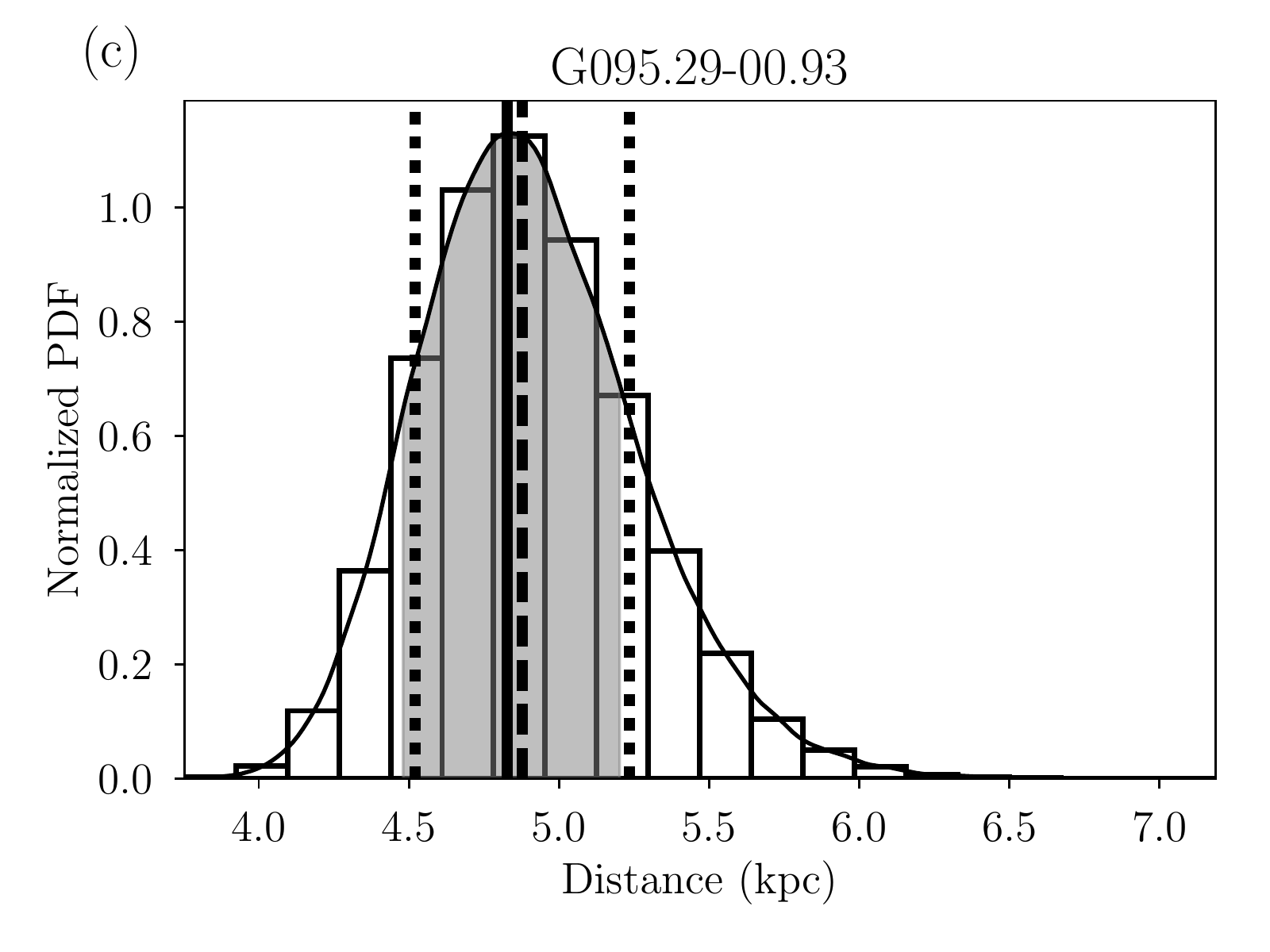}
  \includegraphics[width=0.49\linewidth]{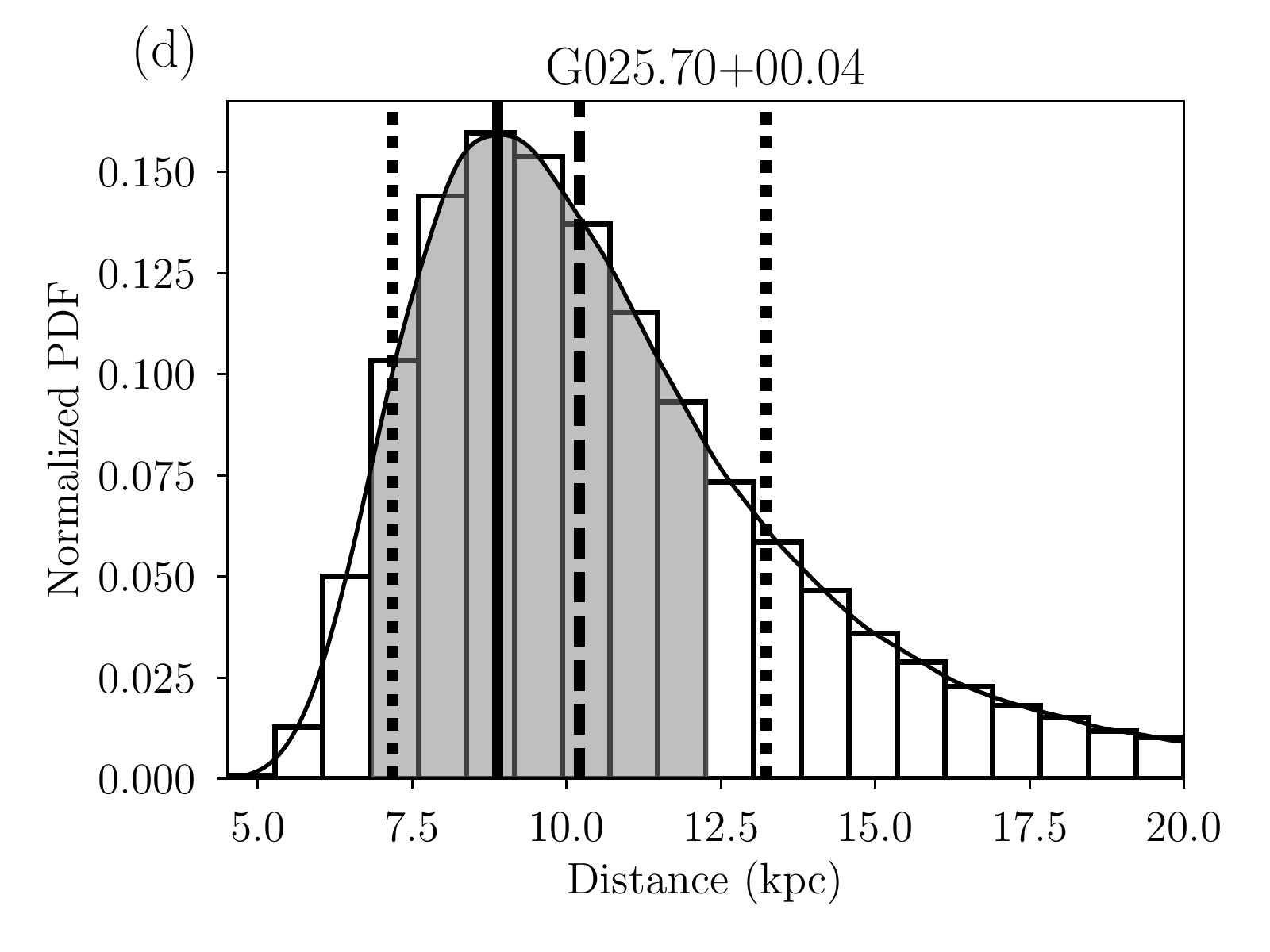}
  \caption{Normalized parallax distance probability distribution
    functions (PDFs) for four HMSFRs. The solid curve is the kernel
    density estimator (KDE) of the distribution; the solid vertical
    line is the peak of the KDE and our assigned parallax
    distance. The dashed vertical line is the parallax distance given
    by Equation~\ref{eq:parallax}. The vertical dotted lines span the
    symmetric uncertainty range in the parallax distance derived by
    propagating the parallax uncertainty through
    Equation~\ref{eq:parallax}. The filled region is the uncertainty
    range derived using the KDE (see text). Panel (a)
    (G049.48\(-\)00.36; W 51 IRS2) has the largest fractional parallax
    uncertainty and thus has the most skewed PDF. Panel (b)
    (G209.00-19.38; Orion Nebula) has the smallest fractional parallax
    uncertainty and has the PDF closest to a Gaussian
    distribution. Panel (c) (G095.29-00.93) has a typical fractional
    parallax uncertainty. Panel (d) (G025.70+00.04) has a large
    fractional parallax uncertainty. It has the largest deviation from
    the Monte Carlo-defined parallax distance and the parallax
    distance derived using Equation~\ref{eq:parallax}.}
  \label{fig:parallax_pdf}
\end{figure*}

We derive a Monte Carlo parallax distance for each HMSFR by
re-sampling the measured parallaxes within their uncertainties,
assuming a Gaussian parallax PDF. We sample the parallax \(10^5\)
times and use Equation~\ref{eq:parallax} to derive the parallax
distance distribution. To approximate the parallax distance PDF, we
fit a kernel density estimator (KDE) to the distribution. We use the
linear combination KDE technique from \citet{jones1993}, which is
accurate even in the presence of physical boundaries such as the
requirement that distances be greater than 0. The parallax distance
PDFs for four sources are shown in Figure~\ref{fig:parallax_pdf}. The
peak of the PDF (i.e. the most likely value) is the parallax
distance. In every case, this distance is smaller than the distance
given by Equation~\ref{eq:parallax}. We derive the uncertainty in the
parallax distance by determining the lower and upper bounds of the PDF
such that 1) the value of the PDF at both bounds is equal and 2) the
integral of the normalized PDF between the bounds is equal to 0.683
(i.e., 68.3\% of the total area under the PDF). This uncertainty is
therefore the 68.3\% confidence interval.

\begin{figure}[ht]
  \centering
  \includegraphics[width=\linewidth]{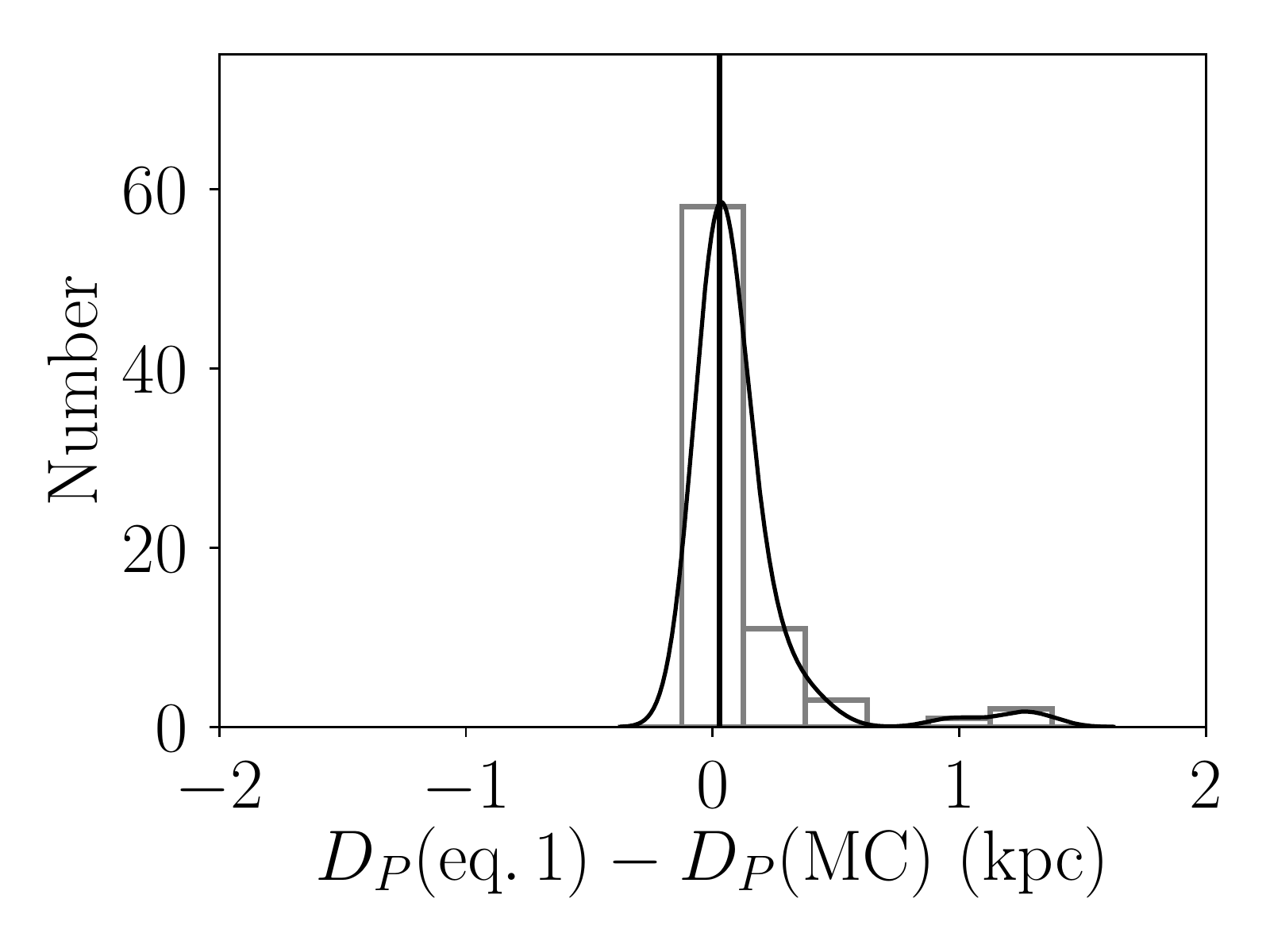}
  \includegraphics[width=\linewidth]{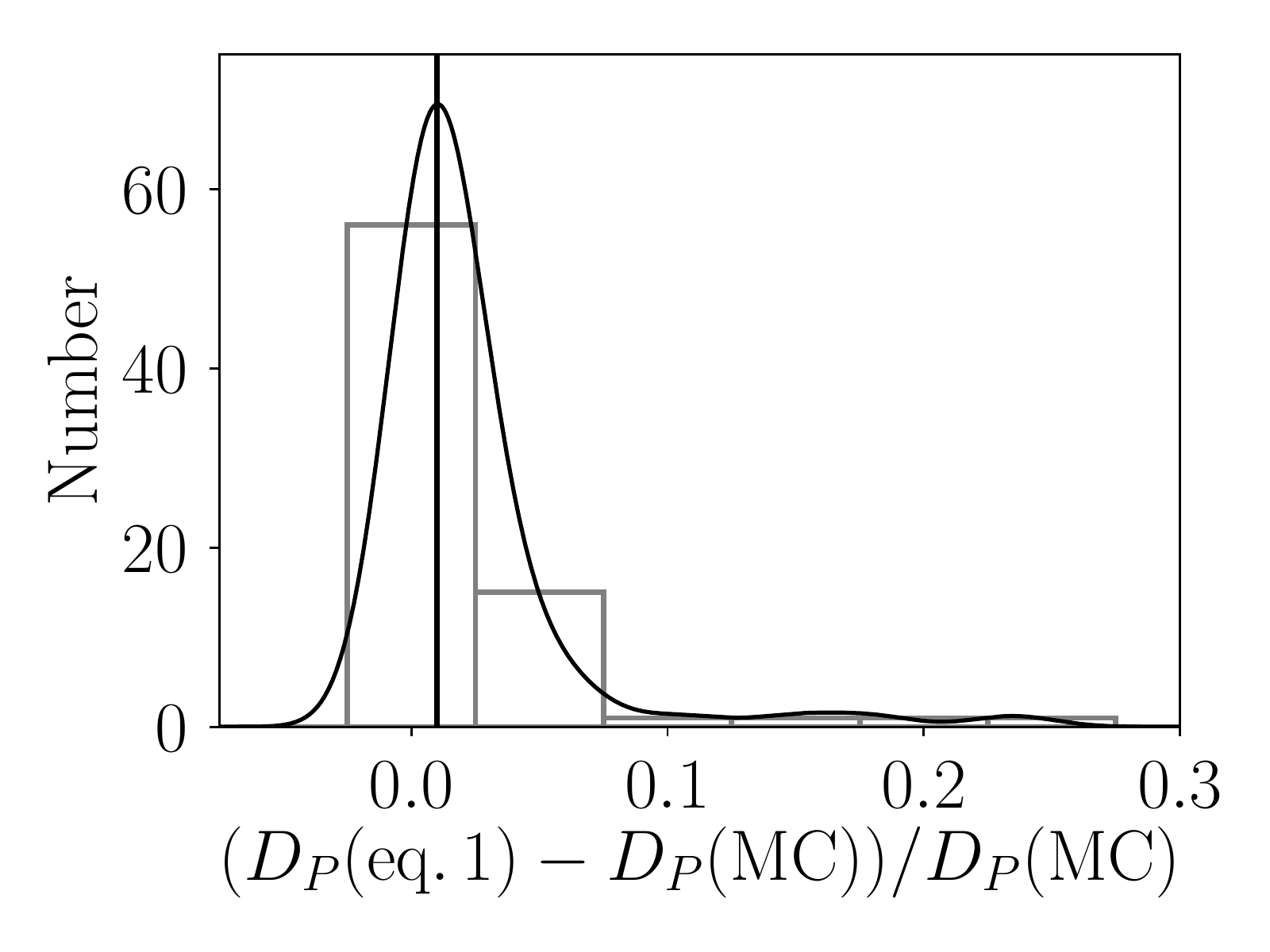}
  \caption{Difference (top) and fractional difference (bottom) between
    parallax distances derived using Equation~\ref{eq:parallax},
    \(D_P({\rm eq.\,1})\), and using the Monte Carlo method,
    \(D_P({\rm MC})\). The solid curve is the KDE fit to the
    difference distribution and the solid vertical line is the median
    of the distribution.}
  \label{fig:para_pdf_para}
\end{figure}

The difference between the parallax distances derived using
Equation~\ref{eq:parallax} and our Monte Carlo-derived parallax
distances is small; the median difference is 0.03 kpc and the largest
difference is 1.32 kpc for G025.70+00.04
(Figure~\ref{fig:parallax_pdf}, panel
(d)). Figure~\ref{fig:para_pdf_para} shows the distribution of
parallax distance differences between these two methods for our HMSFR
sample (Table~\ref{tab:sample}). The majority of objects in our sample
have less than \(0.1\kpc\) difference between the
Equation~\ref{eq:parallax} and Monte Carlo parallax distances.

\section{Kinematic Distances}

A fundamental assumption of the kinematic distance method is that the
chosen GRM, which gives the Galactic orbital speed, \(\Theta\), at all
Galactocentric radii, \(R\), accurately models the Galaxy. Several
different techniques have been employed to derive \(\Theta(R)\), for
example the tangent point method
\citep[e.g.,][]{mcclure-griffiths2007} or using the full phase-space
kinematics of masers associated with HMSFRs
\citep[e.g.,][]{reid2014}. The former method is only reliable in the
inner-Galaxy (within the Solar orbit) whereas the latter method works
across the entire Galactic disk. \citet{reid2016b} demonstrated that
both methods predict similar rotation curves in the inner-Galaxy.

\begin{figure}[ht]
  \centering
  \includegraphics[width=0.8\linewidth]{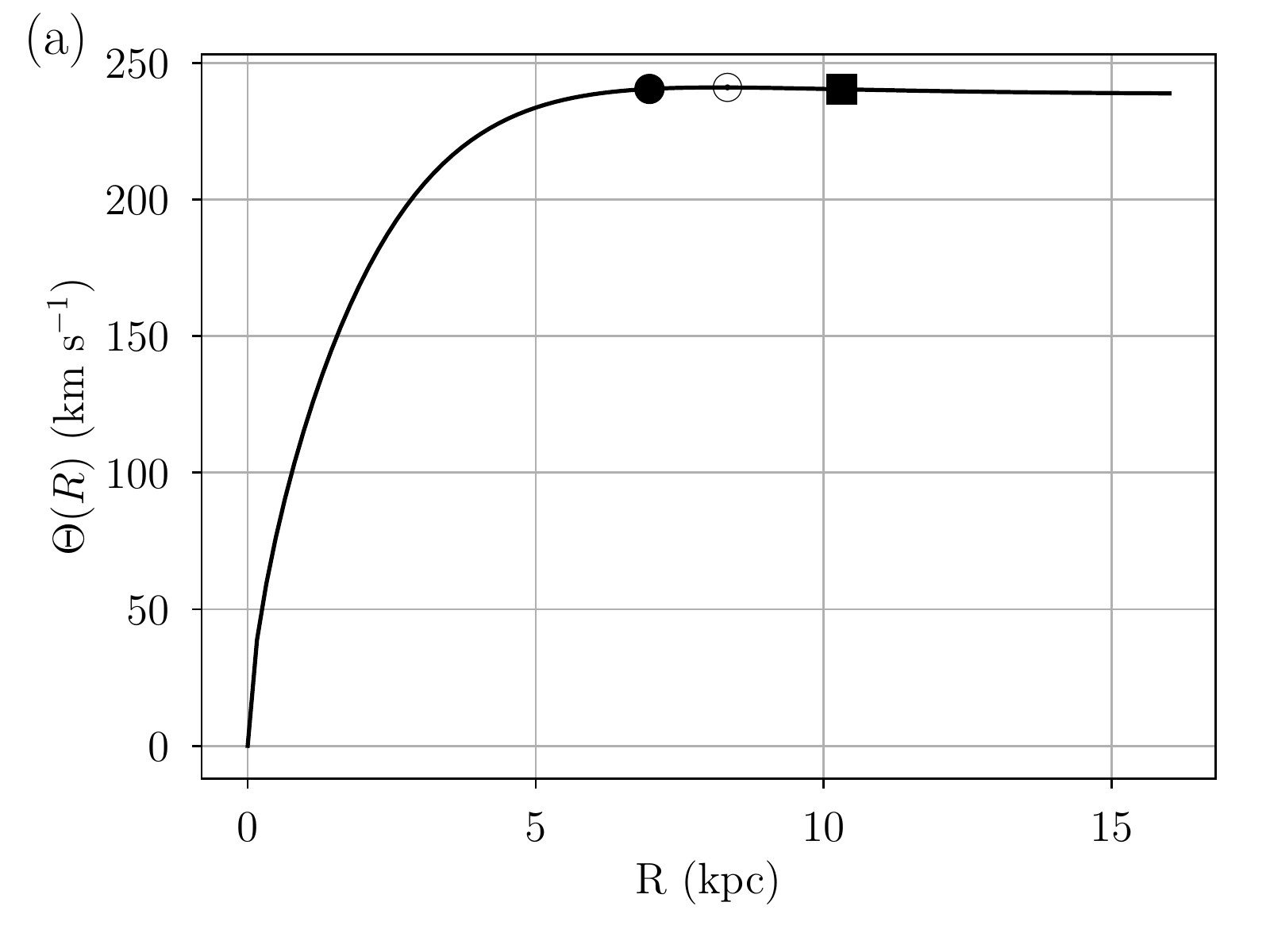} \\
  \includegraphics[width=0.8\linewidth]{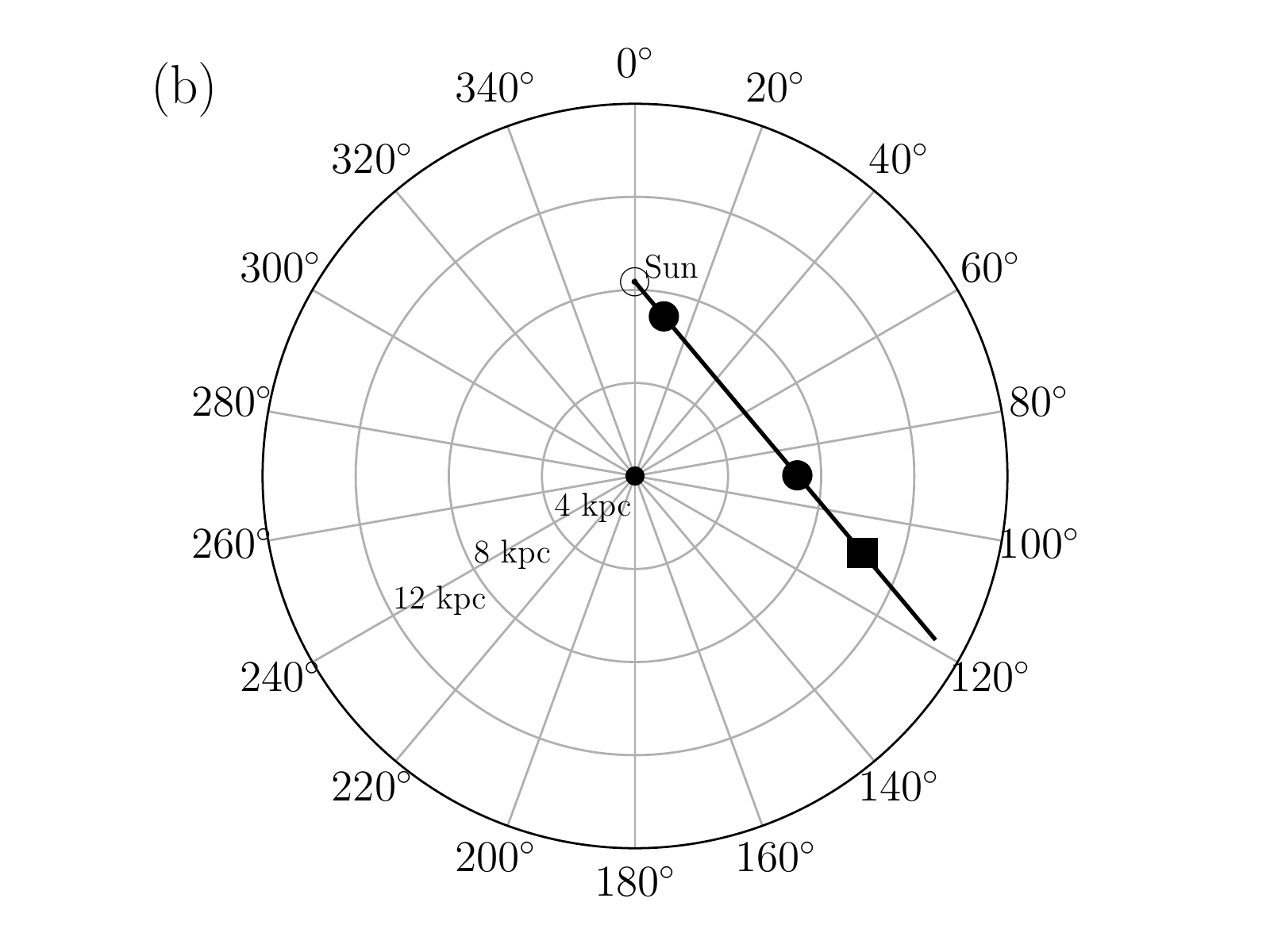} \\
  \includegraphics[width=0.8\linewidth]{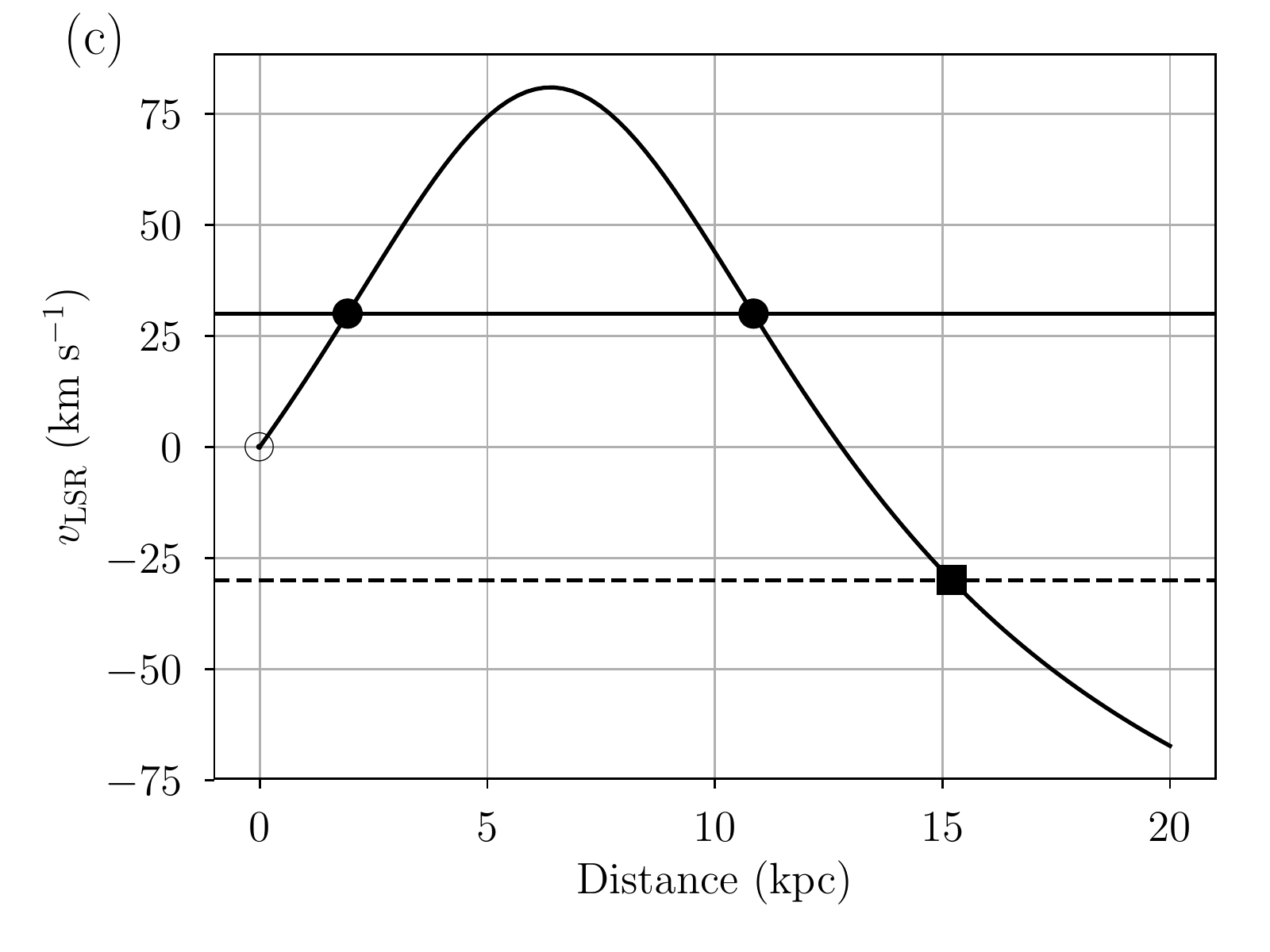}
  \caption{Schematic of the kinematic distance technique. Panel (a) is
    the \citet{reid2014} rotation curve. Panel (b) is a face-on view
    of the Galaxy with the Galactic Center located at the center and
    the Sun located 8.34 kpc in the direction \(\az = 0^\circ\).  The
    concentric circles are 4, 8, and 12 kpc in \(R\) and \(\az\) is
    given in degrees. The solid line is a line of sight through the
    Galaxy with \(\gl = 40^\circ\).  Panel (c) is the LSR velocity
    profile along this line-of-sight. An object with \(V_{\rm LSR} =
    30\kms\) in this direction (solid horizontal line) is an
    inner-Galaxy object and has two possible kinematic distances
    (black circles).  An object with \(V_{\rm LSR} = -30\kms\) (dashed
    horizontal line) is an outer-Galaxy object and has only one
    possible kinematic distance (black square). Open circles show the
    location of the Sun. }
  \label{fig:kinematic_distances}
\end{figure}

The GRM rotation curve is used to transform the Galactic longitude,
Galactic latitude, distance space (\(\gl,b,d\)) to Galactic longitude,
Galactic latitude, LSR velocity space (\(\gl,b,\vlsr\)). A
schematic of the kinematic distance technique is shown in
Figure~\ref{fig:kinematic_distances}.

Many studies have shown that HMSFRs in the Milky Way do not have
perfectly circular orbits; there are significant non-circular motions
due to streaming in the vicinity of the Galactic bars and spiral arms
\citep[e.g.,][]{burton1971,gomez2006,reid2009b,reid2014}. These
streaming motions compromise the accuracy of kinematic distances in a
complicated, uncertain way and are typically not accounted for in the
derivation of kinematic distances.

\begin{figure}[ht]
  \centering
  \includegraphics[width=\linewidth]{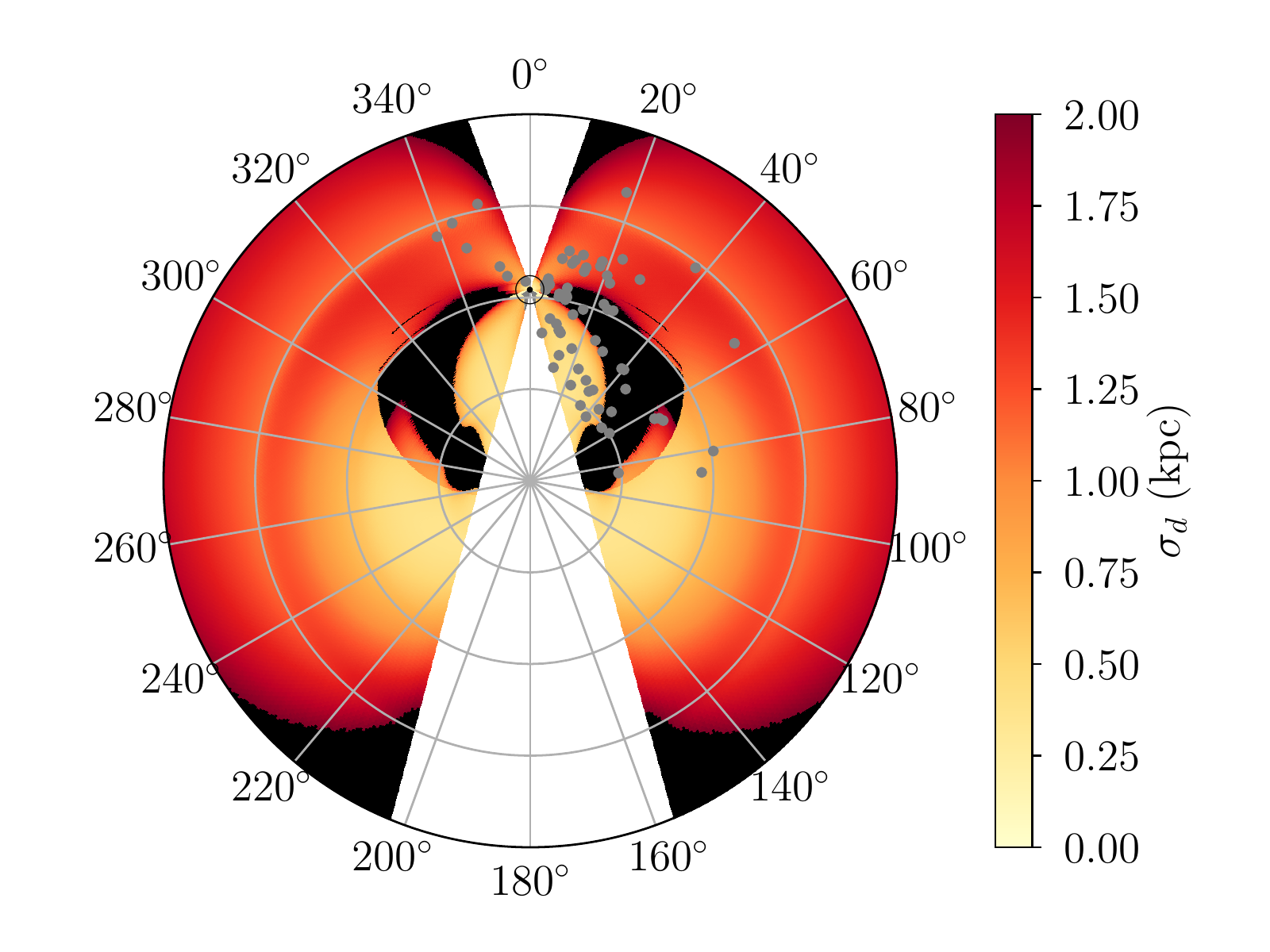} \\
  \includegraphics[width=\linewidth]{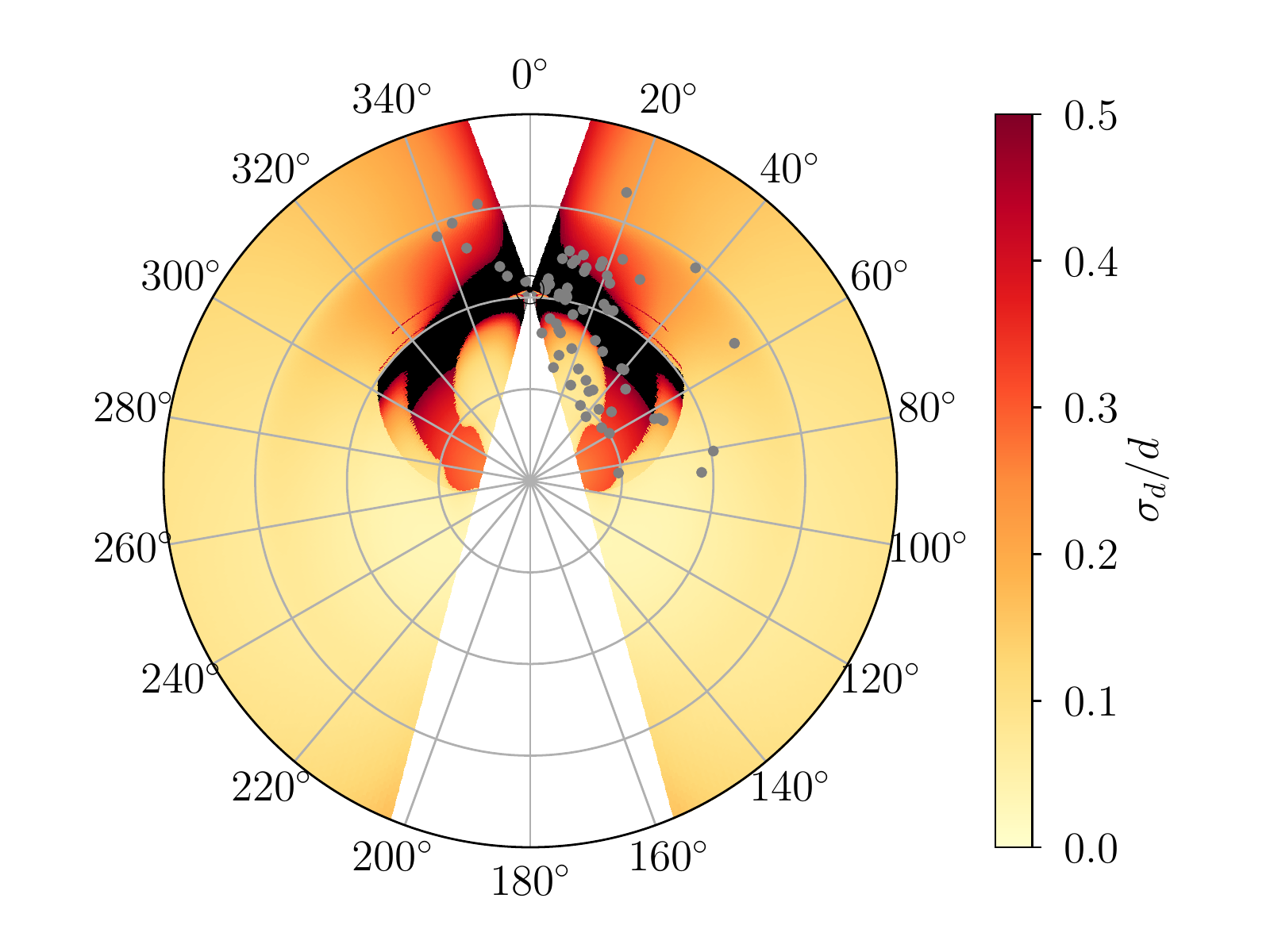}
  \caption{Face-on Galactic view of the A12 kinematic distance
    uncertainty model. The top panel is the absolute distance
    uncertainty and the bottom panel is the fractional distance
    uncertainty. The Galactic Center is located at the origin and the
    Sun is located 8.34 kpc in the direction \(\az = 0^\circ\). The
    concentric circles are 4, 8, and 12 kpc in \(R\) and \(\az\) is
    given in degrees. The color represents the distance
    uncertainty. The regions \(-15^\circ < \gl < 15^\circ\) and
    \(160^\circ < \gl < 200^\circ\) are masked (white) since kinematic
    distances are very inaccurate towards the Galactic Center and
    Galactic Anti-center. The black regions represent distance
    uncertainties greater than \(\sigma_d = 2\kpc\) (top) or
    \(\sigma_d/d = 0.5\) (bottom). The gray points are the HMSFRs in
    our sample.}
  \label{fig:lda_uncertainty}
\end{figure}

A face-on view of the \citet[][hereafter A12]{anderson2012} kinematic
distance uncertainty model is shown in
Figure~\ref{fig:lda_uncertainty}. The A12 model includes uncertainties
that stem from: (1) the variation in kinematic distances when using
different GRMs; (2) the adopted values of the Solar Galactocentric
Radius, \(R_0\), and Solar circular orbit speed, \(\Theta_0\); and,
(3) including a global \(7\kms\) streaming motion uncertainty.  This
streaming motion uncertainty is an estimate of the true global
streaming motion uncertainty which may be between 5 and \(10\kms\)
\citep{burton1966}. They did not, however, consider uncertainties with
the GRMs or in the Solar motion parameters that define the LSR.

Here we discuss three methods for calculating kinematic distances: the
traditional method using the \citet{brand1993} GRM (Method A), the
traditional method using updated Solar motion parameters and the
\citet{reid2014} GRM (Method B), and a new Monte Carlo technique using
the \citet{reid2014} GRM (Method C).

\subsection{Method A: Traditional Method, \citet{brand1993} GRM}

The traditional method for calculating kinematic distances uses a GRM
and the measured position and LSR velocity, (\(\gl,b,\vlsr\)), of an
object to determine the distance(s) that correspond to the measured
LSR velocity.  This is typically accomplished by finding the minimum
difference between the GRM LSR velocity and the measured LSR velocity
(see Figure~\ref{fig:kinematic_distances}).

We derive the Method A kinematic distances for our sample of HMSFRs
using the \citet{brand1993} GRM and the uncertainty model from
A12. This rotation curve and uncertainty model provide the kinematic
distances and distance uncertainties listed in the \textit{WISE}
Catalog. We resolve the KDA by finding the kinematic distance closest
to the parallax distance. If the region has an LSR velocity within
\(20\kms\) of the tangent point velocity, we assign it to the tangent
point. A12 used a similar tangent point strategy, but with a velocity
cutoff of \(10\kms\). Our \(20\kms\) cutoff is more conservative and
is more consistent with the GRM uncertainties discussed in the
following sections.

\subsection{Method B: Updated Solar Motion Parameters, \citet{reid2014} GRM}

In 1985, the LSR was \textit{defined} by the International
Astronomical Union Commission 33 as \(220\kms\) in the direction
\((\ell,b) = (90^\circ,0^\circ)\) with a Solar non-circular motion of
\(20\kms\) in the direction \(\alpha=18^{\rm h}\), \(\delta =
+30^\circ\) (1900) \citep{kerr1986}.  Precessing to the modern epoch
(J2000), the Solar non-circular motion is defined in Galactic
Cartesian coordinates as \(U_\odot^{\rm Std} = 10\kms\) in the
direction of the GC, \(V_\odot^{\rm Std} = 15\kms\) in the direction
of the Solar orbit, and \(W_\odot^{\rm Std} = 7\kms\) in the direction
of the North Galactic Pole. Since this definition was adopted, many
authors have published more accurate derivations of the Solar
non-circular motion parameters. For example, \citet{reid2014} derived
updated Solar motion parameters by fitting a \citet{persic1996}
universal rotation curve to the full phase-space kinematics of a
sample of maser parallaxes and proper motions towards HMSFRs.  The
\citet{persic1996} universal rotation curve is a physically-motivated
GRM, rather than an empirical model, that includes the gravitational
potential of both the disk and halo. The \citet{persic1996} universal
rotation curve is given by
\begin{equation}
  \Theta(R) = a_1 \left[\frac{1.97\beta x^{1.22}}{\left(x^2 + 0.78^2\right)^{1.43}} + \left(1-\beta\right)x^2\frac{1+a_3^2}{x^2 + a_3^2}\right]^{1/2}\label{eq:persic_rotcurve}
\end{equation}
where \(x = R/(a_2R_0)\) and \(\beta = 0.72 +
0.44\log_{10}[(a3/1.5)^5]\). Here, \(a_1\), \(a_2\), and \(a_3\) are
the parameters fit by \citet{reid2014}. These parameters, as well as
the updated Solar motion parameters fit by \citet{reid2014}, are
listed in Table~\ref{tab:reid2014_rotcurve}.

\begin{deluxetable}{lr@{ \(\pm\!\) }l}
\tablewidth{0pt}
\tabletypesize{\normalsize}
\tablecaption{Universal Rotation Curve Parameters from \citet{reid2014} \label{tab:reid2014_rotcurve}}
\tablehead{
\colhead{Parameter} & \multicolumn{2}{c}{Value} 
}
\decimals
\startdata
\(U_\odot^{\rm Rev}\) (km s\(^{-1}\)) & 10.5 & 1.7 \\
\(V_\odot^{\rm Rev}\) (km s\(^{-1}\)) & 14.4 & 6.8 \\
\(W_\odot^{\rm Rev}\) (km s\(^{-1}\)) & 8.9 & 0.9 \\
\(R_0\) (kpc) & 8.31 & 0.16 \\
\(a_1\) (km s\(^{-1}\)) & 241 & 8 \\
\(a_2\) & 0.90 & 0.06 \\
\(a_3\) & 1.46 & 0.16 \\
\enddata

\tablecomments{These rotation curve parameters are derived for the
  \citet{persic1996} universal rotation curve (see
  equation~\ref{eq:persic_rotcurve}).}

\end{deluxetable}

To correct the LSR velocities in our sample for the updated Solar
non-circular motion parameters, we first convert the measured LSR
velocity to a heliocentric velocity via
\begin{equation}
\begin{aligned}
  V_{\rm helio} & = V_{\rm LSR} - \left(U_\odot^{\rm Std}\cos\gl + V_\odot^{\rm Std}\sin\gl\right)\cos b \\
  & - W_\odot^{\rm Std}\sin b.\label{eq:helio}
\end{aligned}
\end{equation}
Next, we use the \citet{reid2014} Solar motion parameters to derive
the revised LSR velocity, \(V_{\rm LSR}^{\rm Rev}\):
\begin{equation}
\begin{aligned}
  V_{\rm LSR}^{\rm Rev} & = V_{\rm helio} + \left(U_\odot^{\rm Rev}\cos\gl + V_\odot^{\rm Rev}\sin\gl\right)\cos b \\
  & + W_\odot^{\rm Rev}\sin b. \label{eq:vlsr}
\end{aligned}
\end{equation}
The uncertainty in this LSR velocity (\(\sigma_V^{\rm Rev}\)) includes
contributions from the uncertainty in the measured LSR velocity
(\(\sigma_V\)) and the uncertainties in the \citet{reid2014} Solar
motion parameters, \(\sigma_{U\odot}^{\rm Rev},\sigma_{V\odot}^{\rm
  Rev},\sigma_{W\odot}^{\rm Rev}\). The combined uncertainty in the
revised LSR velocity is
\begin{equation}
\begin{aligned}
  {\sigma_V^{\rm Rev}}^2 & = \sigma_V^2 + \left(\sigma_{U\odot}^{\rm Rev}\cos\gl\cos b\right)^2 + \left(\sigma_{V\odot}^{\rm Rev}\sin\gl\cos b\right)^2 \\
  & + \left(\sigma_{W\odot}^{\rm Rev}\sin b\right)^2 \label{eq:velocity_uncertainty}
\end{aligned}
\end{equation}
For simplicity, we ignore the cross-terms between the Solar motion
parameter uncertainties. Including these cross-terms would have little
effect since \citet{reid2014} finds that the magnitude of the Pearson
product-moment correlation coefficients between these parameters is
small, ranging between \(0.011\) and \(0.017\).

To compute the Method B kinematic distances to our sample of HMSFRs,
we use the \citet{reid2014} fits to the \citet{persic1996} universal
rotation curve and these revised LSR velocities. As before, we assign
the near or far KDAR by determining which kinematic distance is
closest to the parallax distance. If the HMSFR has an LSR velocity
within \(20\kms\) of the tangent point velocity, we assign it to the
tangent point distance. The Method B kinematic distance uncertainties
are again determined by the A12 kinematic distance uncertainty model.

\begin{figure}[ht]
  \includegraphics[width=\linewidth]{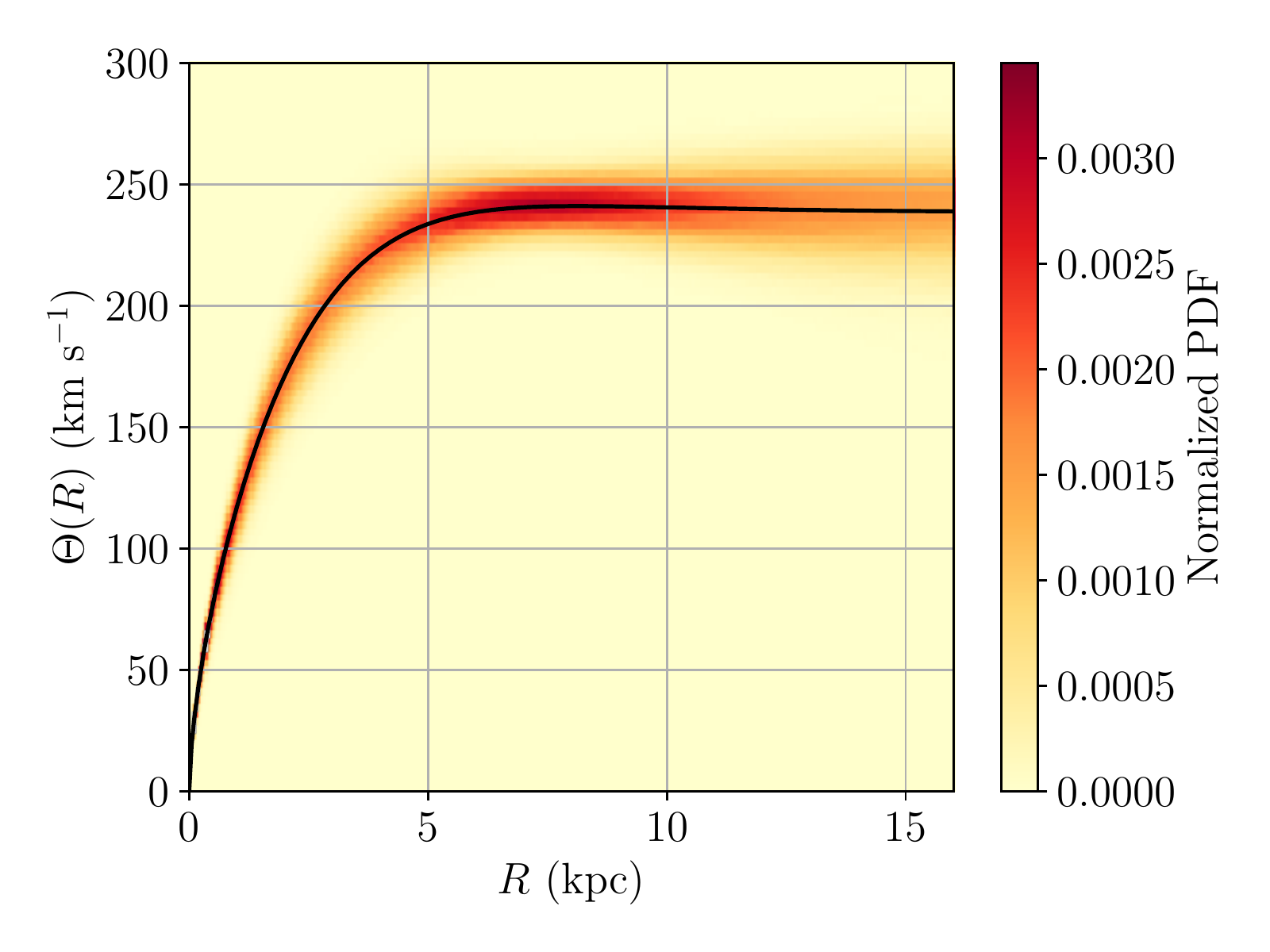}
  \caption{The \citet{reid2014} universal rotation curve. The solid
    line is the nominal rotation curve using the parameters listed in
    Table~\ref{tab:reid2014_rotcurve}. The colors represent the
    probability distribution function (PDF) derived by Monte Carlo
    re-sampling the rotation curve parameters within their
    uncertainties.}
  \label{fig:reid2014_rotcurve}
\end{figure}

\subsection{Method C: Monte Carlo Method, \citet{reid2014} GRM}

Here we develop a method to derive kinematic distances and their
uncertainties in a more statistically robust way. With this method we
re-sample all measured and derived parameters within their
uncertainties and determine the probability distribution function
(PDF) of kinematic distances.

We first correct the measured LSR velocities as described above. We
then re-sample the revised LSR velocities from a normal distribution
centered on the nominal revised LSR velocity, \(\vlsr^{\rm Rev}\),
with a width \(\sigma_V^{\rm Rev}\). The width of this distribution is
the total revised LSR velocity uncertainty, which includes both the
measured uncertainty and the uncertainties in the Solar motion
parameters.

We also re-sample the \citet{reid2014} universal Galactic rotation
curve parameters, including \(R_0\), from a normal distribution
centered on the nominal values and a width equal to the uncertainty
(see Equation~\ref{eq:persic_rotcurve} and
Table~\ref{tab:reid2014_rotcurve}). The variation of this re-sampled
rotation curve is shown in Figure~\ref{fig:reid2014_rotcurve}.  Unlike
A12, we do not add any additional streaming motion uncertainty into
these calculations because the derivation of the \citet{reid2014}
rotation curve inherently includes uncertainties due to streaming
motions.

\begin{figure*}[ht]
  \centering
  \includegraphics[width=0.8\linewidth]{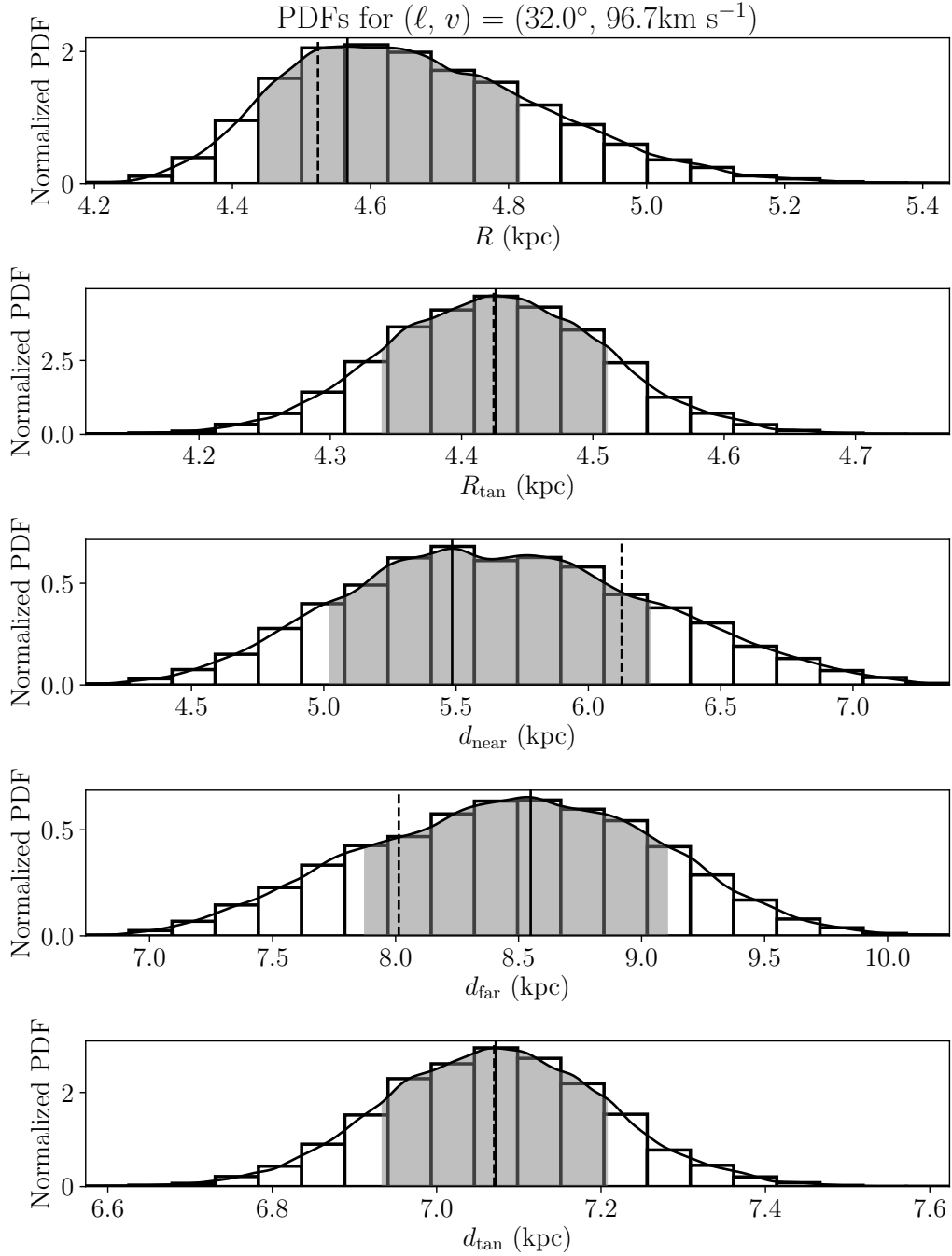}
  \caption{Normalized probability distribution functions (PDFs) of the
    Method C kinematic distances derived for G032.04+00.05,
    \((\gl,V_{\rm LSR}) = (32.0^\circ, 96.7\,\kms)\). Shown from top
    to bottom are the kinematic distance PDFs for: Galactocentric
    radius, \(R\), Galactocentric radius of tangent point, \(R_{\rm
      tan}\), near kinematic distance, \(d_{\rm near}\), far kinematic
    distance, \(d_{\rm far}\), and tangent point kinematic distance,
    \(d_{\rm tan}\). The PDFs are determined by Monte Carlo
    resampling the \citet{reid2014} rotation curve within the
    uncertainties in the rotation curve parameters and then deriving
    the kinematic distances.  The solid curve is the kernel density
    estimation (KDE) derived using the linear combination technique
    from \citet{jones1993}. The dashed vertical line is the distance
    derived using the ``traditional'' kinematic techniques whereas the
    solid vertical line is the peak of the KDE. The gray region is the
    \(68.3\%\) confidence interval.}
  \label{fig:pdf_example_0lag_0stream}
\end{figure*}

By re-sampling the above parameters \(10^5\) times for each HMSFR, we
derive the kinematic distance PDF for each object (see example for
G032.04+00.05 in Figure~\ref{fig:pdf_example_0lag_0stream}).  Each
panel in Figure~\ref{fig:pdf_example_0lag_0stream} represents one
distance we derive: the Galactocentric radius, \(R\), the
Galactocentric radius of the tangent point, \(R_{\rm tan}\), the near
kinematic distance, \(d_{\rm near}\), the far kinematic distance,
\(d_{\rm far}\), and the tangent point kinematic distance, \(d_{\rm
  tan}\). The shapes of the PDFs are determined by the uncertainties
in the LSR velocities, the Galactocentric radius of the Solar orbit,
and the parameters of the rotation curve model.

We fit each PDF with a kernel density estimator (KDE) derived using
the linear combination technique from \citet{jones1993}.  The peak of
this KDE is the most probable kinematic distance, and the width and
shape of the KDE describe the range of possible kinematic
distances. Similar to how we defined parallax distances, we define the
Method C kinematic distance as the peak of the KDE. The uncertainty in
this distance is the 68.3\% confidence interval.

We resolve the KDAR in the same way as in the previous two methods. If
the object has a velocity in excess of the magnitude of the tangent
point velocity, then the uncertainty in the tangent point distance is
the formal Monte Carlo uncertainty (i.e., the 68.3\% confidence
interval).  If the object's velocity is smaller than the magnitude of
the tangent point velocity but still within \(20\kms\), then the
tangent point distance uncertainty is the total range from the near
distance to the far distance. In this velocity range, traditional KDAR
techniques are inaccurate and the object could be anywhere between the
near and far kinematic distance (A12).

\begin{figure*}[ht]
  \centering
  \includegraphics[width=0.45\linewidth]{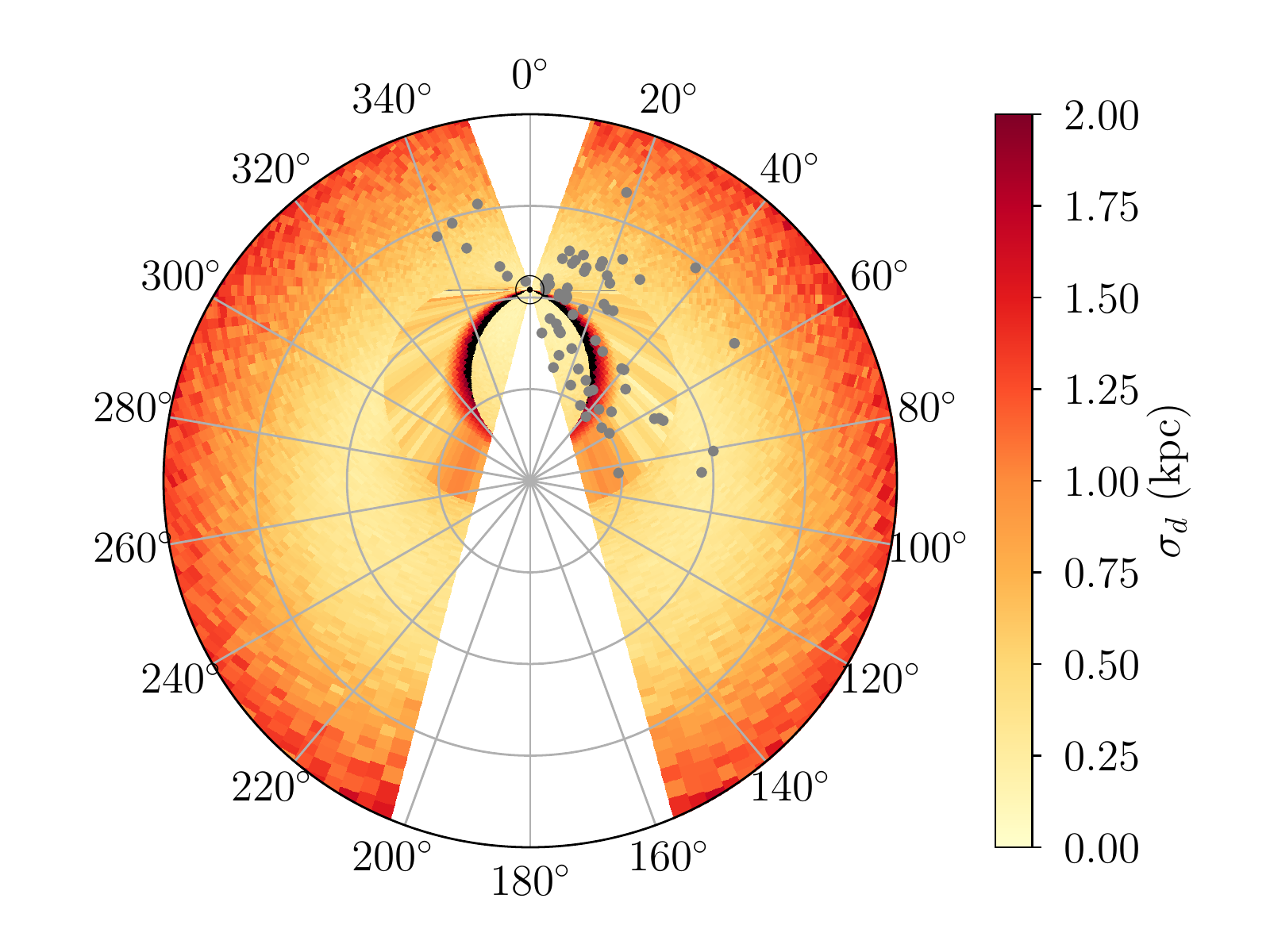}
  \includegraphics[width=0.45\linewidth]{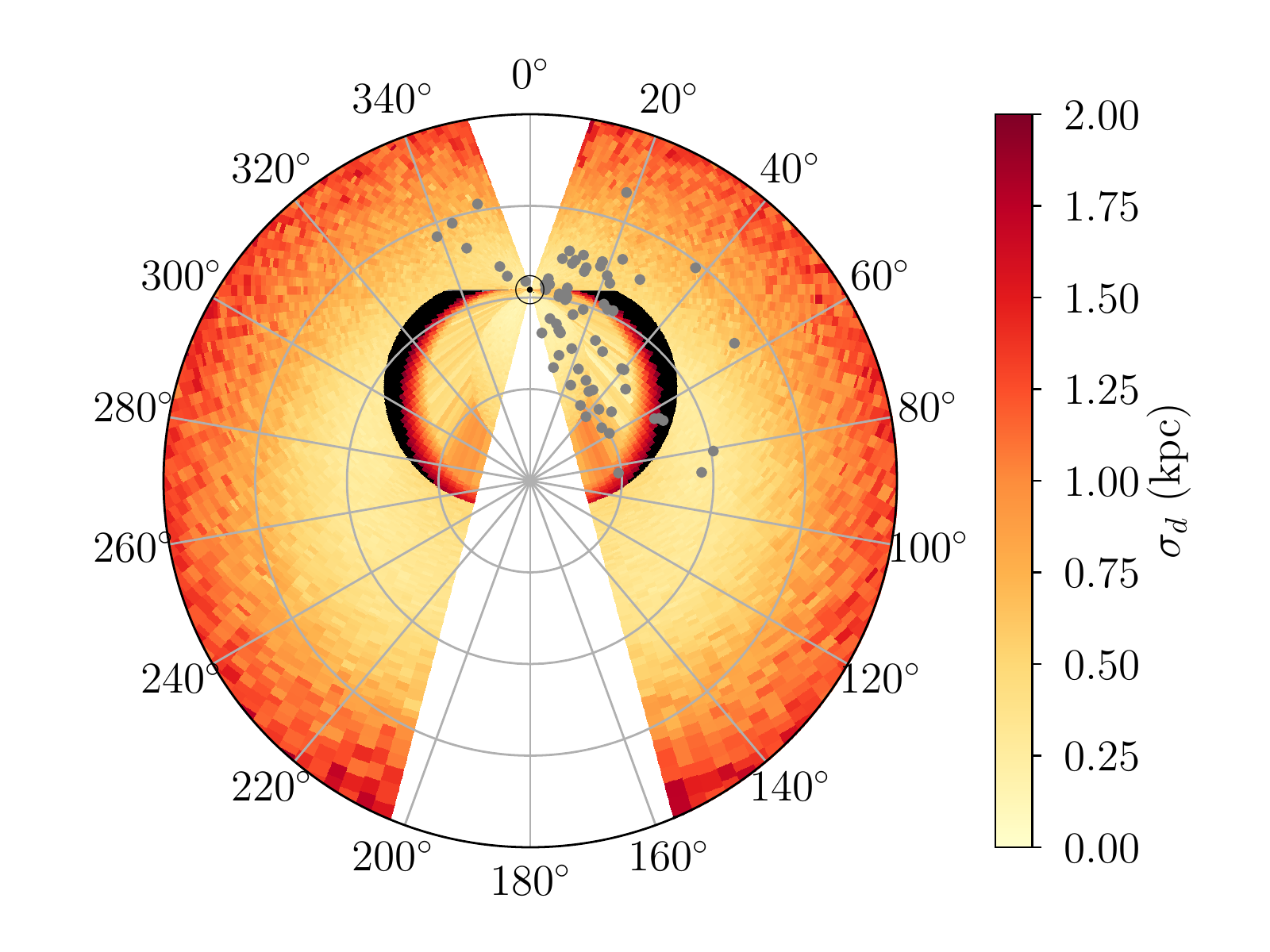} \\
  \includegraphics[width=0.45\linewidth]{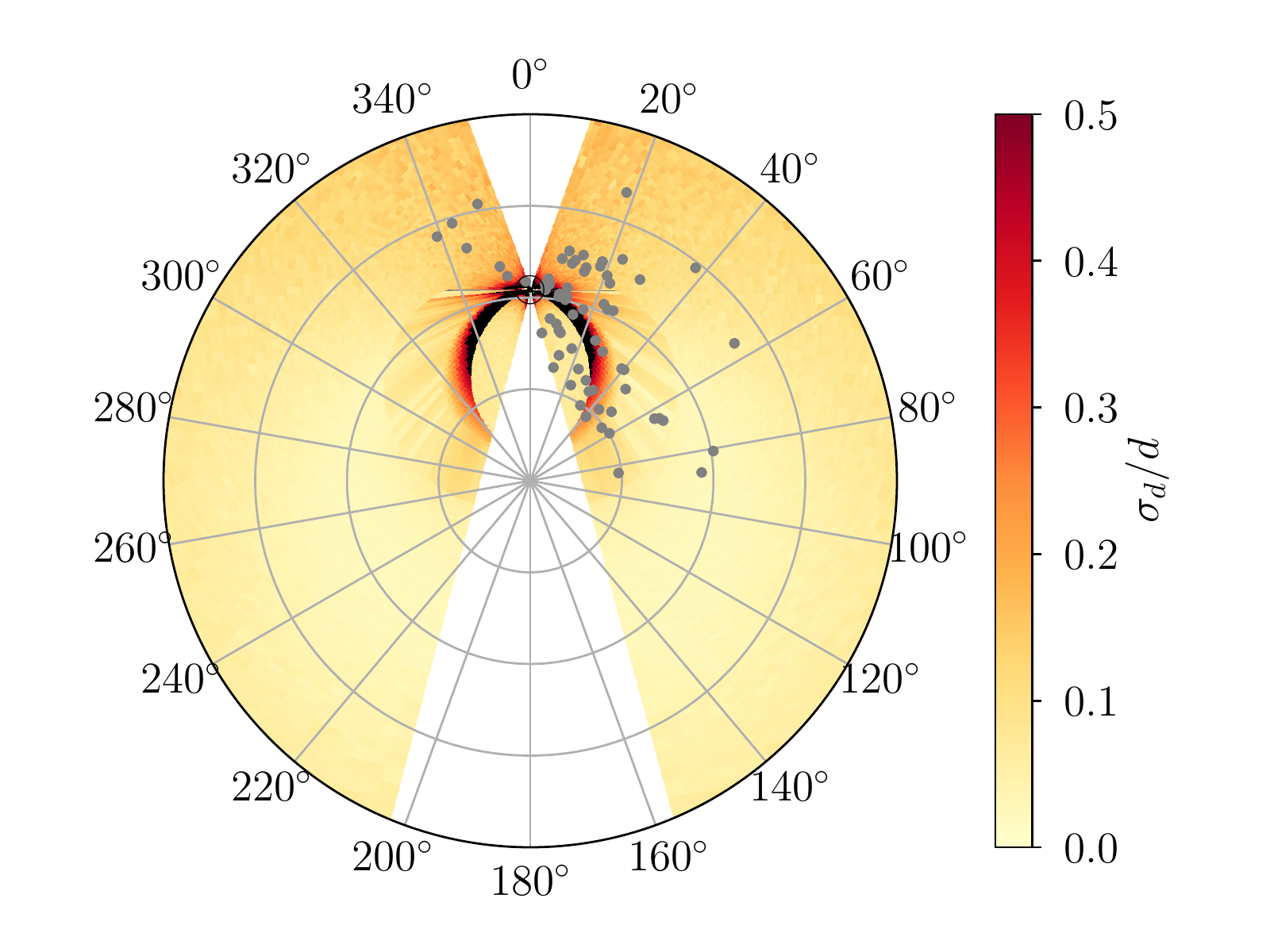}
  \includegraphics[width=0.45\linewidth]{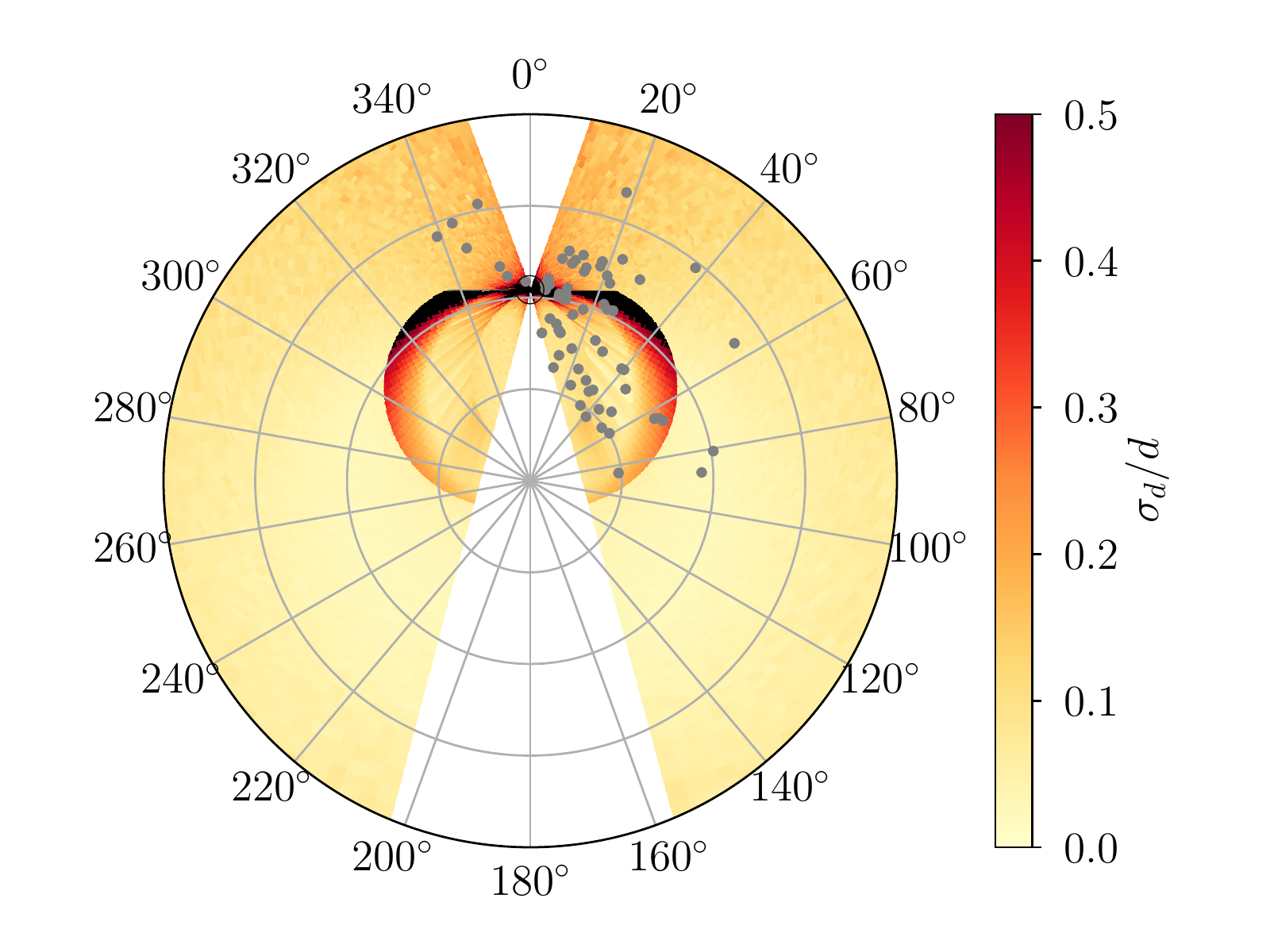} \\
  \caption{Face-on Galactic view of the Monte Carlo kinematic distance
    uncertainties. These uncertainties are not symmetric; the left
    panels are the uncertainty in the negative direction (toward the
    Sun) and the right panels are the uncertainty in the positive
    direction (away from the Sun). The top panels are the absolute
    distance uncertainties and the bottom panels are the fractional
    distance uncertainties. The Galactic Center is located at the
    origin and the Sun is located 8.34 kpc in the direction \(\az =
    0^\circ\). The concentric circles are 4, 8, and 12 kpc in \(R\)
    and \(\az\) is given in degrees. The color represents the distance
    uncertainty. The regions \(-15^\circ < \gl < 15^\circ\) and
    \(160^\circ < \gl < 200^\circ\) are masked (white) since kinematic
    distances are very inaccurate towards the Galactic Center and
    Galactic Anti-center. The black regions represent distance
    uncertainties greater than \(\sigma_d = 2\kpc\) (top) or
    \(\sigma_d/d = 0.5\) (bottom). The gray points are the HMSFRs in
    our sample.}
  \label{fig:pdf_uncertainty}
\end{figure*}

A face-on Galactic view of the Monte Carlo kinematic distance
uncertainties is shown in Figure~\ref{fig:pdf_uncertainty}. To
construct these maps we use the Monte Carlo technique to compute the
kinematic distance and distance uncertainties in bins of \(2^\circ\)
in Galactic longitude and \(2\kms\) in velocity, with \(10^{4}\) Monte
Carlo samples in each bin.  Since the kinematic distance uncertainties
derived in this method are not symmetric, we show the uncertainties in
both the positive direction (away from the Sun) and the negative
direction (toward the Sun).

\startlongtable
\begin{deluxetable*}{llrcrcrc}
\tablewidth{0pt}
\tabletypesize{\normalsize}
\tablecaption{Monte Carlo Parallax and Derived Kinematic Distances\label{tab:final_distances}}
\tablehead{
\colhead{Name} & \colhead{\(D_P\)} & \colhead{\(D_A\)} & \colhead{KDAR\(_A\)} & \colhead{\(D_B\)} & \colhead{KDAR\(_B\)} & \colhead{\(D_C\)} & \colhead{KDAR\(_C\)} \\ 
\colhead{} & \colhead{(kpc)} & \colhead{(kpc)} & \colhead{} & \colhead{(kpc)} & \colhead{} & \colhead{(kpc)} & \colhead{}
}
\decimals
\startdata
G015.03\(-\)00.67 & \(1.96^{+0.15}_{-0.11}\) & \(2.47\pm0.60\) & N & \(2.33\pm0.57\) & N & \(2.39^{+0.31}_{-0.41}\) & N \\
G016.58\(-\)00.05 & \(3.54^{+0.33}_{-0.27}\) & \(4.60\pm0.36\) & N & \(4.56\pm0.36\) & N & \(4.49^{+0.50}_{-0.40}\) & N \\
G023.00\(-\)00.41 & \(4.53^{+0.38}_{-0.35}\) & \(5.01\pm0.38\) & N & \(4.96\pm0.38\) & N & \(4.80^{+0.56}_{-0.32}\) & N \\
G023.44\(-\)00.18 & \(5.50^{+1.25}_{-0.94}\) & \(5.71\pm0.45\) & N & \(7.65\pm0.60\) & T & \(7.65^{+1.91}_{-2.04}\) & T \\
G023.65\(-\)00.12 & \(3.12^{+0.45}_{-0.39}\) & \(5.11\pm0.39\) & N & \(5.07\pm0.39\) & N & \(4.92^{+0.56}_{-0.36}\) & N \\
G023.70\(-\)00.19 & \(6.05^{+0.93}_{-0.93}\) & \(4.68\pm0.37\) & N & \(4.58\pm0.37\) & N & \(4.58^{+0.45}_{-0.49}\) & N \\
G025.70+00.04 & \(8.88^{+3.34}_{-1.91}\) & \(9.78\pm0.46\) & F & \(9.50\pm0.45\) & F & \(9.60^{+0.58}_{-0.62}\) & F \\
G027.36\(-\)00.16 & \(6.79^{+3.05}_{-1.69}\) & \(5.52\pm0.48\) & N & \(7.41\pm0.65\) & T & \(5.16^{+0.75}_{-0.32}\) & N \\
G028.86+00.06 & \(7.16^{+1.15}_{-0.89}\) & \(7.44\pm0.83\) & T & \(7.30\pm0.82\) & T & \(7.30^{+1.73}_{-1.80}\) & T \\
G029.86\(-\)00.04 & \(6.02^{+0.94}_{-0.68}\) & \(7.37\pm1.01\) & T & \(7.23\pm1.00\) & T & \(7.24^{+1.46}_{-1.52}\) & T \\
G029.95\(-\)00.01 & \(5.12^{+0.63}_{-0.46}\) & \(7.36\pm0.92\) & T & \(7.23\pm0.90\) & T & \(7.24^{+1.56}_{-1.64}\) & T \\
G031.28+00.06 & \(4.05^{+0.95}_{-0.54}\) & \(7.26\pm2.40\) & T & \(7.13\pm2.35\) & T & \(7.10^{+0.91}_{-0.94}\) & T \\
G031.58+00.07 & \(4.64^{+0.87}_{-0.58}\) & \(7.24\pm2.59\) & T & \(7.10\pm2.55\) & T & \(7.11^{+1.59}_{-1.66}\) & T \\
G032.04+00.05 & \(5.16^{+0.24}_{-0.20}\) & \(7.20\pm2.57\) & T & \(7.07\pm2.52\) & T & \(7.07^{+1.48}_{-1.59}\) & T \\
G033.64\(-\)00.22 & \(6.40^{+0.83}_{-0.66}\) & \(3.88\pm0.43\) & N & \(3.65\pm0.41\) & N & \(3.56^{+0.48}_{-0.35}\) & N \\
G034.39+00.22 & \(1.53^{+0.15}_{-0.10}\) & \(3.71\pm0.44\) & N & \(3.48\pm0.41\) & N & \(3.37^{+0.54}_{-0.41}\) & N \\
G035.02+00.34 & \(2.30^{+0.22}_{-0.22}\) & \(3.42\pm0.44\) & N & \(3.19\pm0.41\) & N & \(3.17^{+0.47}_{-0.43}\) & N \\
G035.19\(-\)00.74 & \(2.17^{+0.24}_{-0.21}\) & \(2.06\pm0.50\) & N & \(1.93\pm0.47\) & N & \(1.87^{+0.59}_{-0.46}\) & N \\
G035.20\(-\)01.73 & \(3.14^{+0.56}_{-0.42}\) & \(2.82\pm0.46\) & N & \(2.62\pm0.43\) & N & \(2.60^{+0.35}_{-0.35}\) & N \\
G037.43+01.51 & \(1.88^{+0.07}_{-0.08}\) & \(2.75\pm0.48\) & N & \(2.56\pm0.45\) & N & \(2.50^{+0.41}_{-0.33}\) & N \\
G043.16+00.01 & \(10.93^{+0.94}_{-0.79}\) & \(11.79\pm0.71\) & F & \(11.52\pm0.70\) & F & \(11.40^{+0.65}_{-0.42}\) & F \\
G043.79\(-\)00.12 & \(6.01^{+0.19}_{-0.19}\) & \(3.09\pm0.60\) & N & \(2.85\pm0.55\) & N & \(9.16^{+0.92}_{-0.72}\) & F \\
G043.89\(-\)00.78 & \(7.82^{+1.67}_{-1.06}\) & \(6.13\pm1.12\) & T & \(6.01\pm1.10\) & T & \(6.01^{+2.50}_{-2.40}\) & T \\
G045.07+00.13 & \(7.95^{+0.37}_{-0.28}\) & \(6.00\pm3.04\) & T & \(5.89\pm2.98\) & T & \(5.91^{+1.96}_{-1.86}\) & T \\
G045.45+00.05 & \(8.14^{+1.30}_{-1.16}\) & \(5.96\pm1.40\) & T & \(5.85\pm1.37\) & T & \(5.84^{+2.31}_{-2.22}\) & T \\
G048.60+00.02 & \(10.65^{+0.63}_{-0.53}\) & \(10.02\pm0.75\) & F & \(9.83\pm0.73\) & F & \(9.83^{+0.58}_{-0.54}\) & F \\
G049.19\(-\)00.33 & \(5.27^{+0.21}_{-0.20}\) & \(5.55\pm2.57\) & T & \(5.45\pm2.55\) & T & \(5.44^{+0.97}_{-0.95}\) & T \\
G049.48\(-\)00.36 & \(4.15^{+2.40}_{-1.10}\) & \(5.52\pm2.91\) & T & \(5.42\pm2.91\) & T & \(5.43^{+1.37}_{-1.41}\) & T \\
G049.48\(-\)00.38 & \(5.38^{+0.32}_{-0.29}\) & \(5.52\pm2.85\) & T & \(5.42\pm2.84\) & T & \(5.42^{+1.29}_{-1.15}\) & T \\
G052.10+01.04 & \(3.60^{+1.27}_{-0.72}\) & \(5.22\pm3.44\) & T & \(5.12\pm3.45\) & T & \(5.15^{+1.94}_{-1.91}\) & T \\
G059.78+00.06 & \(2.16^{+0.10}_{-0.10}\) & \(4.28\pm4.32\) & T & \(4.20\pm2.82\) & T & \(4.20^{+2.14}_{-2.16}\) & T \\
G069.54\(-\)00.97 & \(2.46^{+0.08}_{-0.08}\) & \(2.97\pm5.57\) & T & \(2.91\pm5.87\) & T & \(2.91^{+2.05}_{-2.09}\) & T \\
G074.03\(-\)01.71 & \(1.59^{+0.05}_{-0.04}\) & \(2.34\pm4.15\) & T & \(2.29\pm0.98\) & T & \(2.30^{+2.12}_{-2.05}\) & T \\
G075.29+01.32 & \(9.21^{+0.47}_{-0.43}\) & \(10.91\pm1.18\) & F & \(9.75\pm1.02\) & F & \(9.61^{+1.09}_{-0.93}\) & F \\
G075.76+00.33 & \(3.48^{+0.31}_{-0.26}\) & \(2.09\pm0.50\) & T & \(2.05\pm0.47\) & T & \(2.06^{+3.30}_{-2.05}\) & T \\
G075.78+00.34 & \(3.73^{+0.49}_{-0.42}\) & \(2.09\pm0.82\) & T & \(2.05\pm0.73\) & T & \(2.04^{+2.54}_{-2.03}\) & T \\
G076.38\(-\)00.61 & \(1.29^{+0.10}_{-0.08}\) & \(2.00\pm0.67\) & T & \(1.96\pm0.62\) & T & \(1.97^{+2.67}_{-1.96}\) & T \\
G078.12+03.63 & \(1.63^{+0.08}_{-0.08}\) & \(1.75\pm0.60\) & T & \(1.72\pm0.55\) & T & \(1.72^{+2.85}_{-1.71}\) & T \\
G078.88+00.70 & \(3.27^{+0.33}_{-0.25}\) & \(1.64\pm0.54\) & T & \(1.61\pm0.51\) & T & \(1.61^{+2.67}_{-1.60}\) & T \\
G079.73+00.99 & \(1.33^{+0.14}_{-0.10}\) & \(1.51\pm0.65\) & T & \(1.49\pm0.59\) & T & \(1.48^{+2.43}_{-1.47}\) & T \\
G079.87+01.17 & \(1.60^{+0.08}_{-0.06}\) & \(1.49\pm0.57\) & T & \(1.47\pm0.52\) & T & \(1.47^{+2.63}_{-1.46}\) & T \\
G080.79\(-\)01.92 & \(1.60^{+0.13}_{-0.13}\) & \(1.36\pm0.65\) & T & \(1.33\pm0.58\) & T & \(1.33^{+2.42}_{-1.32}\) & T \\
G080.86+00.38 & \(1.45^{+0.09}_{-0.08}\) & \(1.35\pm0.64\) & T & \(1.32\pm0.58\) & T & \(1.32^{+2.36}_{-1.31}\) & T \\
G081.75+00.59 & \(1.49^{+0.09}_{-0.07}\) & \(1.22\pm0.65\) & T & \(1.20\pm0.57\) & T & \(1.20^{+2.33}_{-1.19}\) & T \\
G081.87+00.78 & \(1.28^{+0.08}_{-0.07}\) & \(1.20\pm3.65\) & T & \(1.18\pm4.01\) & T & \(1.18^{+1.44}_{-1.17}\) & T \\
G090.21+02.32 & \(0.67^{+0.02}_{-0.02}\) & \(2.05\pm2.90\) & F & \(1.48\pm1.49\) & F & \(2.23^{+0.84}_{-1.14}\) & F \\
G092.67+03.07 & \(1.62^{+0.06}_{-0.05}\) & \(2.05\pm2.03\) & F & \(1.52\pm1.31\) & F & \(2.23^{+1.11}_{-1.31}\) & F \\
G094.60\(-\)01.79 & \(3.49^{+0.43}_{-0.35}\) & \(6.36\pm1.45\) & F & \(5.48\pm1.23\) & F & \(5.30^{+1.04}_{-0.82}\) & F \\
G095.29\(-\)00.93 & \(4.81^{+0.42}_{-0.32}\) & \(5.46\pm1.44\) & F & \(4.68\pm1.22\) & F & \(4.72^{+0.84}_{-0.99}\) & F \\
G097.53+03.18 & \(7.28^{+1.07}_{-0.86}\) & \(9.02\pm1.27\) & F & \(7.61\pm1.07\) & F & \(7.28^{+1.33}_{-0.85}\) & F \\
G100.37\(-\)03.57 & \(3.42^{+0.14}_{-0.10}\) & \(4.79\pm1.42\) & F & \(4.06\pm1.18\) & F & \(4.03^{+1.24}_{-1.24}\) & F \\
G105.41+09.87 & \(0.88^{+0.06}_{-0.04}\) & \(1.63\pm1.36\) & F & \(1.21\pm0.94\) & F & \(1.22^{+0.81}_{-0.81}\) & F \\
G107.29+05.63 & \(0.76^{+0.07}_{-0.06}\) & \(1.64\pm1.30\) & F & \(1.24\pm0.92\) & F & \(1.19^{+0.83}_{-0.69}\) & F \\
G108.18+05.51 & \(0.75^{+0.11}_{-0.08}\) & \(1.60\pm1.28\) & F & \(1.21\pm0.90\) & F & \(1.17^{+0.74}_{-0.65}\) & F \\
G108.20+00.58 & \(4.24^{+0.60}_{-0.47}\) & \(5.38\pm1.37\) & F & \(4.48\pm1.11\) & F & \(4.18^{+1.13}_{-0.68}\) & F \\
G108.47\(-\)02.81 & \(3.23^{+0.11}_{-0.10}\) & \(5.89\pm1.28\) & F & \(4.91\pm1.02\) & F & \(4.77^{+1.01}_{-0.84}\) & F \\
G108.59+00.49 & \(2.49^{+0.22}_{-0.18}\) & \(5.67\pm1.32\) & F & \(4.72\pm1.06\) & F & \(4.75^{+0.81}_{-1.05}\) & F \\
G109.87+02.11 & \(0.69^{+0.04}_{-0.04}\) & \(1.12\pm1.27\) & F & \(0.79\pm0.81\) & F & \(0.81^{+0.67}_{-0.67}\) & F \\
G111.23\(-\)01.23 & \(3.38^{+0.57}_{-0.49}\) & \(5.63\pm1.27\) & F & \(4.67\pm1.01\) & F & \(4.54^{+1.33}_{-1.13}\) & F \\
G111.25\(-\)00.76 & \(3.37^{+0.20}_{-0.17}\) & \(4.58\pm1.36\) & F & \(3.80\pm1.10\) & F & \(3.63^{+0.99}_{-0.78}\) & F \\
G111.54+00.77 & \(2.63^{+0.13}_{-0.11}\) & \(6.06\pm1.14\) & F & \(5.01\pm0.92\) & F & \(4.88^{+1.00}_{-0.92}\) & F \\
G121.29+00.65 & \(0.93^{+0.03}_{-0.04}\) & \(2.28\pm1.13\) & F & \(1.83\pm0.88\) & F & \(1.86^{+0.63}_{-0.73}\) & F \\
G122.01\(-\)07.08 & \(2.17^{+0.11}_{-0.10}\) & \(5.00\pm1.18\) & F & \(4.10\pm0.92\) & F & \(3.92^{+1.02}_{-0.73}\) & F \\
G123.06\(-\)06.30 & \(2.77^{+0.27}_{-0.22}\) & \(2.91\pm1.19\) & F & \(2.37\pm0.95\) & F & \(2.19^{+0.84}_{-0.60}\) & F \\
G123.06\(-\)06.30 & \(2.36^{+0.14}_{-0.12}\) & \(2.81\pm1.18\) & F & \(2.29\pm0.94\) & F & \(2.35^{+0.54}_{-0.71}\) & F \\
G133.94+01.06 & \(1.95^{+0.04}_{-0.04}\) & \(4.93\pm1.19\) & F & \(3.97\pm0.95\) & F & \(3.81^{+0.96}_{-0.69}\) & F \\
G134.62\(-\)02.19 & \(2.41^{+0.11}_{-0.10}\) & \(3.95\pm1.25\) & F & \(3.19\pm0.98\) & F & \(3.07^{+0.85}_{-0.77}\) & F \\
G135.27+02.79 & \(5.97^{+0.39}_{-0.43}\) & \(9.37\pm1.96\) & F & \(7.32\pm1.53\) & F & \(7.34^{+1.26}_{-1.51}\) & F \\
G209.00\(-\)19.38 & \(0.41^{+0.01}_{-0.01}\) & \(0.44\pm0.84\) & F & \(0.24\pm0.45\) & F & \(0.07^{+0.64}_{-0.06}\) & F \\
G211.59+01.05 & \(4.39^{+0.13}_{-0.15}\) & \(6.49\pm1.82\) & F & \(5.06\pm1.41\) & F & \(4.79^{+1.45}_{-1.16}\) & F \\
G229.57+00.15 & \(4.47^{+0.33}_{-0.27}\) & \(4.77\pm1.11\) & F & \(3.83\pm0.88\) & F & \(3.48^{+1.39}_{-1.09}\) & F \\
G232.62+00.99 & \(1.67^{+0.11}_{-0.10}\) & \(2.01\pm1.04\) & F & \(1.59\pm0.79\) & F & \(1.52^{+0.58}_{-0.53}\) & F \\
G236.81+01.98 & \(3.31^{+0.24}_{-0.18}\) & \(4.22\pm1.25\) & F & \(3.43\pm0.98\) & F & \(3.19^{+1.13}_{-0.70}\) & F \\
G239.35\(-\)05.06 & \(1.16^{+0.09}_{-0.08}\) & \(2.02\pm1.07\) & F & \(1.59\pm0.81\) & F & \(1.62^{+0.52}_{-0.66}\) & F \\
G240.31+00.07 & \(4.68^{+0.44}_{-0.49}\) & \(7.11\pm1.30\) & F & \(5.75\pm1.04\) & F & \(5.75^{+1.03}_{-1.13}\) & F \\
\enddata
\end{deluxetable*}

\section{Kinematic Distance Uncertainty}

We assess the accuracy of kinematic distances by comparing the
parallax and kinematic distances for each of the three kinematic
distance methods.  Table~\ref{tab:final_distances} lists the derived
distances for our sample: the Monte Carlo parallax distance, \(D_P\),
the kinematic distances using each of the three methods, \(D_A\),
\(D_B\), and \(D_C\), and their associated KDARs: KDAR\(_A\),
KDAR\(_B\), and KDAR\(_C\).

Here we investigate the differences between the parallax distances and
kinematic distances and compare those differences to the kinematic and
parallax distance uncertainties. For each kinematic distance method we
generate six figures: (1) a histogram of the difference between the
kinematic distance and the parallax distance; (2) a histogram of the
fractional distance difference; (3) a scatter plot of the distance
difference as a function of the parallax distance; (4) a scatter plot
of the distance difference minus the median difference as a function
of the parallax distance; (5) a cumulative distribution function (CDF)
of the ratio of the distance difference to the uncertainty in the
distance difference; and (6) a CDF of the ratio of the distance
difference minus the median difference to the difference uncertainty.

The distance difference histograms reveal any systematic differences
between the kinematic and parallax distances. The scatter plots
uncover correlations between the distance difference and the parallax
distance. Finally, the CDFs characterize the accuracy of the kinematic
and parallax distance uncertainties; if the kinematic and parallax
distance uncertainties are random and an accurate representation of
the data, the CDF should follow a normal distribution.

\begin{figure*}[ht]
  \centering
  \includegraphics[width=0.42\linewidth]{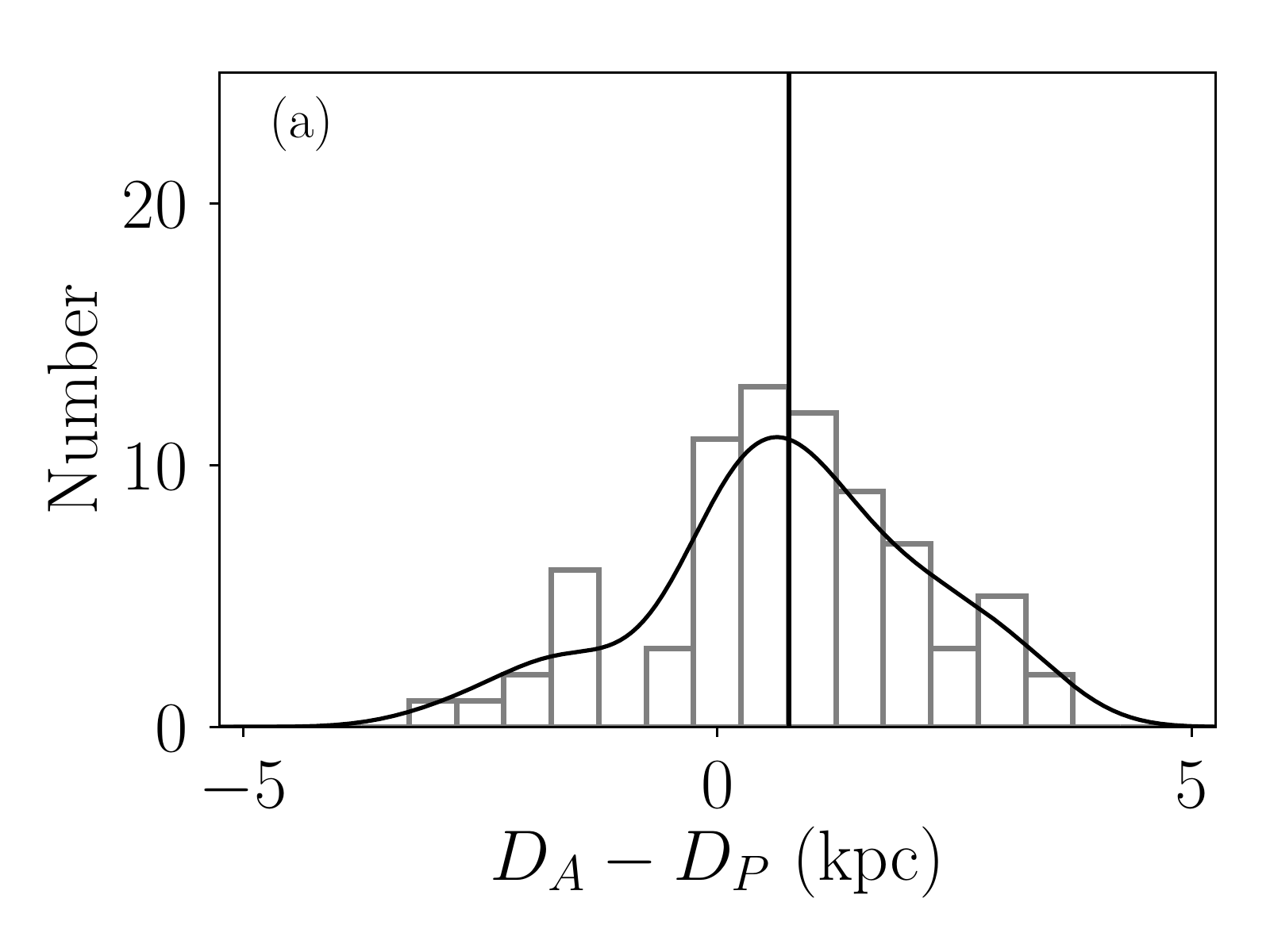}
  \includegraphics[width=0.42\linewidth]{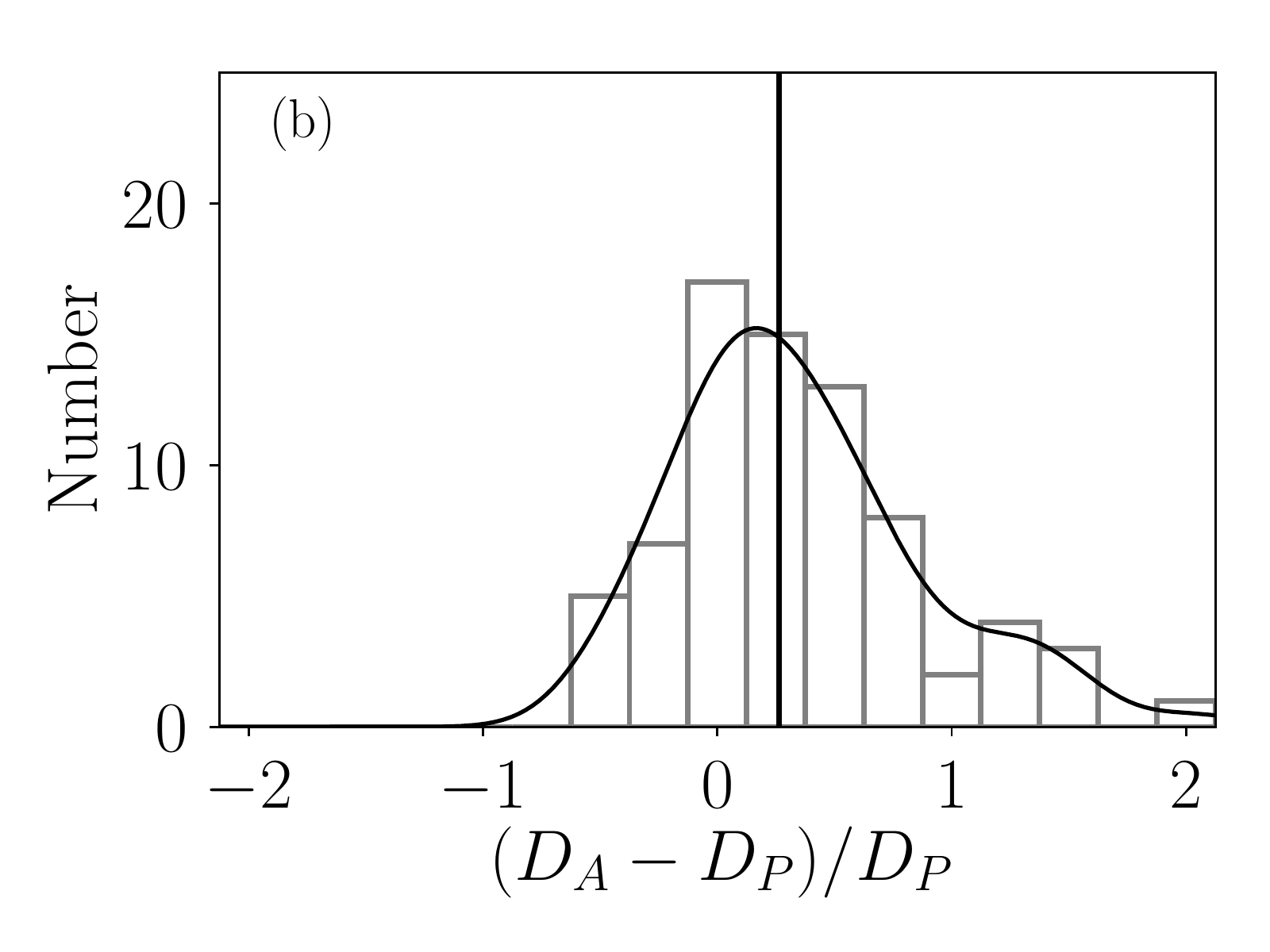} \\
  \includegraphics[width=0.42\linewidth]{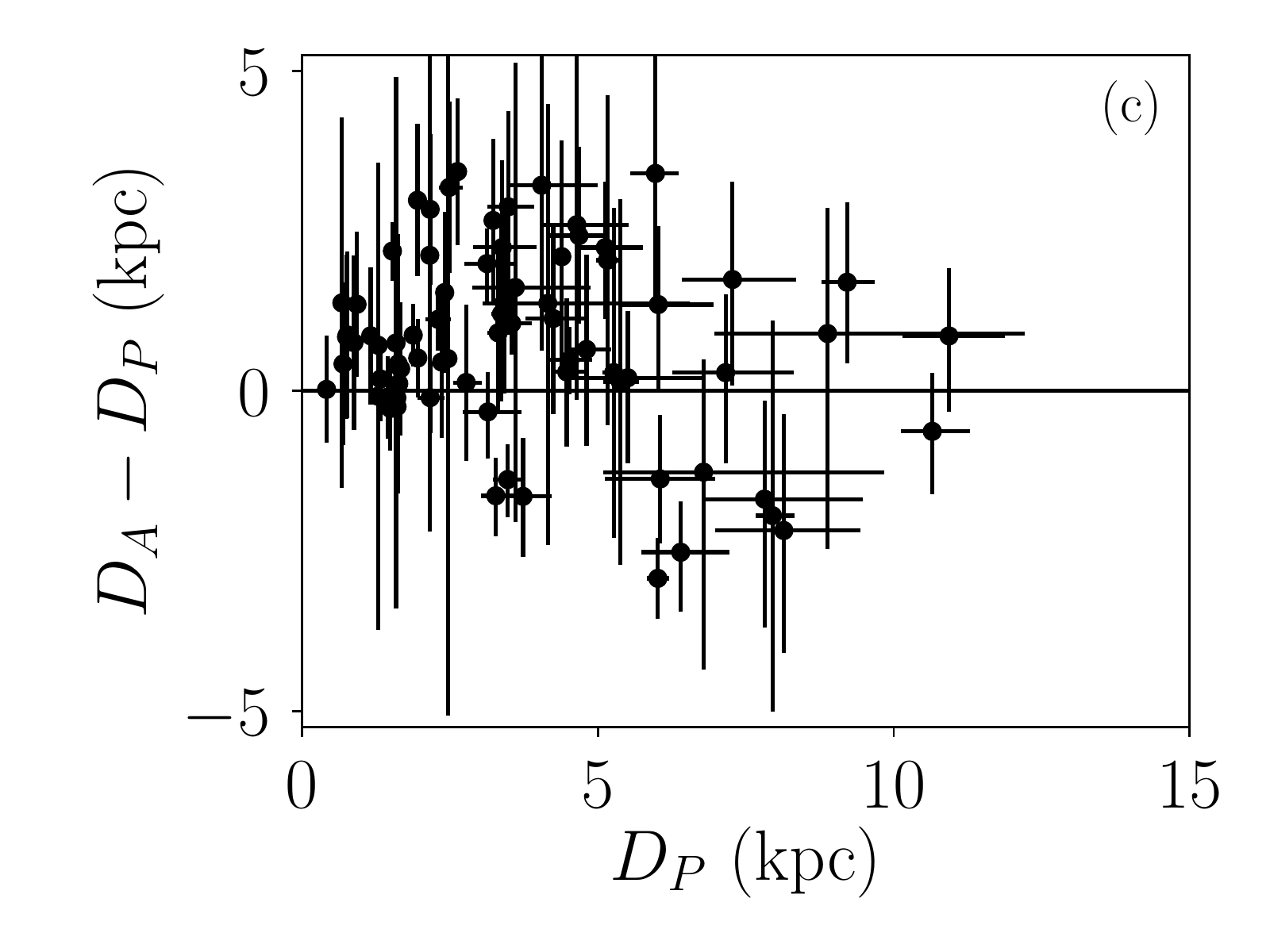}
  \includegraphics[width=0.42\linewidth]{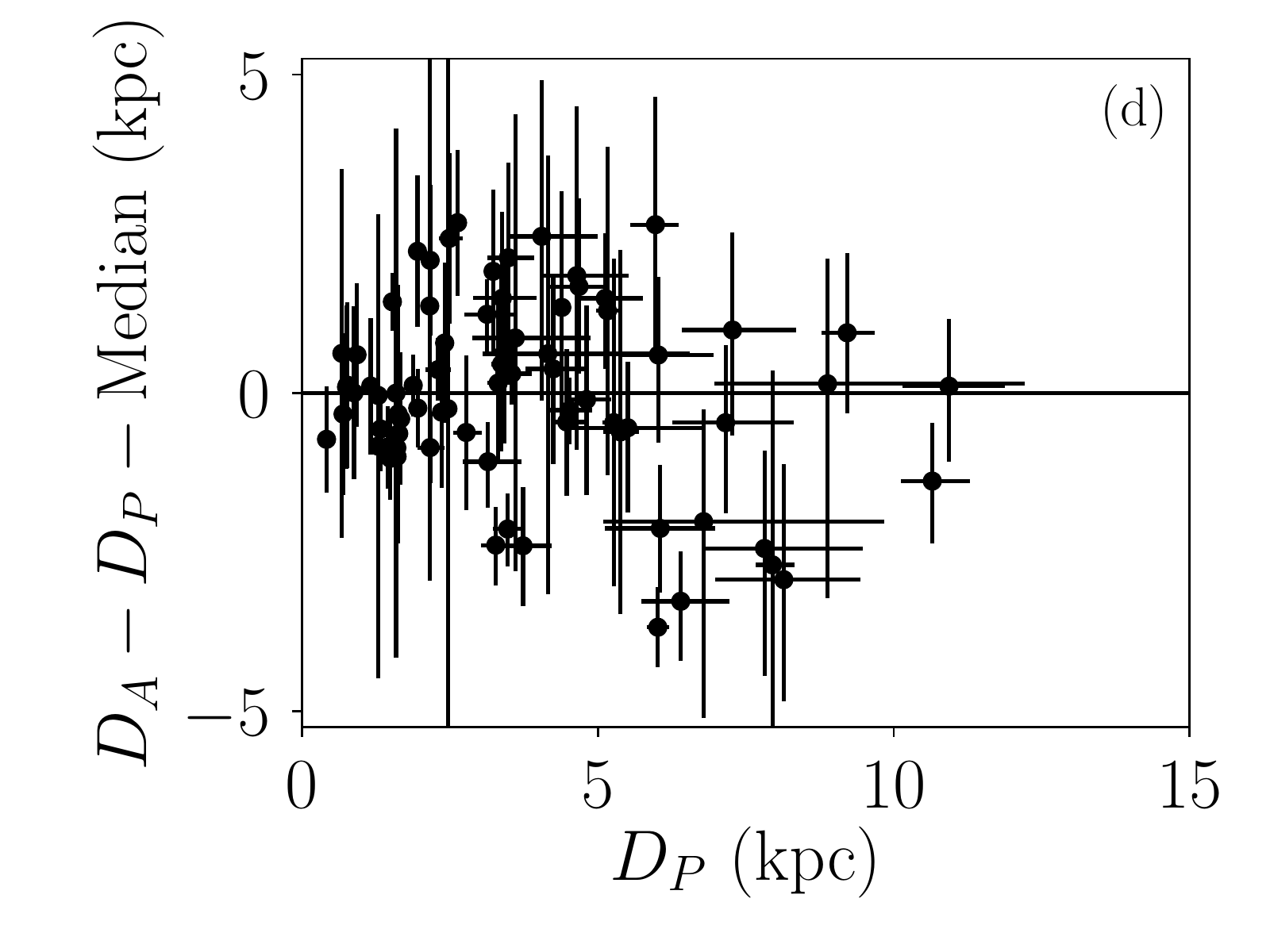} \\
  \includegraphics[width=0.42\linewidth]{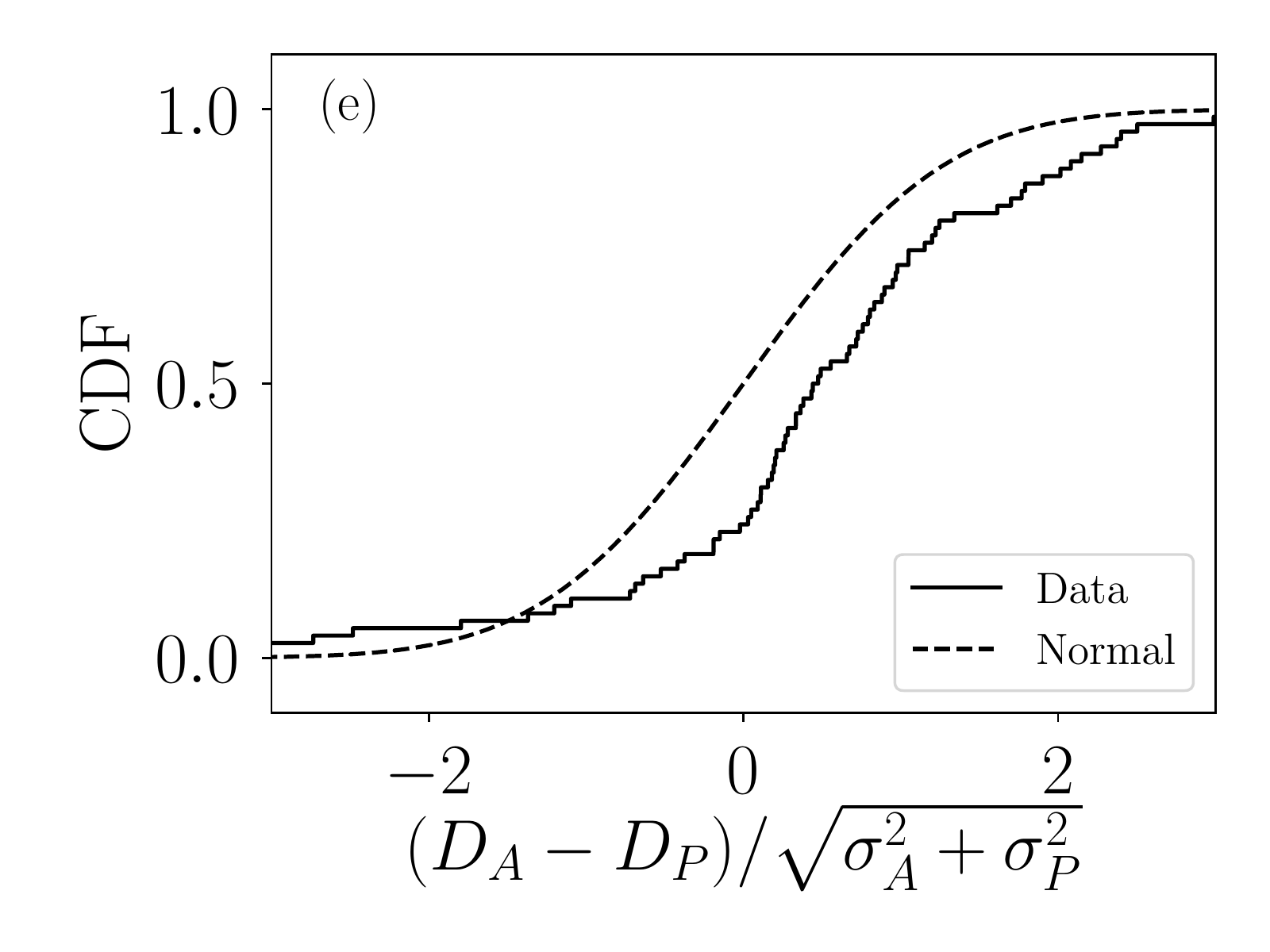}
  \includegraphics[width=0.42\linewidth]{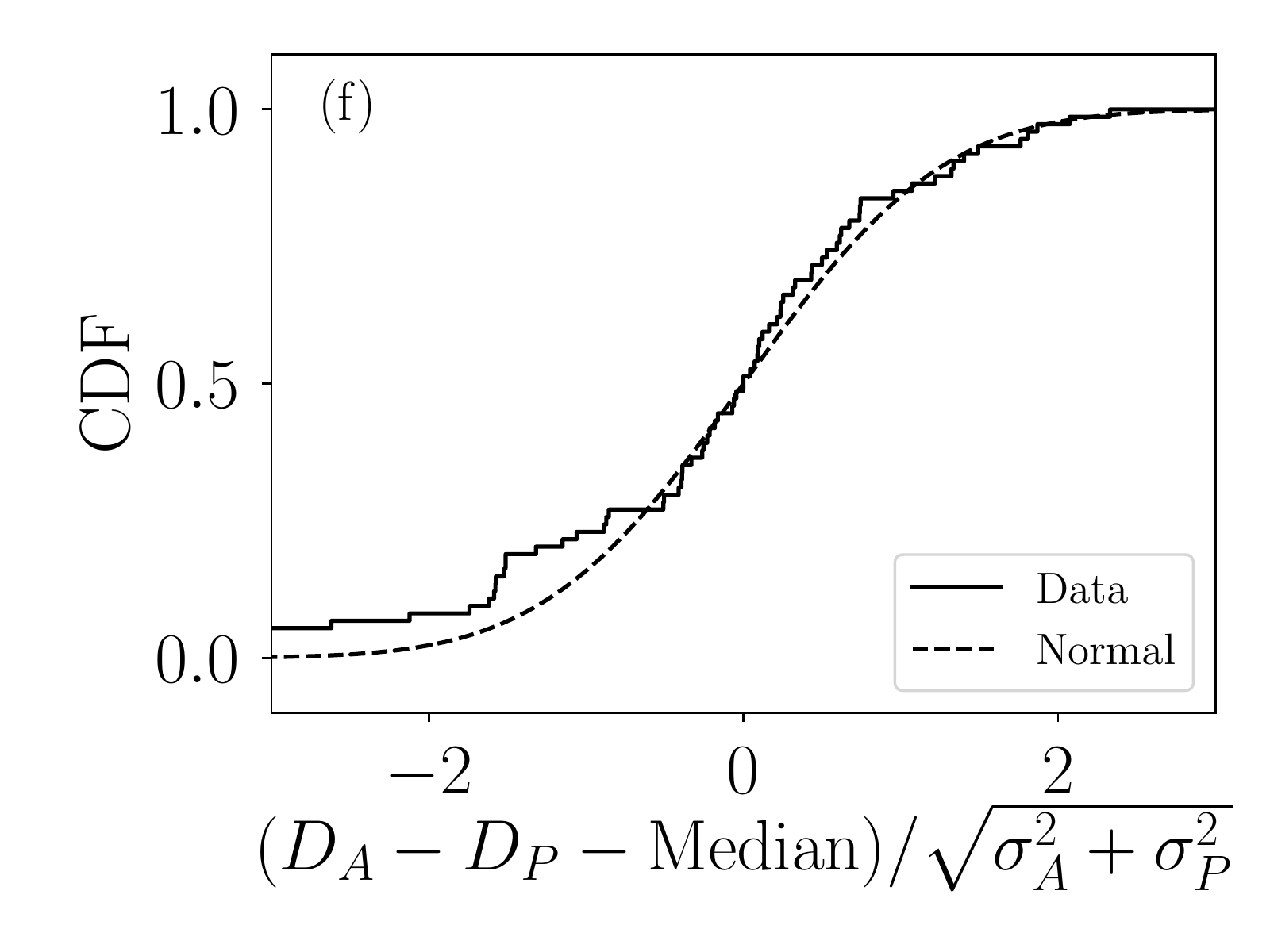}
  \caption{Difference between parallax distances and Method A
    kinematic distances. Panel (a): histogram of distance
    difference. The solid curve is the KDE fit to the difference
    distribution and the solid vertical line is the median of the
    distribution. Panel (b): histogram of the fractional distance
    difference. Panel (c): scatter plot of distance difference as a
    function of parallax distance. Panel (d): scatter plot of the
    distance difference minus the median distance difference as a
    function of parallax distance. Panel (e): cumulative distribution
    function (CDF) of the ratio of the distance differences to the
    distance difference uncertainties. Panel (f): CDF of the ratio of
    the distance differences minus the median difference to the
    distance difference uncertainties. The dashed curve in Panels (e)
    and (f) is the expected CDF for a normal distribution centered on
    zero. The CDF does not go to 0 on the left nor to 1 on the right
    because there is at least one source beyond the limits of the
    abscissas.}
  \label{fig:orig_diff}
\end{figure*}

\begin{figure*}[ht]
  \centering
  \includegraphics[width=0.42\linewidth]{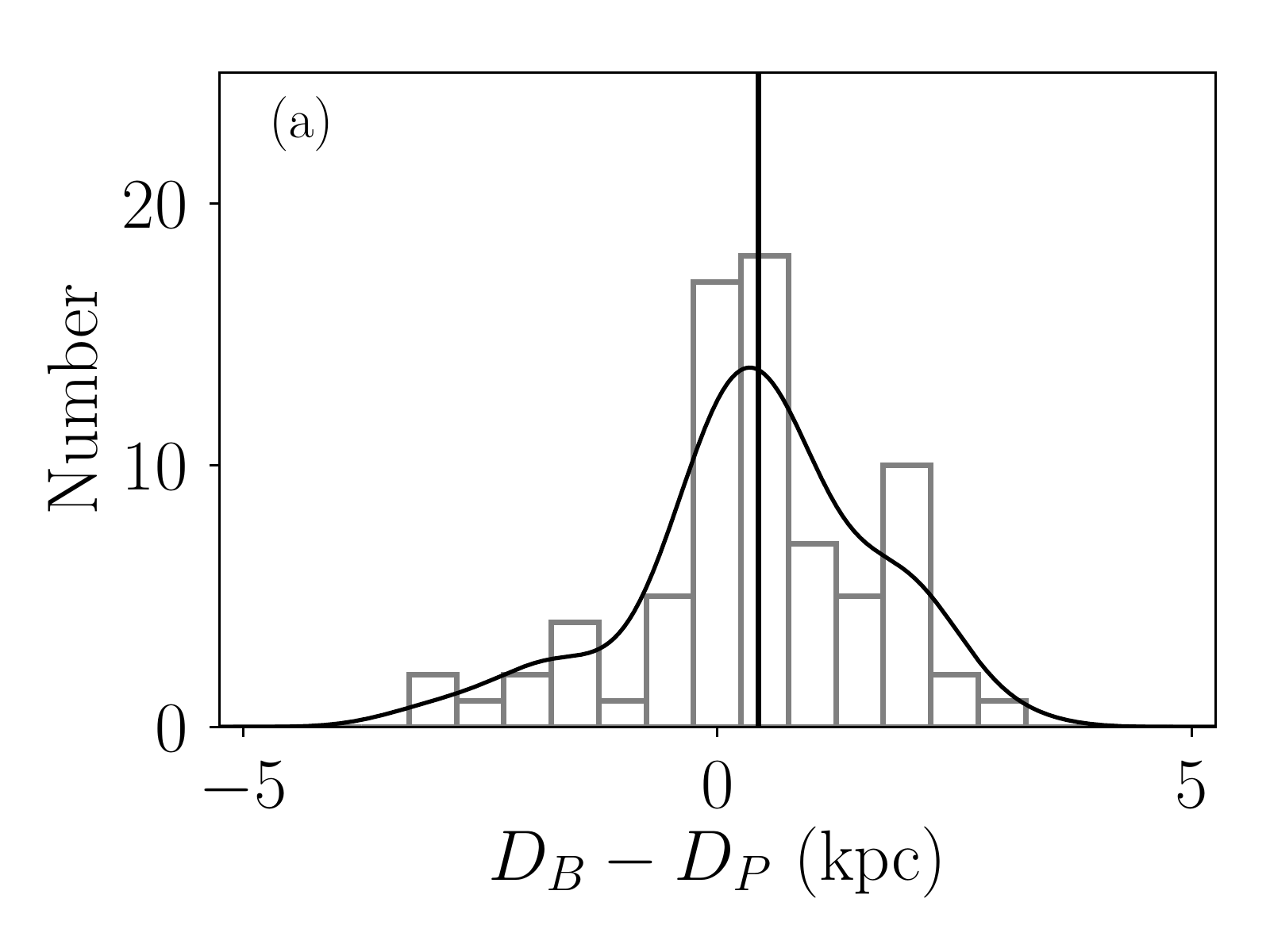}
  \includegraphics[width=0.42\linewidth]{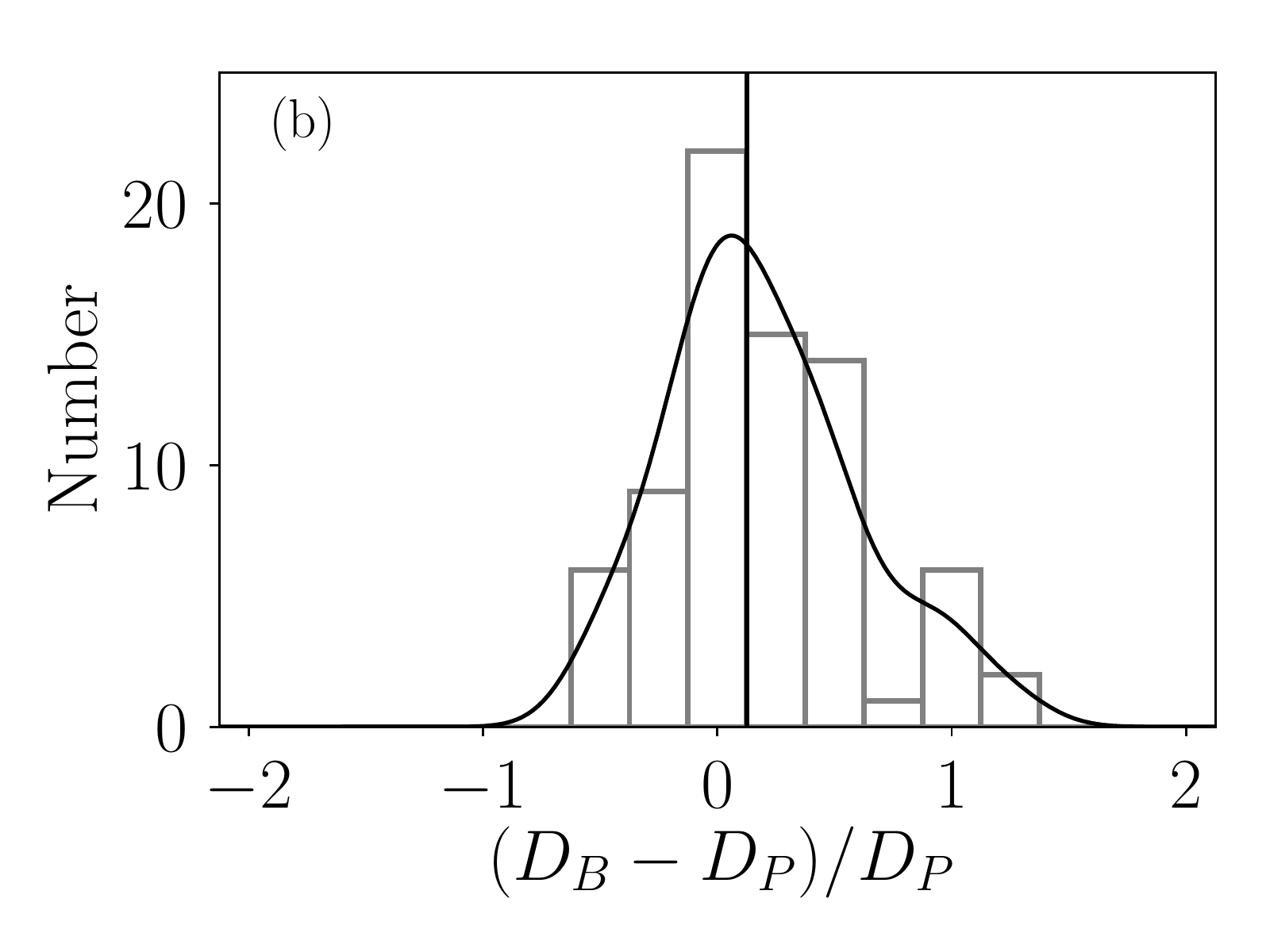} \\
  \includegraphics[width=0.42\linewidth]{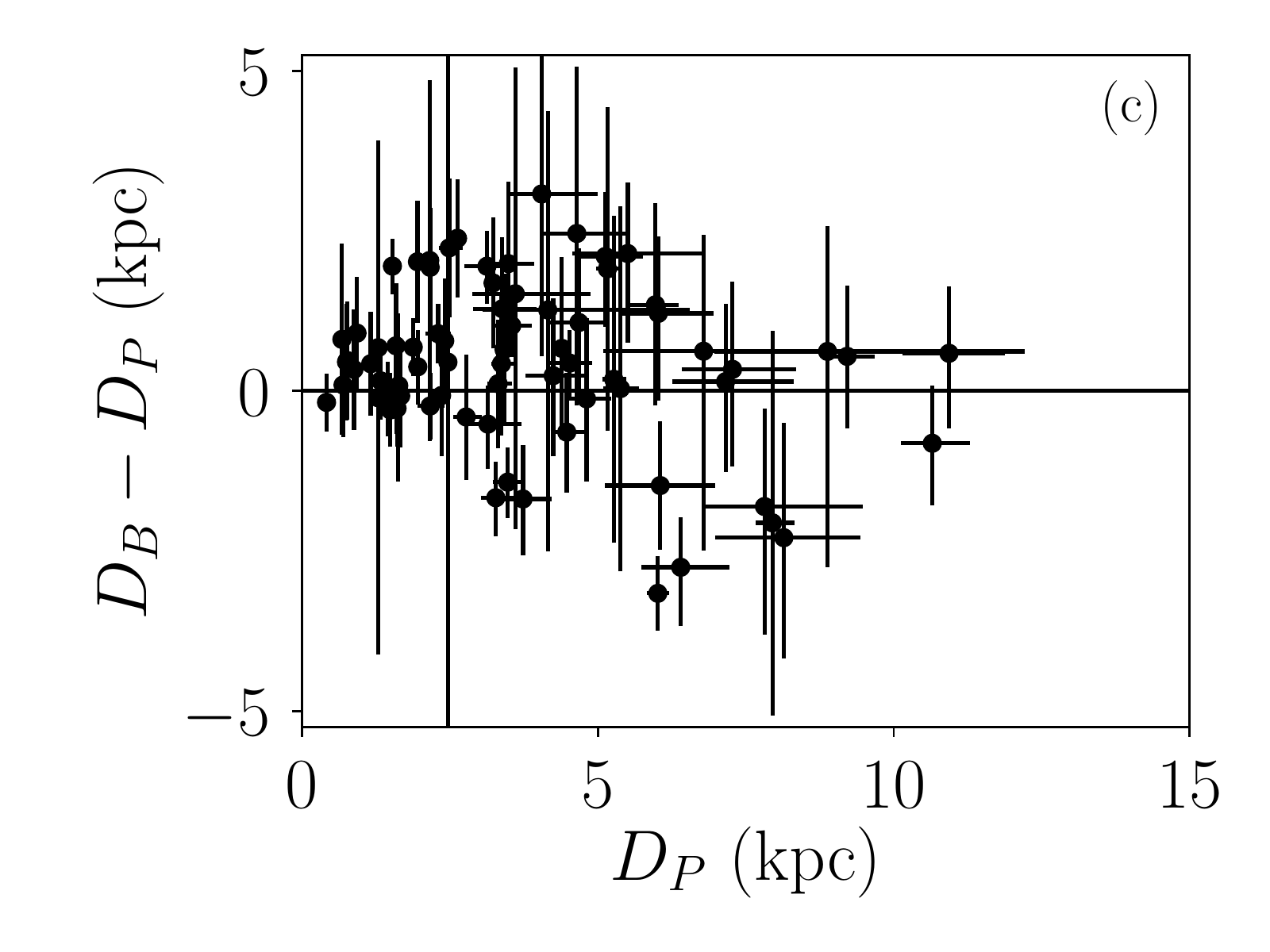}
  \includegraphics[width=0.42\linewidth]{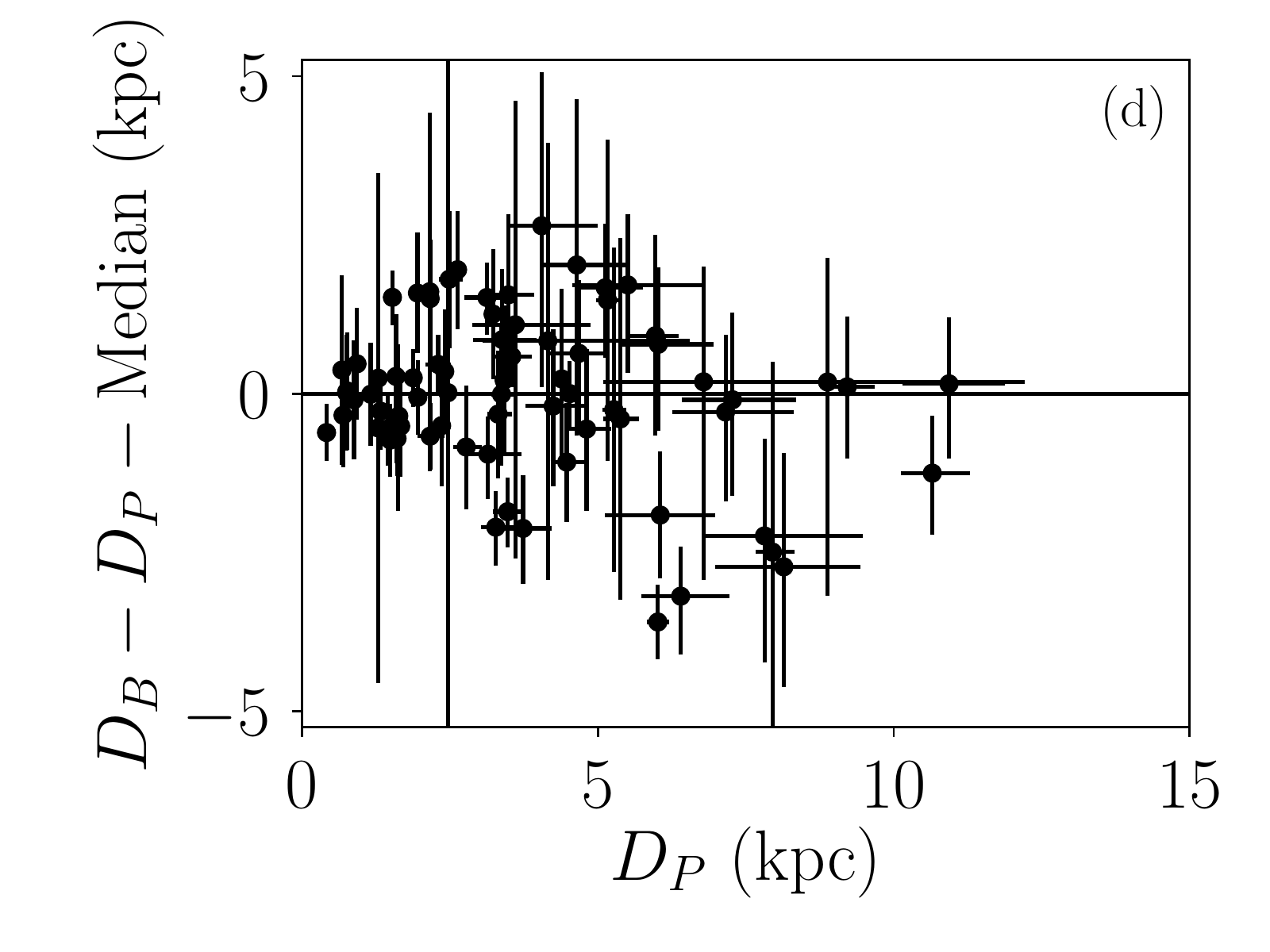} \\
  \includegraphics[width=0.42\linewidth]{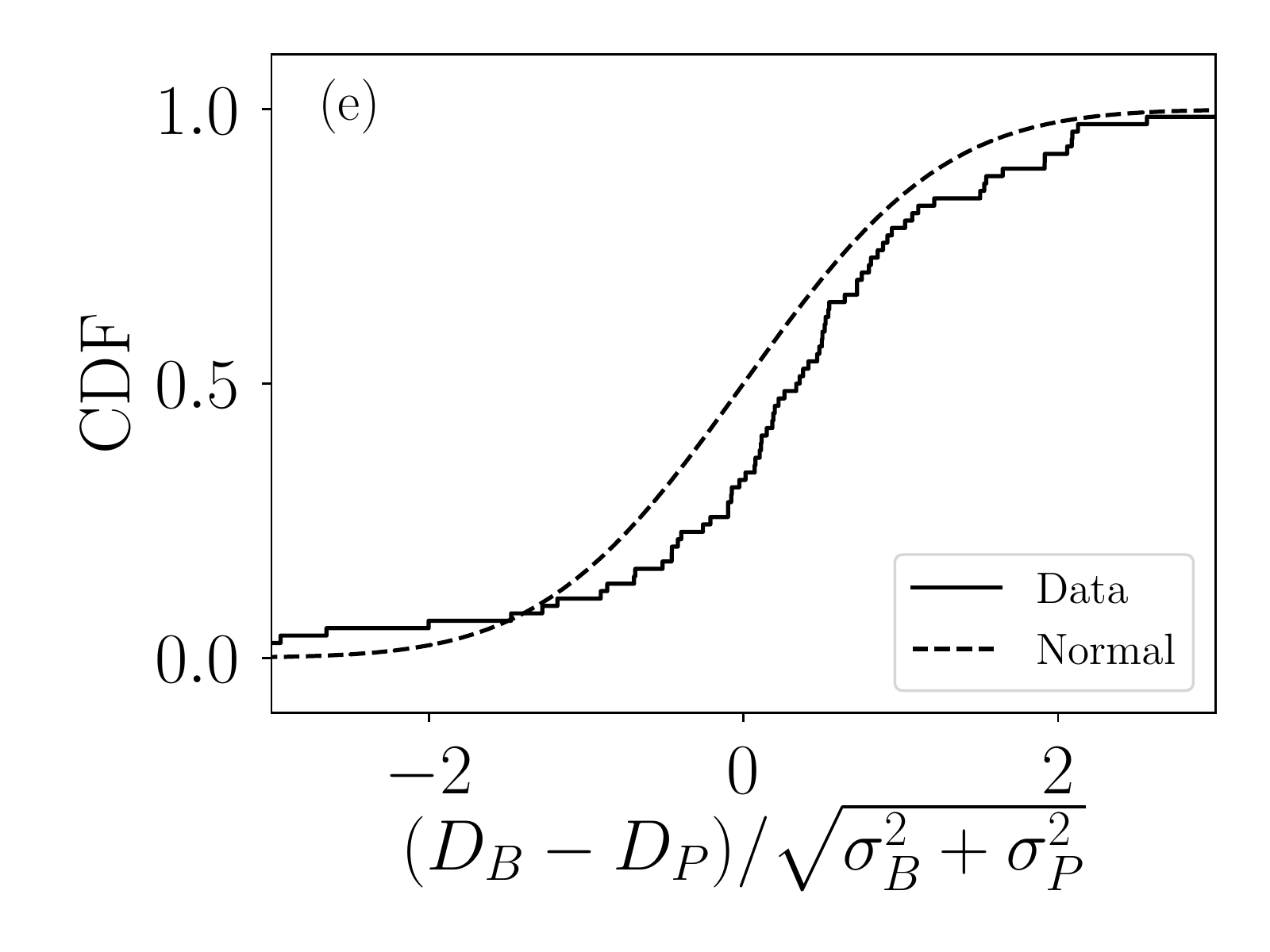}
  \includegraphics[width=0.42\linewidth]{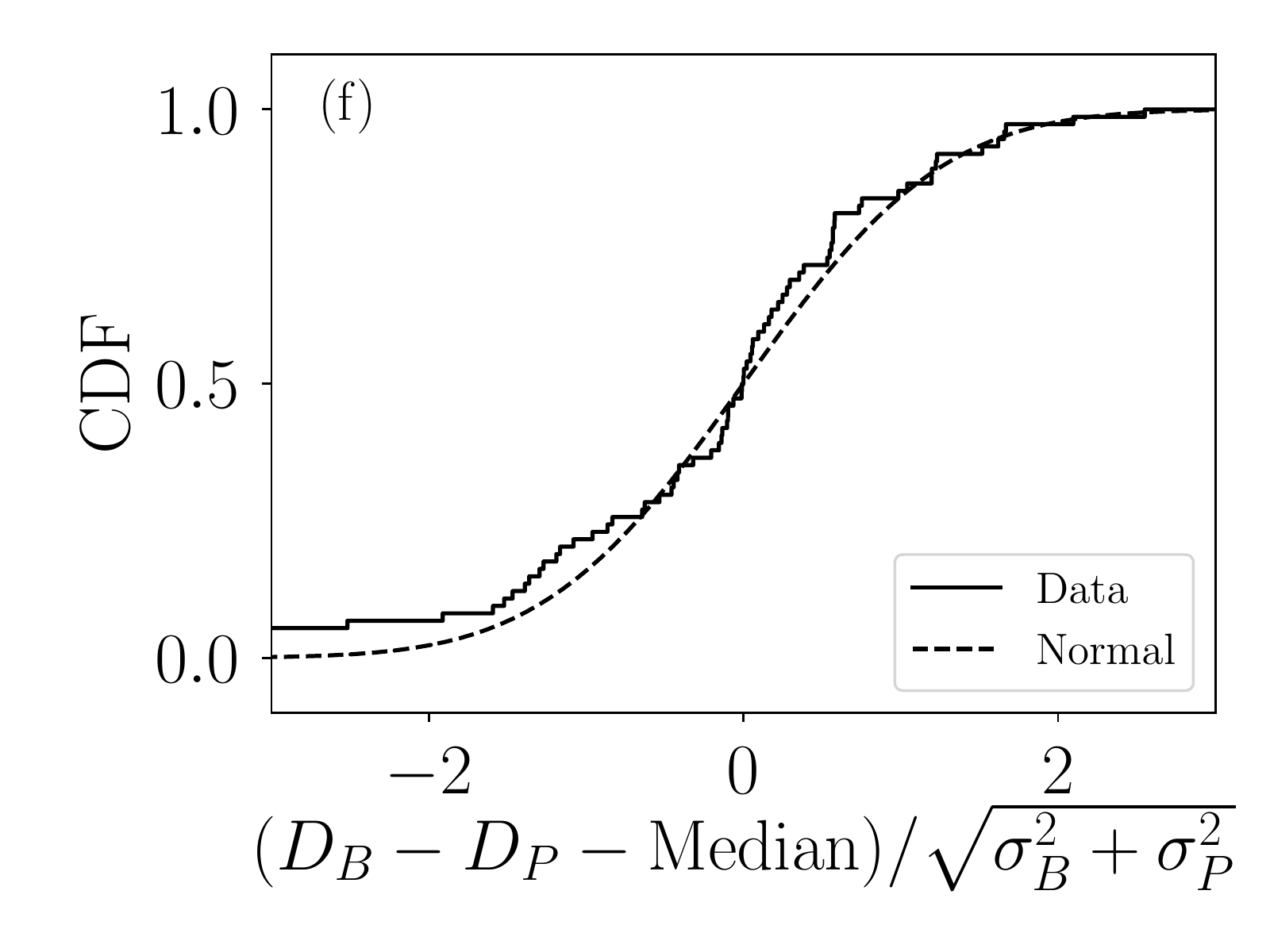}
  \caption{Same as Figure~\ref{fig:orig_diff} but using Method B
    kinematic distances.}
  \label{fig:reid_diff}
\end{figure*}

\begin{figure*}[ht]
  \centering
  \includegraphics[width=0.42\linewidth]{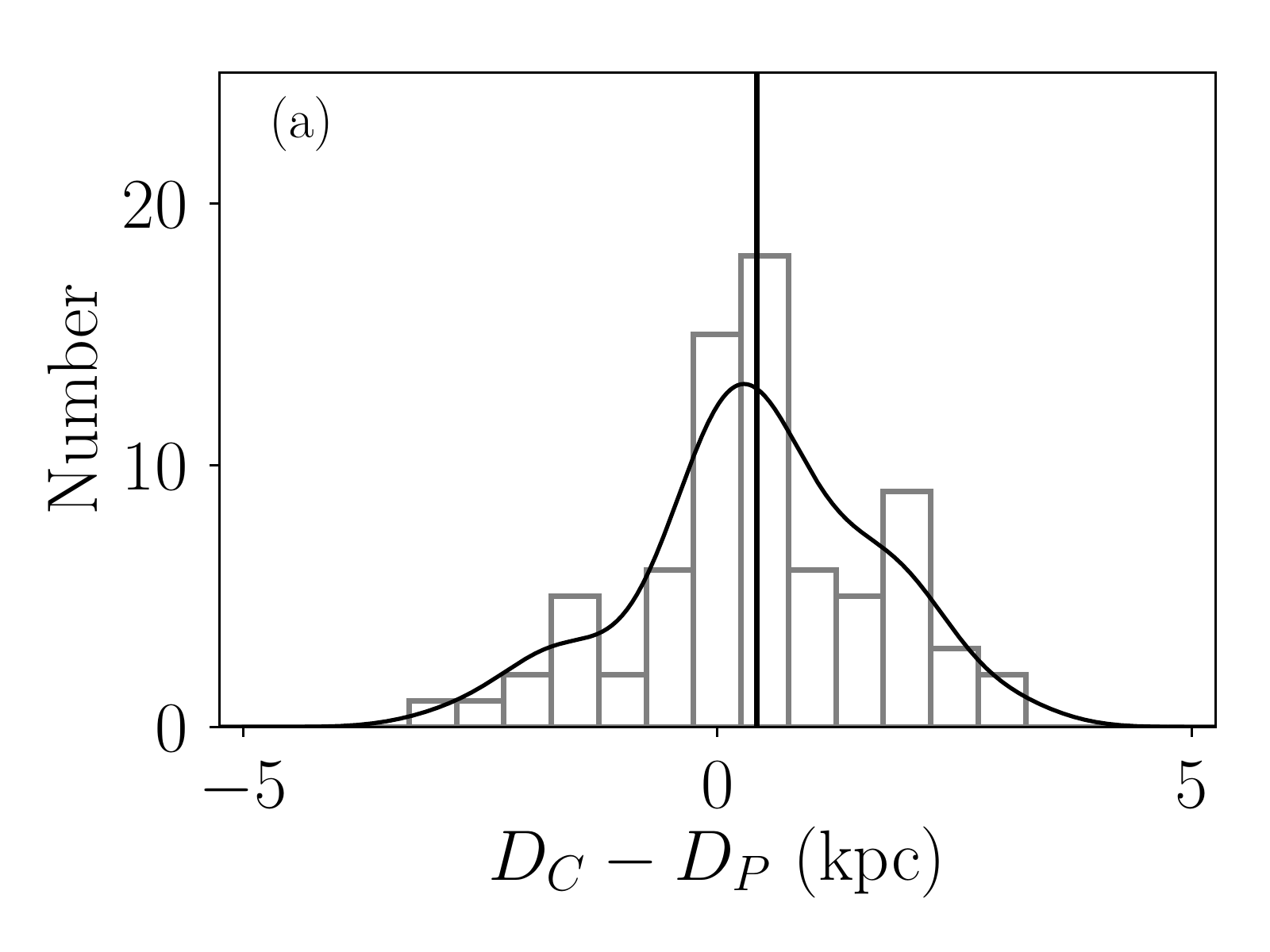}
  \includegraphics[width=0.42\linewidth]{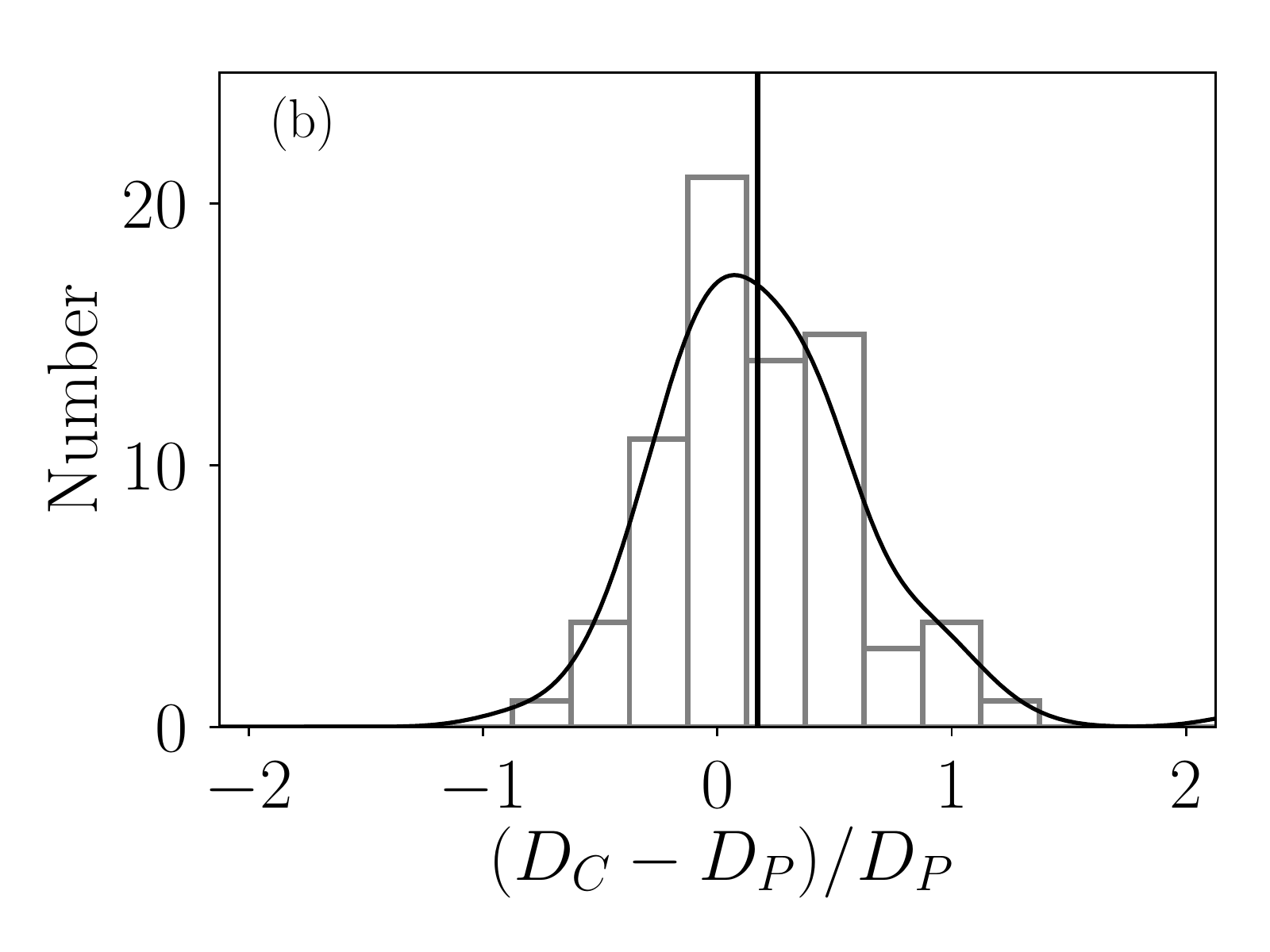} \\
  \includegraphics[width=0.42\linewidth]{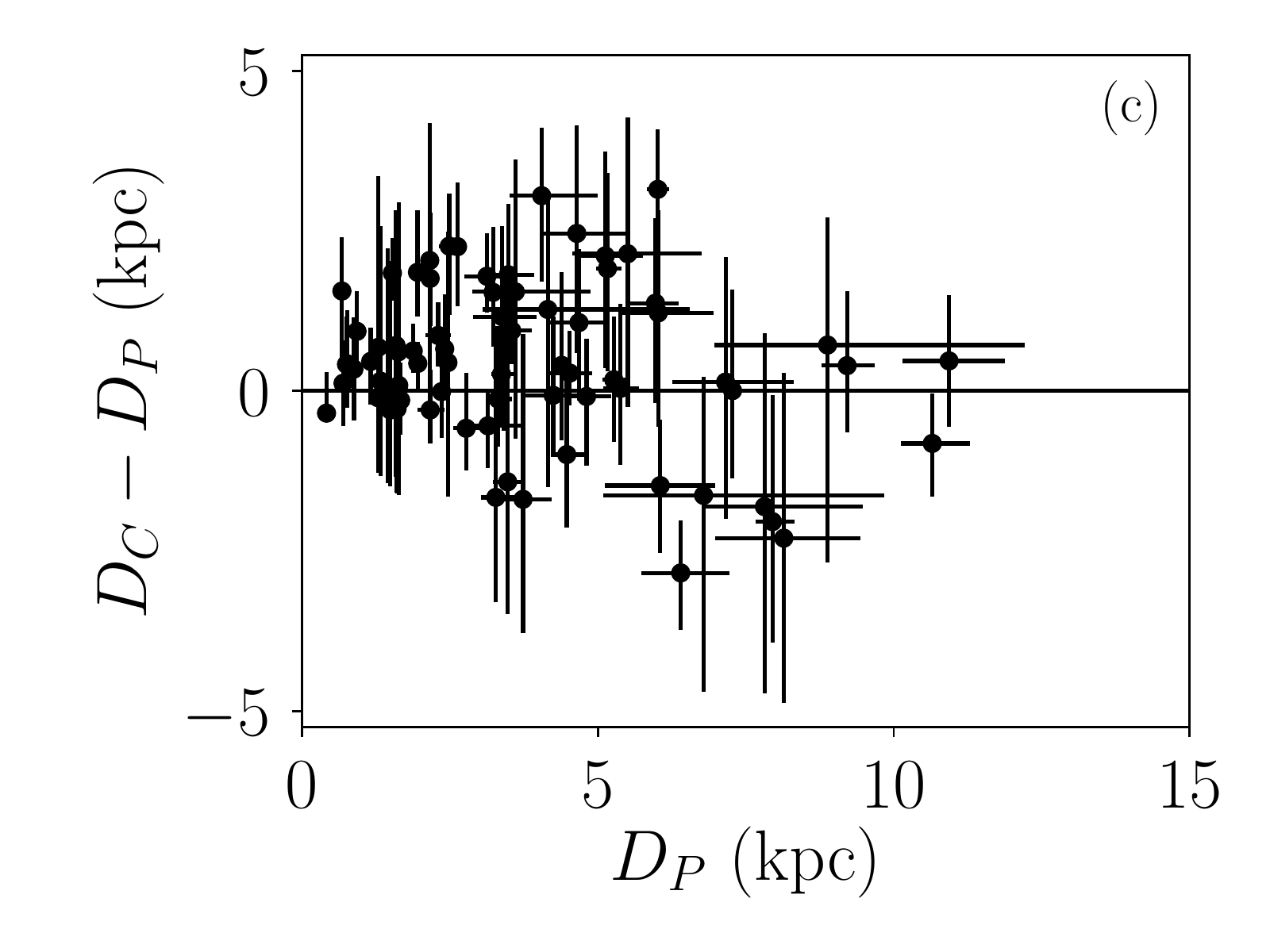}
  \includegraphics[width=0.42\linewidth]{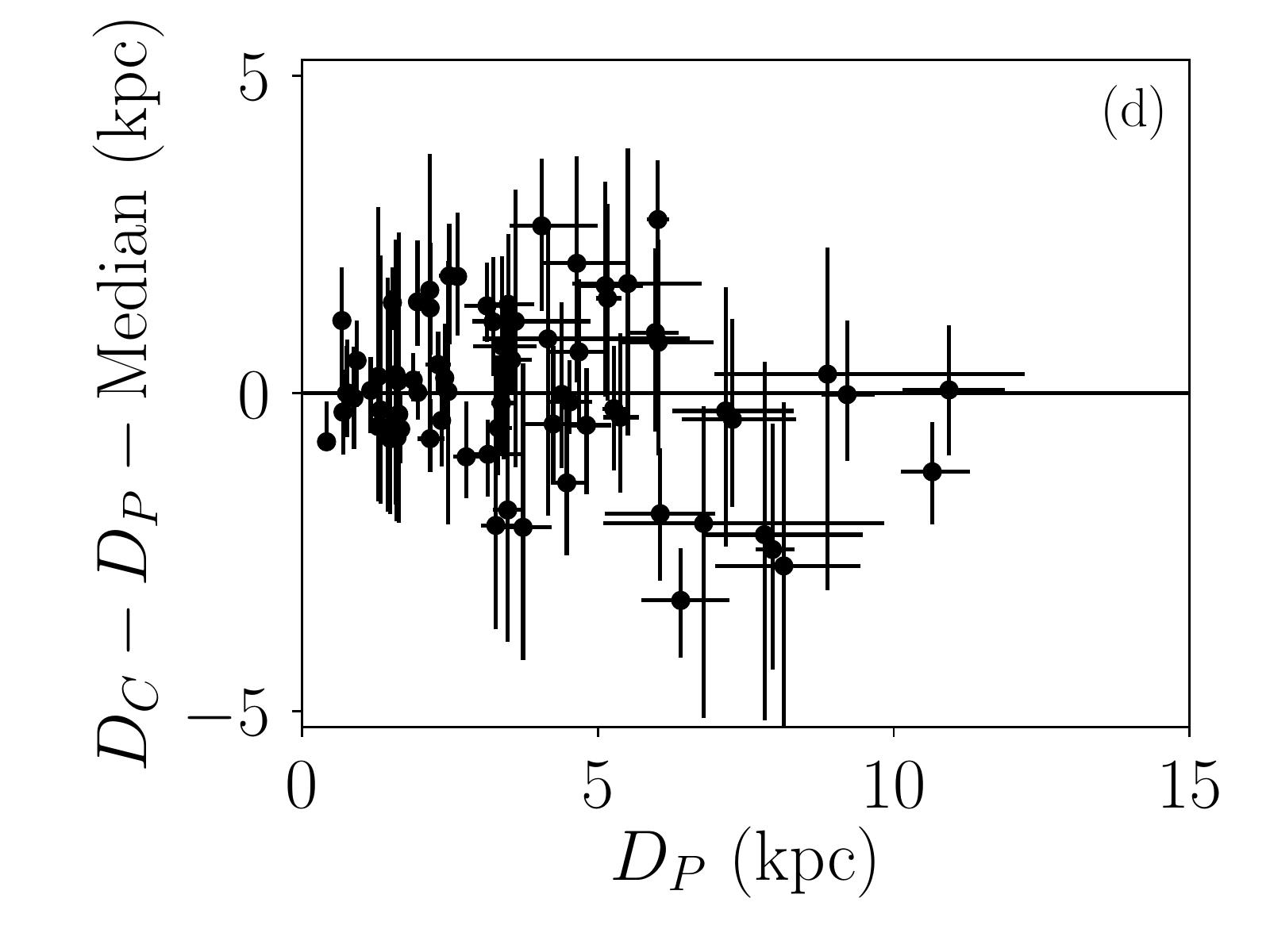} \\
  \includegraphics[width=0.42\linewidth]{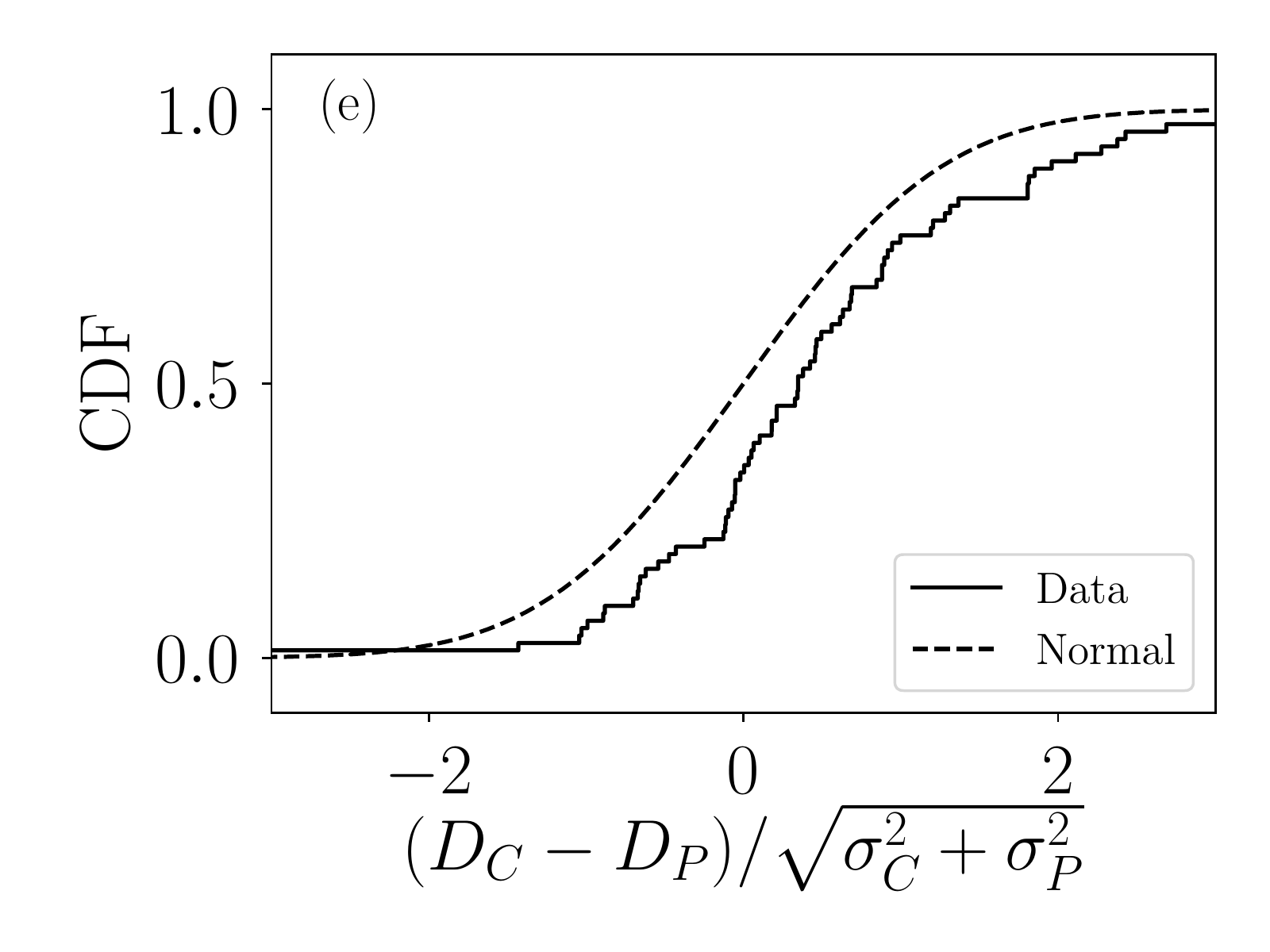}
  \includegraphics[width=0.42\linewidth]{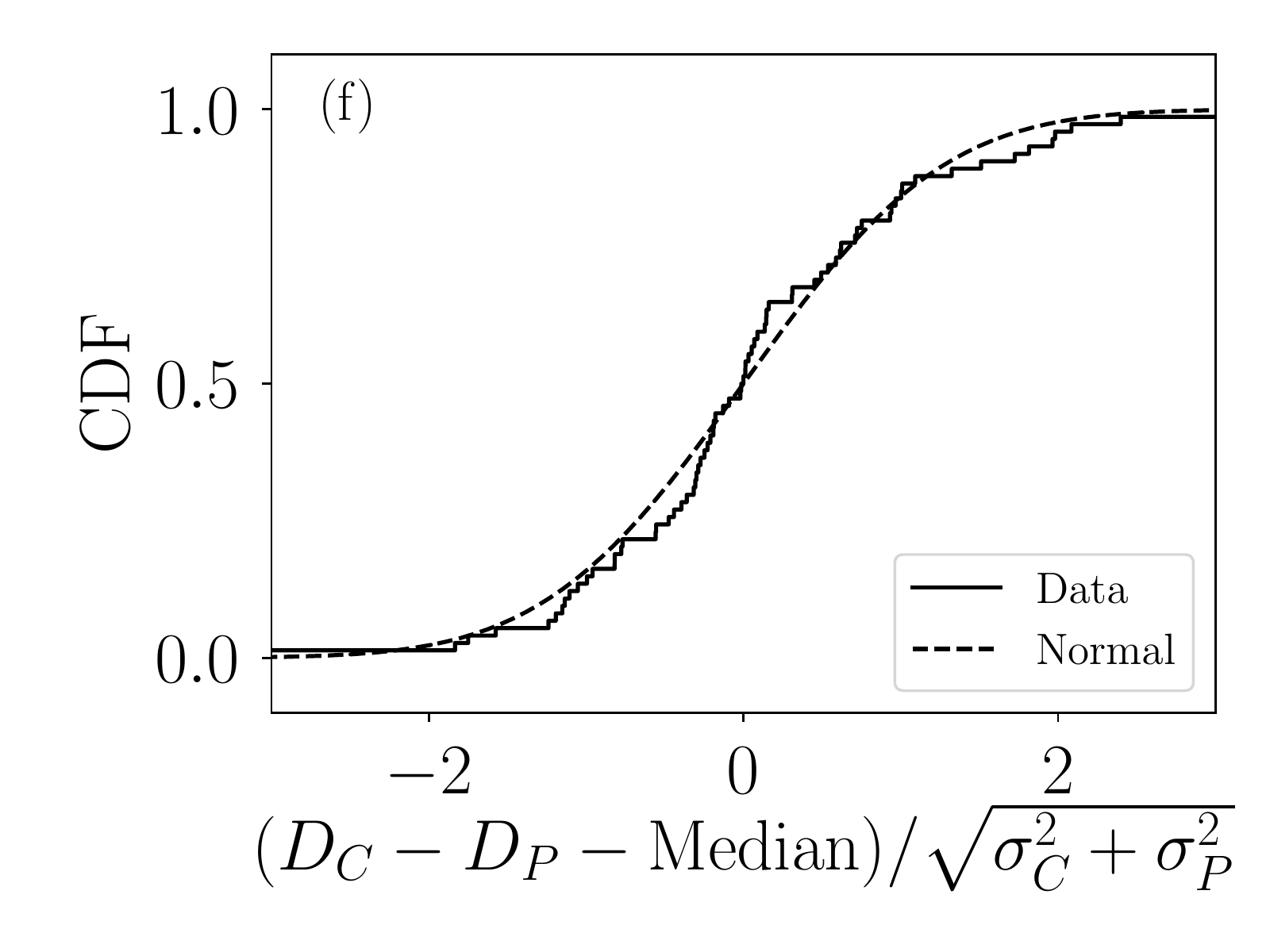}
  \caption{Same as Figure~\ref{fig:orig_diff} but using Method C
    kinematic distances.}
  \label{fig:pdf_diff}
\end{figure*}

We first compare the parallax distances to the Method A kinematic
distances, \(D_A\). The distance differences are shown in
Figure~\ref{fig:orig_diff}. We compute the mean, median, and standard
deviation of the distance difference (i.e., \(D_A-D_P\)), the absolute
distance difference (i.e., \(|D_A-D_P|\)), the fractional distance
difference (i.e., \((D_A-D_P)/D_P\)), and the absolute fractional
distance difference (i.e., \(|D_A-D_P|/D_P\)).  These values are
listed in Table~\ref{tab:final_stats}. The fractional distance
difference distribution in Panel (b) has a long tail towards larger
kinematic distances.

After subtracting the median offset, the kinematic distance
uncertainties from the A12 model fit the differences between the
kinematic and parallax distances well. Panel (f) of
Figure~\ref{fig:orig_diff} shows that the ratio of the distance
difference (minus the median difference) to the difference uncertainty
follows a normal distribution, indicating that the kinematic and
parallax distance uncertainties accurately represent the random errors
in the distances. The K--S statistic for this distribution is \(0.121\)
which corresponds to a p--value of \(0.203\). Panel (d), however,
shows that some of the error bars are large even when the difference
between the kinematic and parallax distance is small. This implies
that the kinematic distance uncertainty model is over-predicting the
kinematic distance uncertainties in some cases. The median Method A
kinematic distance uncertainty (i.e., \(\sigma_A/D_A\)) is \(28.0\%\).

Next, we compare the parallax distances to the Method B kinematic
distances, \(D_B\). The differences between these two distances are
shown in Figure~\ref{fig:reid_diff}, and the mean, median, and
standard deviation statistics are listed in
Table~\ref{tab:final_stats}.  The mean and median distance differences
are significantly smaller than those found using Method A. The
fractional distance difference distribution is both centered closer to
zero and narrower than the Method A distribution. The tail towards
larger fractional differences is not nearly as long using this method.

Once again, the A12 kinematic distance uncertainty model seems to
accurately represent the typical differences between the parallax and
kinematic distances. The K--S statistic for the median-corrected CDF
is \(0.081\) with a p--value of \(0.716\), thus strongly implying that
the uncertainties are sampled from a normal distribution. The
kinematic distance errors are, however, large for sources with
kinematic distances and parallax distances in good agreement. The
median Method B kinematic distance uncertainty is the same as with
method A at \(28.0\%\).

Finally, we compare the parallax distances to the Method C kinematic
distances, \(D_C\).  The distance differences using this method are
shown in Figure~\ref{fig:pdf_diff} and the mean, median, and standard
deviation statistics are in Table~\ref{tab:final_stats}. These
statistics and distributions are nearly identical to those found using
Method B.

The kinematic distance uncertainties derived using the Monte Carlo
method (Method C) are just as accurate as those given by the A12
kinematic distance uncertainty model (Methods A and B). Panel (f) of
Figure~\ref{fig:pdf_diff} shows that the kinematic distance
uncertainties follow a normal distribution with a K--S statistic of
\(0.083\) (p--value is \(0.681\)). This distribution and K--S
statistic are nearly the same as that of Method B, yet the distance
uncertainties are not assigned based on a model but rather derived
based on the data and GRM. The median Method C kinematic distance
uncertainty is slightly smaller than that of Method B at
\(25.8\%\). More than half (56\%) of the Method C kinematic distance
uncertainties are smaller than the Method B uncertainties. Despite
these smaller error bars, panel (f) of Figure~\ref{fig:pdf_diff} shows
that these kinematic distance uncertainties fit the data just as well
as the A12 model used in Method A and B.

Table~\ref{tab:final_stats} summarizes the aforementioned results for
the three kinematic distance methods. The median absolute distance
difference is nearly \({\sim}40\%\) smaller using Methods B and C,
with a \({\sim}12\%\) smaller standard deviation. The median-corrected
K--S statistic is about \(30\%\) smaller using Methods B and C, and
nearly identical between Methods B and C. This suggests that the Monte
Carlo-derived kinematic distance uncertainties (Method C) are just as
accurate as the A12 kinematic distance uncertainty model (Method
B). The median kinematic distance uncertainty, \(\sigma_D/D\), is the
smallest using Method C.

\begin{deluxetable*}{lccc}
\tablewidth{0pt}
\tabletypesize{\normalsize}
\tablecaption{Distance Difference Statistics\label{tab:final_stats}}
\tablehead{
  \colhead{} & \colhead{Method A} & \colhead{Method B} & \colhead{Method C}
}
\startdata
\sidehead{\(D-D_P\) (kpc)}
Median & 0.75 & 0.43 & 0.42 \\
Mean & 0.74 & 0.40 & 0.42 \\
Std. Dev. & 1.42 & 1.24 & 1.23 \\
\hline
\sidehead{\(|D-D_P|\) (kpc)}
Median & 1.13 & 0.68 & 0.71 \\
Mean & 1.29 & 1.00 & 1.01 \\
Std. Dev. & 0.95 & 0.83 & 0.83 \\
\hline
\sidehead{\((D-D_P)/D_P\) (percent)}
Median & 26 & 13 & 17 \\ 
Mean & 35 & 20 & 21 \\
Std. Dev. & 52 & 40 & 46 \\
\hline
\sidehead{\(|D-D_P|/D_P\) (percent)}
Median & 33 & 24 & 26 \\
Mean & 46 & 34 & 36 \\
Std. Dev. & 43 & 30 & 36 \\
\hline
\sidehead{\(\sigma_D/D\) (percent)}
Median & 28.0 & 28.0 & 25.8 \\
\hline
\sidehead{\((D-D_P-{\rm Median})/\sqrt{\sigma_D^2 + \sigma_P^2}\)}
K--S statistic & 0.121 & 0.081 & 0.083 \\
K--S p--value & 0.203 & 0.716 & 0.681 \\
\enddata
\end{deluxetable*}

\section{Kinematic Distance Ambiguity (KDA)\label{sec:kdar}}

Thus far we resolved the KDA by assigning the kinematic distance
closest to the parallax distance, or by assigning objects within
\(20\kms\) of the tangent point velocity to the tangent point
distance. The \textit{WISE} Catalog of Galactic \hii\ Regions
\citep{anderson2014} contains the KDAR for 34 of our sources
determined using a variety of KDAR techniques. Here we compare our
parallax-based KDARs to the \textit{WISE} Catalog KDARs.

\begin{figure}[ht]
  \centering
  \includegraphics[width=\linewidth]{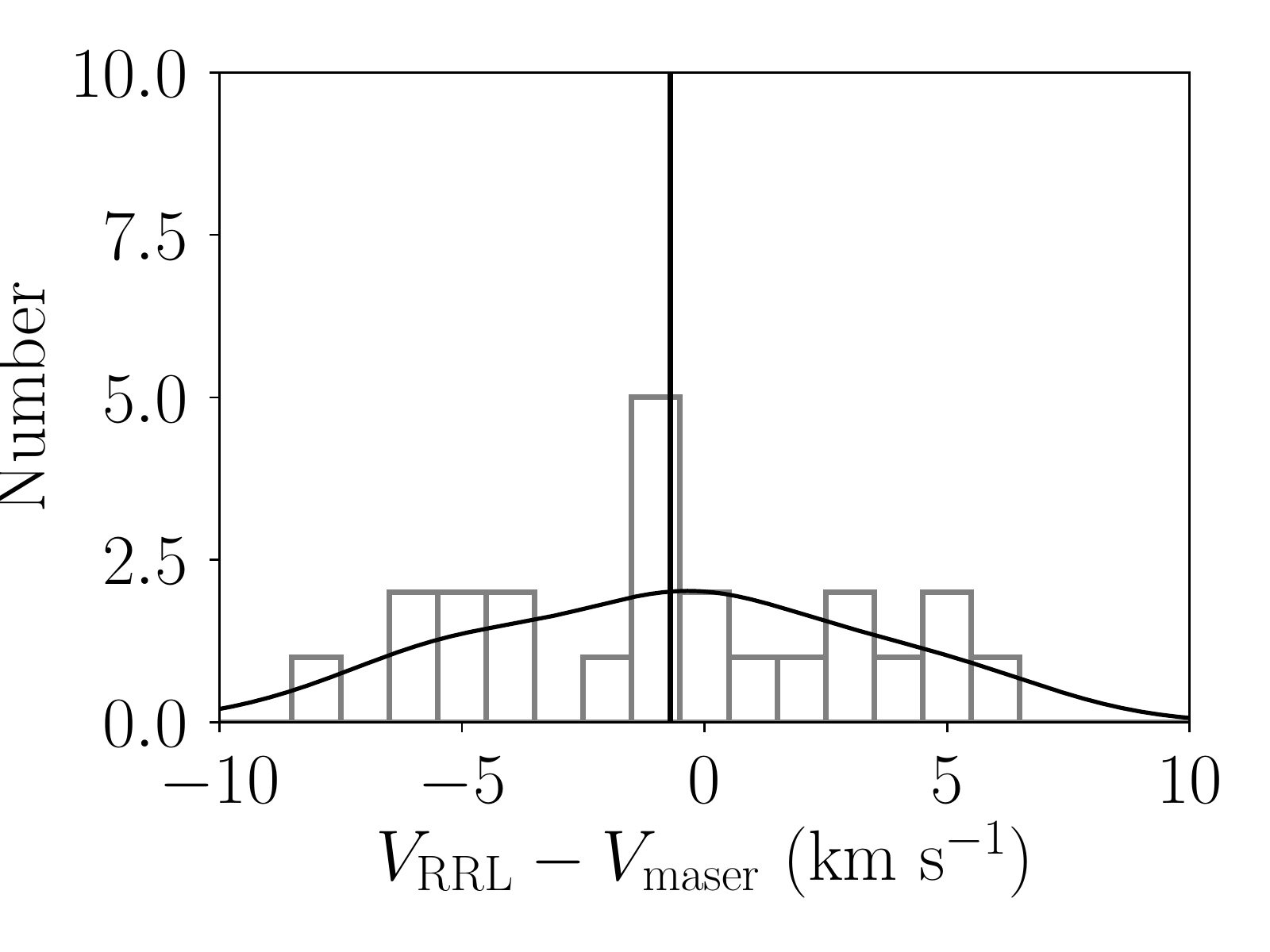} \\
  \includegraphics[width=\linewidth]{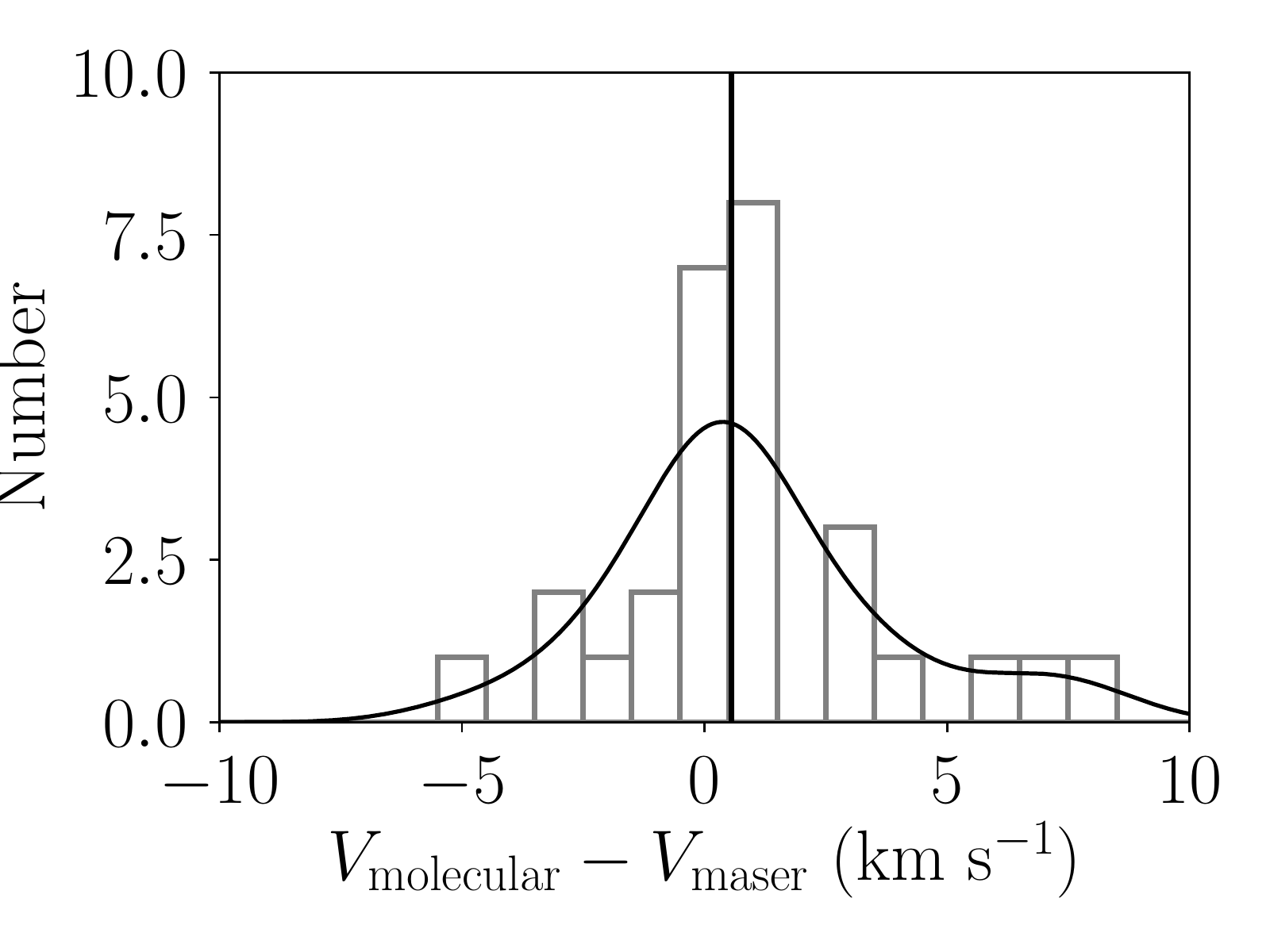}
  \caption{Difference between RRL and maser LSR velocity (top) and
    molecular line and maser LSR velocity (bottom). The solid curve is
    the KDE fit to each distribution, and the vertical line is the
    median.}
  \label{fig:velocity}
\end{figure}

\begin{figure*}[ht]
  \centering
  \includegraphics[width=0.45\linewidth]{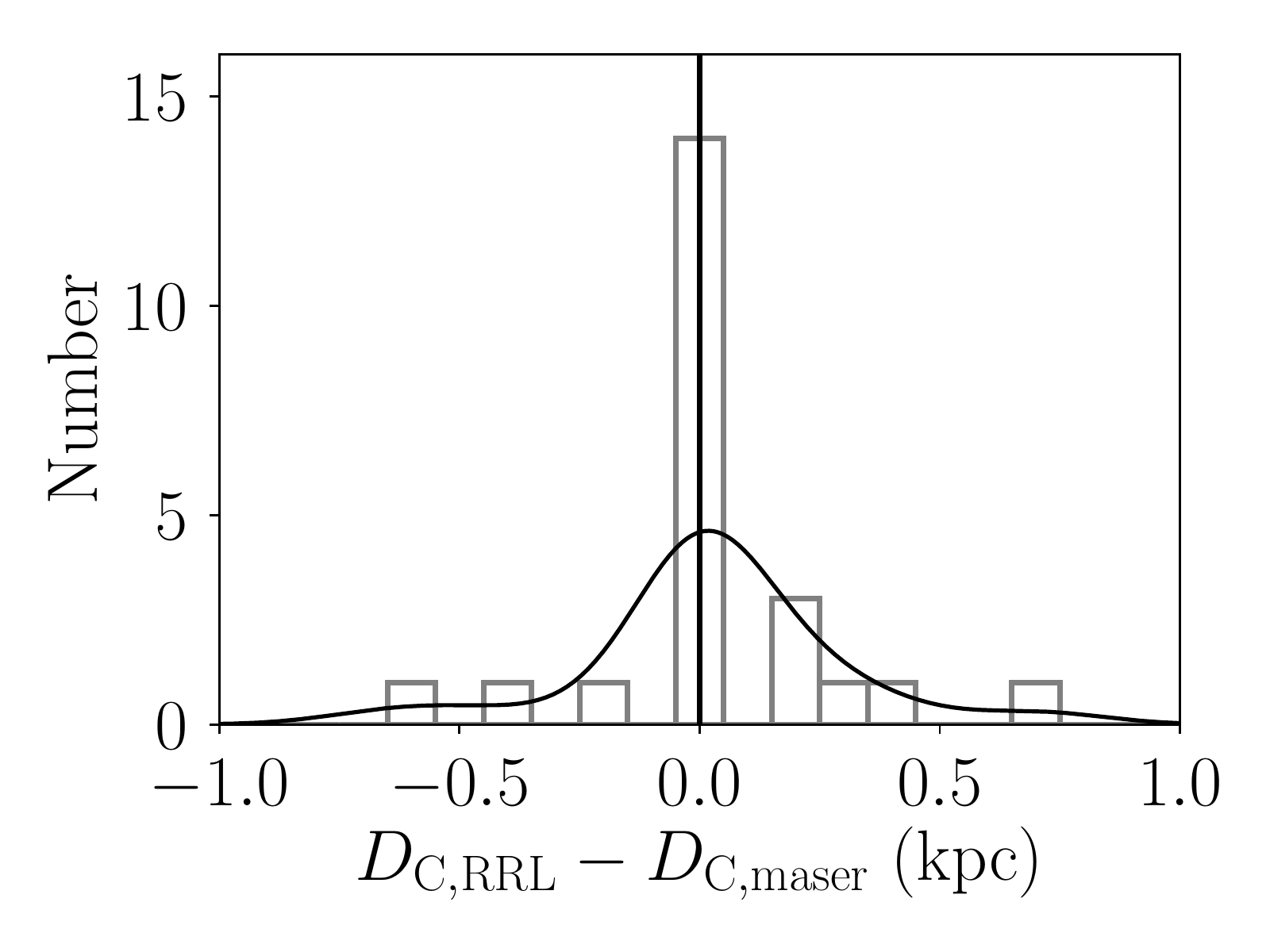} 
  \includegraphics[width=0.45\linewidth]{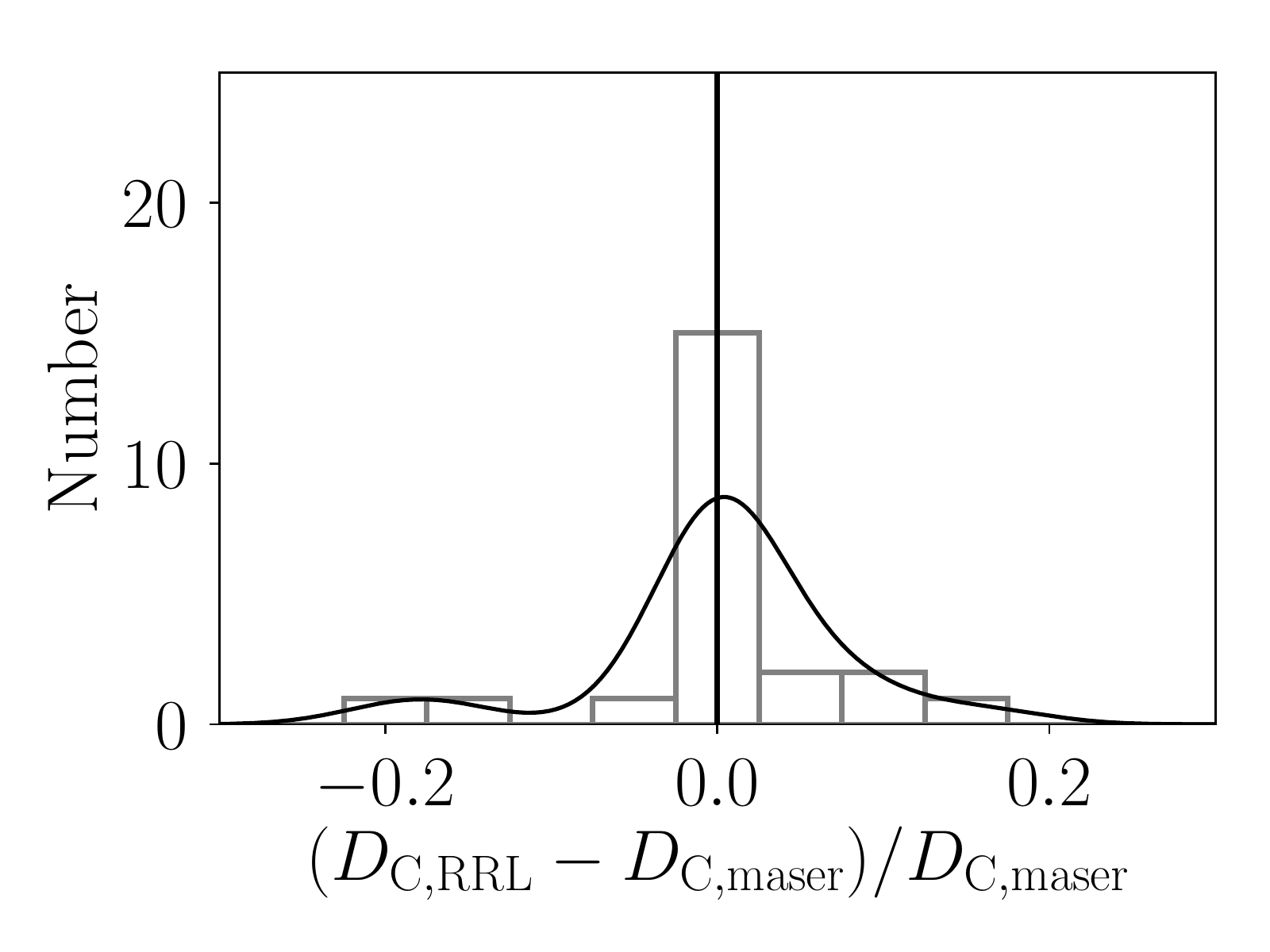} \\
  \includegraphics[width=0.45\linewidth]{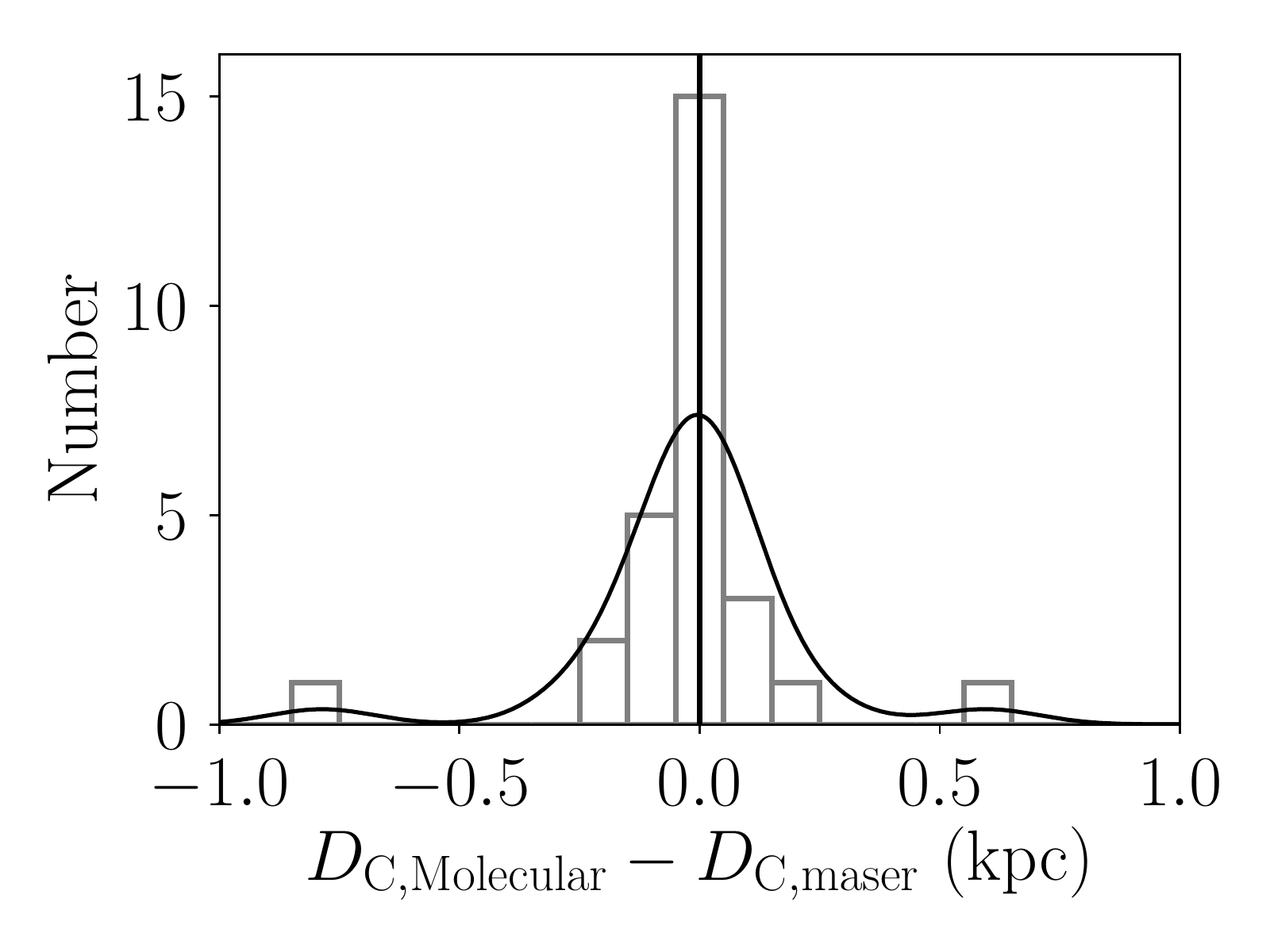}
  \includegraphics[width=0.45\linewidth]{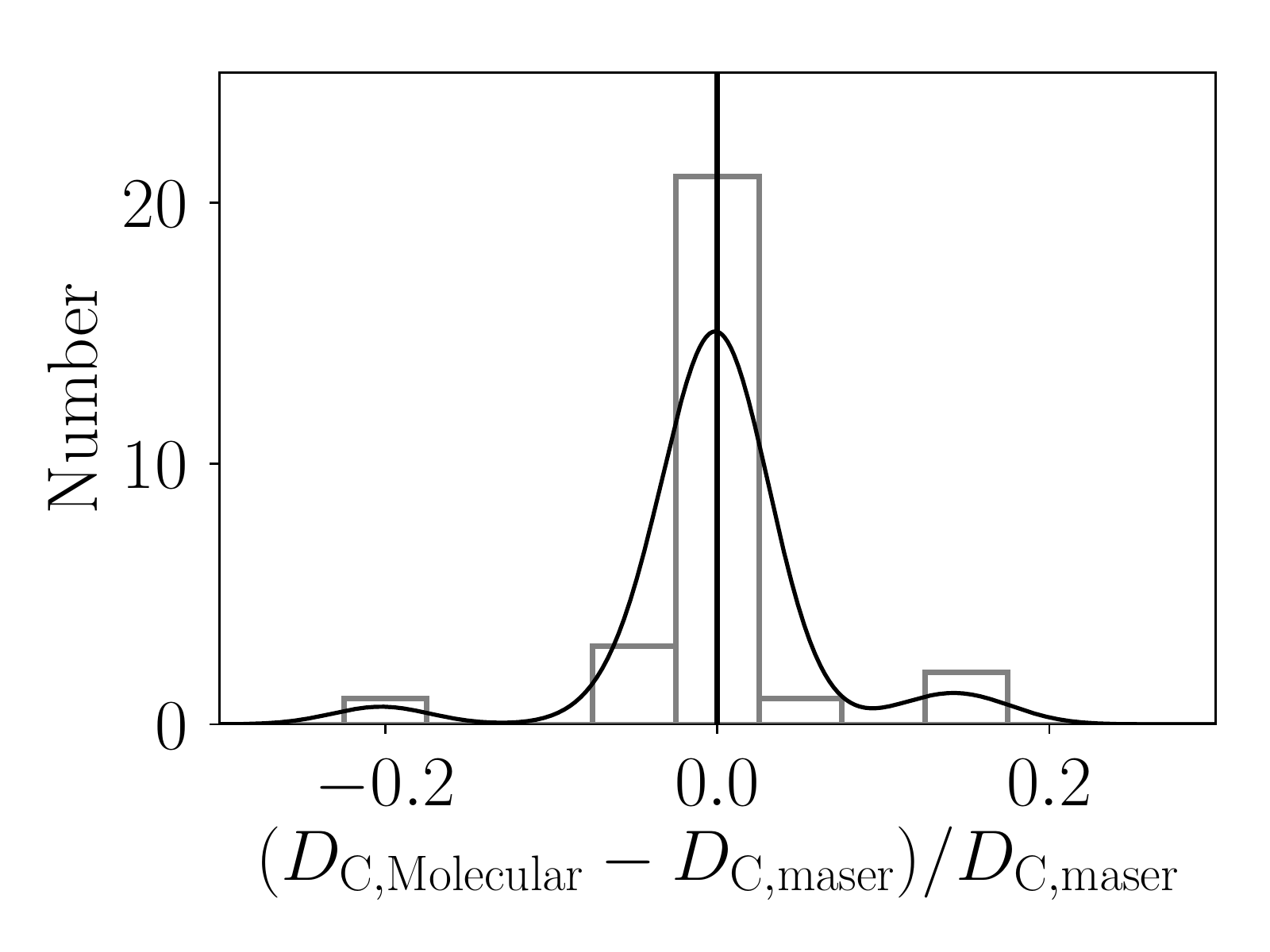}
  \caption{Difference between the Method C kinematic distances derived
    using the RRL and maser LSR velocities (top) and molecular line
    and maser LSR velocities (bottom). The absolute difference is
    shown in the left panels and the fractional difference is shown in
    the right panels. The solid curve is the KDE fit to each
    distribution, and the vertical line is the median.}
  \label{fig:distance_comp}
\end{figure*}

We first compare the LSR velocities of non-maser transitions in the
\textit{WISE} Catalog to the maser velocities from
\citet{reid2014}. The \textit{WISE} Catalog contains RRL velocities
and/or non-maser molecular spectral line velocities for 34 HMSFRs: 6
regions with only RRL velocities, 11 regions with only molecular line
velocities, and 17 regions with both. Since the RRL emission comes
from the ionized gas of the HMSFR and the non-maser molecular line
emission comes from molecular clouds associated with the HMSFR, the
LSR velocities of these transitions need not be the same as that of
the maser emission, which originates within the molecular envelope of
the high mass stars.  Figure~\ref{fig:velocity} shows the difference
between the RRL and maser velocities and the difference between the
molecular line velocities and maser velocities. The median difference
is \(-0.70\kms\) with a standard deviation of \(3.83\kms\) for RRL
velocities and \(0.55\kms\) with a standard deviation of \(2.73\kms\)
for molecular line velocities. These distributions are consistent with
the expected \({\sim}10\kms\) difference between maser spot emission
region motions and bulk gas motions
\citep[e.g.,][]{reid2009b,reid2014}. The difference in LSR velocities
corresponds to differences in kinematic distances.  The differences
between the Method C kinematic distances derived using the RRL,
molecular, and maser velocities are shown in
Figure~\ref{fig:distance_comp}. The median difference is \(0\kpc\) for
both the RRL and maser distance difference and the molecular and maser
distance difference, with standard deviations of \(0.25\kpc\) and
\(0.21\kpc\), respectively. The maximum fractional difference is
\(20\%\) in both cases, which implies that the choice of LSR velocity
tracer has a moderate impact on the derived kinematic distance.

If we limit our sample to inner-Galaxy \textit{WISE} Catalog objects
more than \(20\kms\) from the tangent point velocity using the
\citet{reid2014} GRM, there are 9 HMSFRs. Of these, the KDAs are
resolved using: \hi\ emission/absorption and self-absorption
experiments based on RRL velocities \citep[2
  objects;][]{anderson2009a,anderson2012}, \hi\ self-absorption
experiments based on molecular line velocities \citep[2
  objects;][]{urquhart2012,roman-duval2009}, and H\(_2\)CO absorption
experiments \citep[4 objects;][]{araya2002,watson2003,sewilo2004}.
One object is a visible \hii\ region and thus likely located at the
near distance.

Based on the KDAR determined using the Method C kinematic distance
method and selecting the distance closest to the parallax distance,
the \textit{WISE} Catalog has incorrect KDARs for 3 of our sample
objects: one source (G034.39+00.22) using an \hi\ self-absorption
experiment based on RRL velocities \citep{anderson2009a} and two
sources (G023.70-00.19, G035.02+00.34) using H\(_2\)CO absorption
experiments \citep{watson2003,sewilo2004}. The \citet{anderson2009a}
KDAR resolution for G034.39+00.22 was determined only with
\hi\ self-absorption techniques, and, as the authors show in that
paper, \hi\ self-absorption techniques are much less reliable than
\hi\ emission/absorption techniques. Too, this object had a low
confidence \hi\ self-absorption detection (quality factor B in that
paper).  The H\(_2\)CO absorption spectra for the other two sources
are marginal detections. The absorption feature for (G023.70$-$00.19)
is on the wing of the RRL \citep{sewilo2004}, and the absorption
feature for G035.02+00.34 is weak and \({\sim}5\kms\) beyond the
tangent point velocity \citep{watson2003}. This sample size is too
small to make any definitive conclusions about the accuracy of the
KDAR techniques. Authors using the \textit{WISE} Catalog KDARs should
investigate the original KDAR work to assess the quality of the
distance resolution.

\section{Discussion}

Based on the results of this analysis, we recommend the following
prescription for deriving kinematic distances: (1) correct the
measured LSR velocity using the \citet{reid2014} Solar motion
parameters and Equations~\ref{eq:helio} and \ref{eq:vlsr}; (2) use the
corrected LSR velocity and the Monte Carlo method (Method C) to derive
the kinematic distances and uncertainties; and (3) use only the
highest quality KDARs from the \textit{WISE} Catalog (if available) to
resolve the kinematic distance ambiguity. The \textit{Python} code we
used to calculate the Monte Carlo kinematic distances is publicly
available and may be utilized through an online
tool\footnote{\url{http://doi.org/10.5281/zenodo.1166001}}
\citep{kdutils2017}.

Changing the method used to derive kinematic distances may have
important implications. When applying kinematic distances to Galactic
morphological or metallicity structure analyses, it is important to
consider the kinematic distance uncertainties and inaccuracies in the
KDAR techniques. For example, \citet{koo2017} recently re-analyzed the
Leiden/Argentine/Bonn \hi\ 21 cm line all-sky survey
\citep{hartmann1997,arnal2000,bajaja2005,kalberla2005} to characterize
the spiral structure in the outer Galaxy. They derived kinematic
distances to their \hi\ features to produce a face-on map of the
\hi\ distribution beyond the Solar orbit. Even though there is no KDA
in this part of the Galaxy, their kinematic distances will be affected
by the uncertainties discussed here. Their results, determining the
pitch angles of the spiral features for example, may change
significantly if they use the Monte Carlo method to derive the
kinematic distances of their \hi\ features.

Monte Carlo kinematic distances will also affect the interpretation of
Galactic metallicity structure. For example, \citet{balser2015}
recently discovered azimuthal variations in the radial metallicity
gradient of the Milky Way inferred by the electron temperatures of
Galactic \hii\ regions. They used the \citet{reid2014} rotation curve
to derive their kinematic distances and the A12 kinematic distance
uncertainty model to assign distance uncertainties. After resampling
their \hii\ region distances within the A12 uncertainties, they
determined that the azimuthal metallicity gradient variations were
statistically significant. The \citet{balser2015} result may be
affected by the results of this analysis. Not only will the kinematic
distances for their sample of \hii\ regions change slightly, the
uncertainties will change as well.  These changes will affect the
statistical significance of their result.

This new kinematic distance method will affect all distance estimation
techniques that rely, at least in part, on kinematic distances. For
example, \citet{reid2016a} used a Bayesian distance estimation method
to derive the distance to HMSFRs. The priors in their method included
the parallax distance (if available), the kinematic distance with
equal weight given to both the near and far kinematic distance, the
Galactic latitude, and a spiral arm model of the Galaxy. Instead of
using a Gaussian kinematic distance PDF, future Bayesian analyses
should use the full Monte Carlo kinematic distance PDF.

\begin{figure}[ht]
  \centering
  \includegraphics[width=\linewidth]{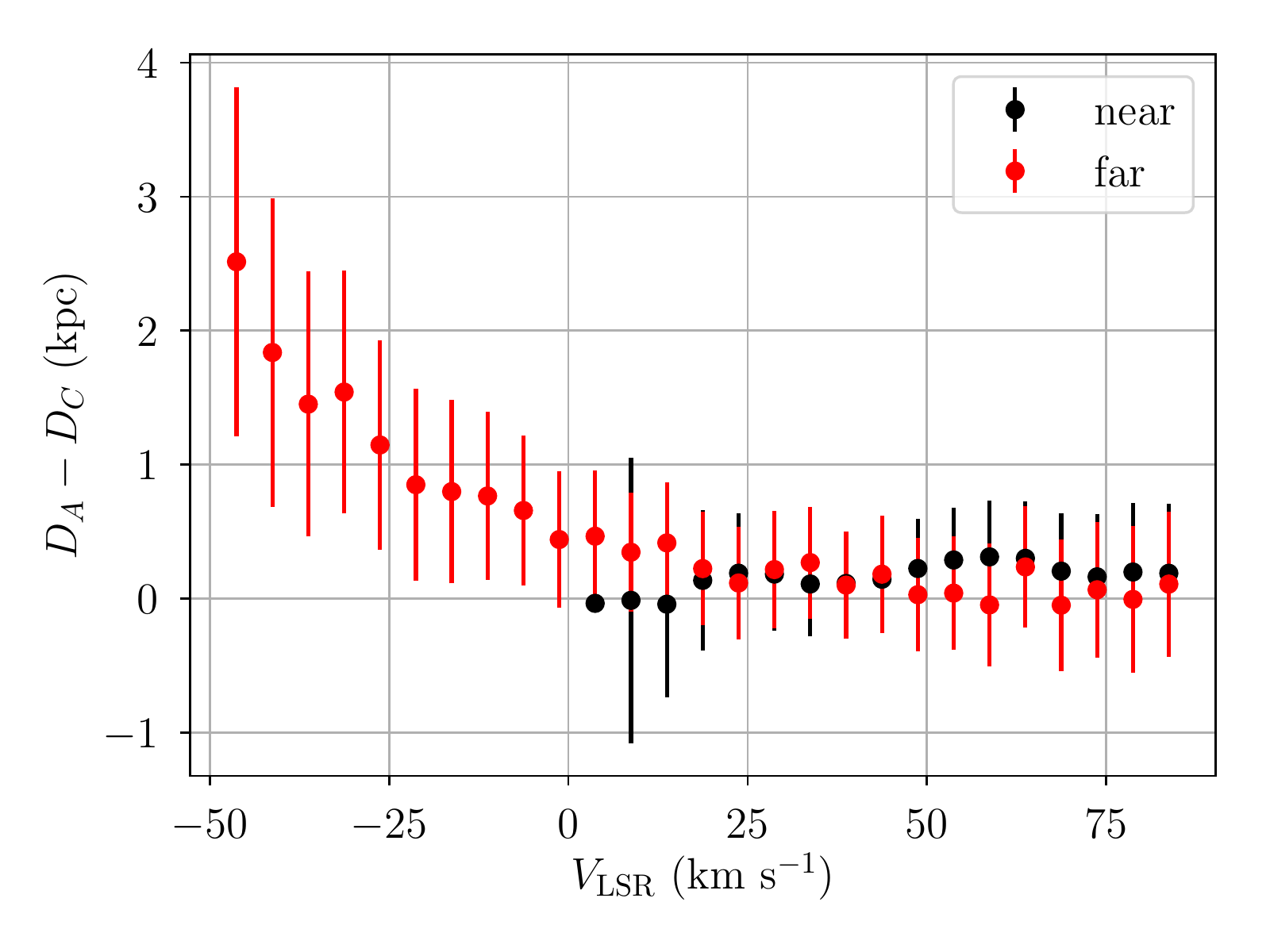}
  \includegraphics[width=\linewidth]{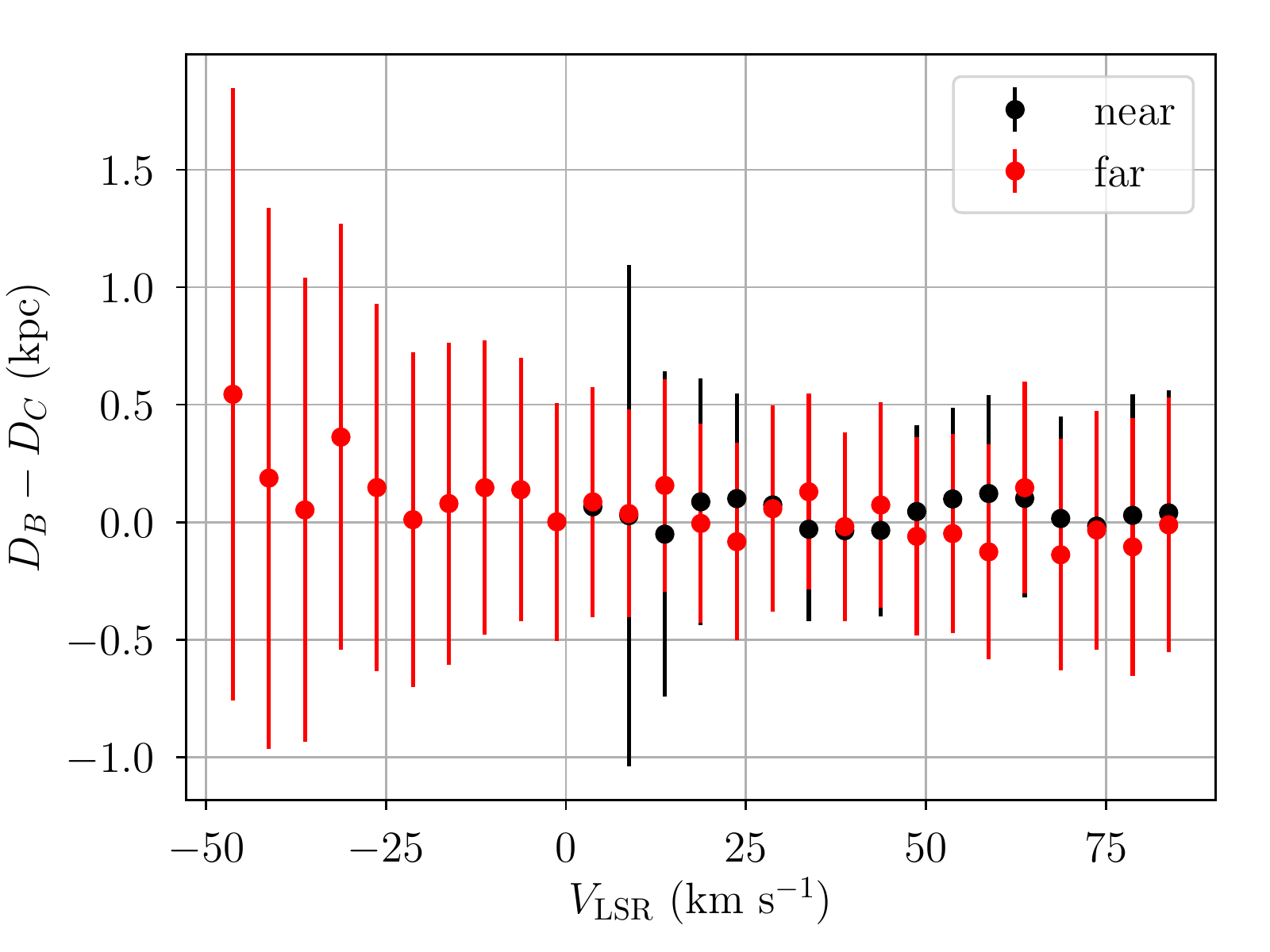}
  \caption{Difference between Method A and Method C kinematic
    distances (top) and Method B and Method C kinematic distances
    (bottom) as a function of LSR velocity in the direction
    \(\ell=30^\circ\). The black points correspond to near kinematic
    distances and the red points correspond to far kinematic
    distances. The error bars are the combined uncertainties of both
    the A12 kinematic distance uncertainty model (Methods A and B) and
    the Method C Monte Carlo uncertainty. We exclude \(20\kms\) near
    the tangent point for clarity.}
  \label{fig:30long_dist}
\end{figure}

The difference between Method A and Method C kinematic distances is
fairly large, whereas the difference between Method B and Method C is
small. Figure~\ref{fig:30long_dist} shows the difference between the
Method A and C distances as well as the Method B and C distances for
LSR velocities along \(\ell=30^{\circ}\). We choose this line-of-sight
because it crosses both the inner and outer Galaxy through most of the
Galactic disk. The difference between Method A and C is \(<0.5\kpc\)
within the Solar orbit, and approaches \(3\kpc\) at a distance of
\(20\kpc\). This discrepancy is caused by the variations in the GRMs
used by each method. The difference between Method B and C, however,
is small (\(\lsim0.5\kpc\)) across the Galaxy since both methods use
the same GRM.

\begin{figure}[ht]
  \centering
  \includegraphics[width=\linewidth]{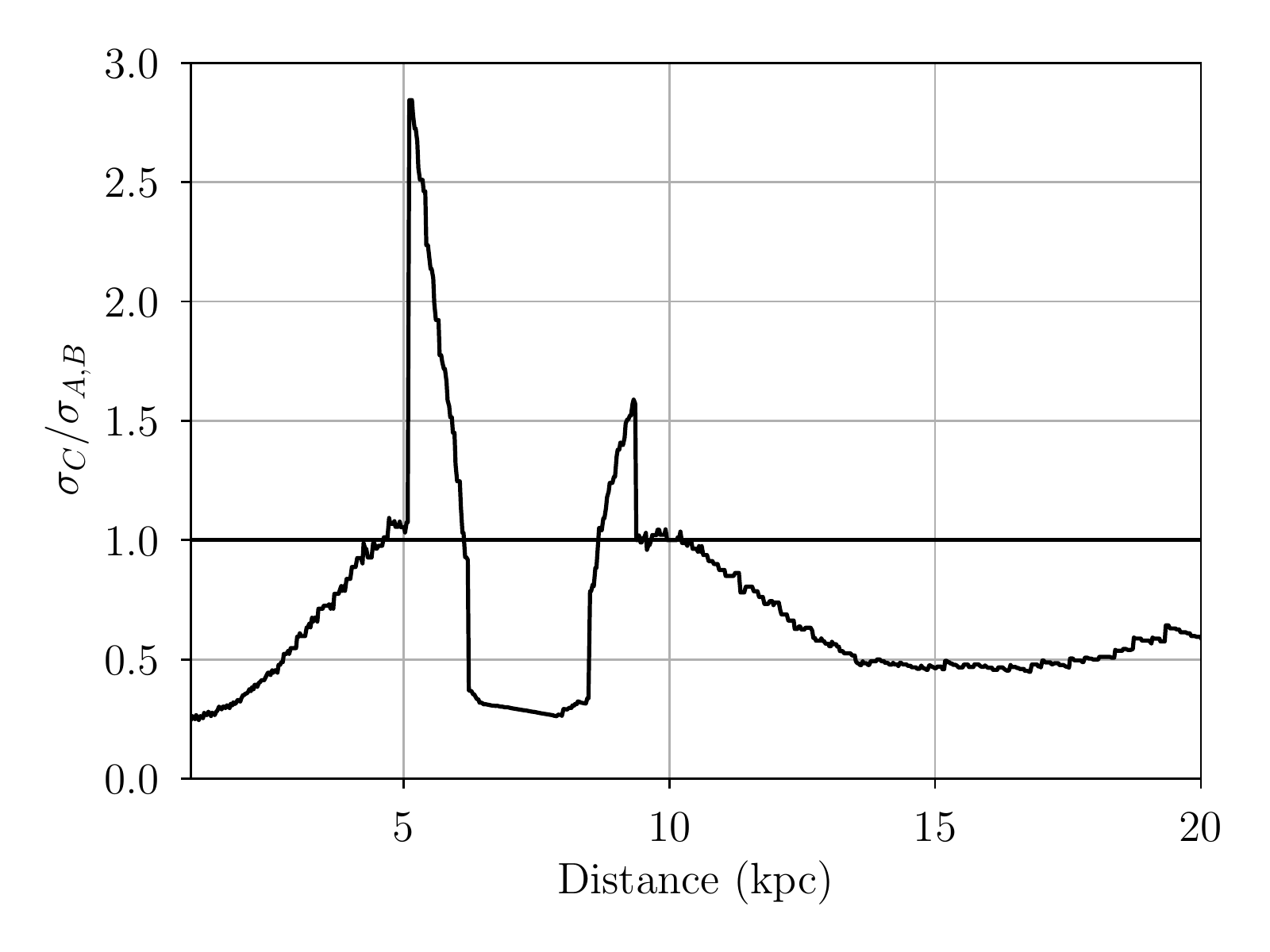}
  \caption{Ratio of the Method C Monte Carlo kinematic distance
    uncertainty to the A12 kinematic distance uncertainty model
    (Methods A and B) as a function of distance in the direction
    \(\ell=30^\circ\). The solid vertical line indicates a ratio of
    one where \(\sigma_C = \sigma_{A,B}\). The spikes near \(5\kpc\)
    and \(9\kpc\) are at the boundaries of the ``tangent point
    region,'' defined where the LSR velocity is within \(20\kms\) of
    the tangent point velocity.}
  \label{fig:kd_dist_err_ratio}
\end{figure}

The largest distinction between the different kinematic distance
methods is the magnitude of the uncertainties. In
Figure~\ref{fig:kd_dist_err_ratio} we show the ratio of the Method C
kinematic distance uncertainty to those of Methods A and B (the A12
model) along \(\ell=30^\circ\). Except near the tangent point, the
Method C kinematic distance uncertainties are smaller than the A12
model uncertainties. At a distance of \(15\kpc\), the Model C
uncertainty is half of the A12 model uncertainty. The spikes near
\(5\kpc\) and \(9\kpc\) are located at the boundaries of the ``tangent
point region'' (within \(20\kms\) of the tangent point
velocity). Here, the Method C kinematic distance uncertainties are
much larger than the A12 model uncertainties.

\begin{figure}[ht]
  \centering
  \includegraphics[width=\linewidth]{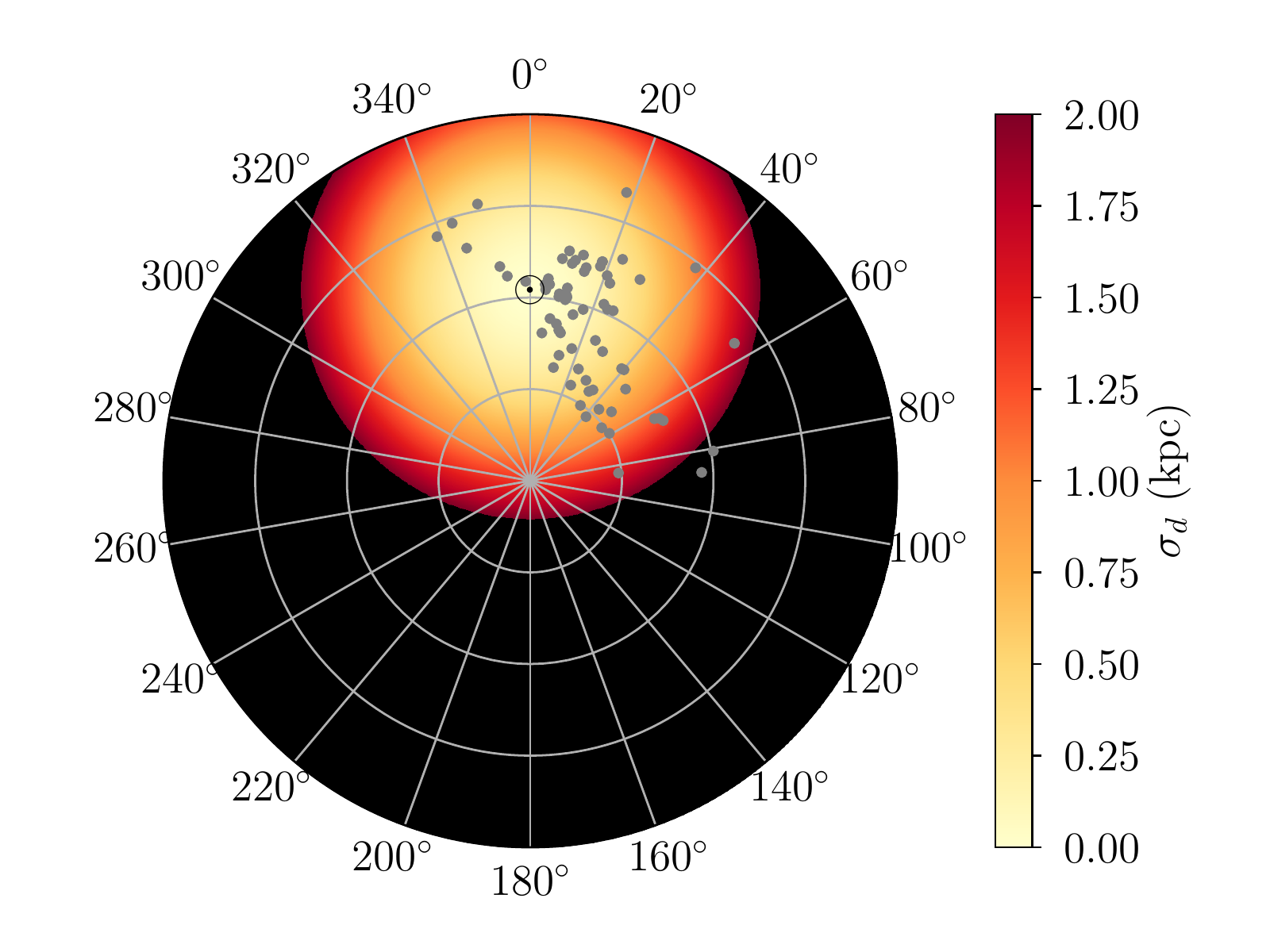}
  \includegraphics[width=\linewidth]{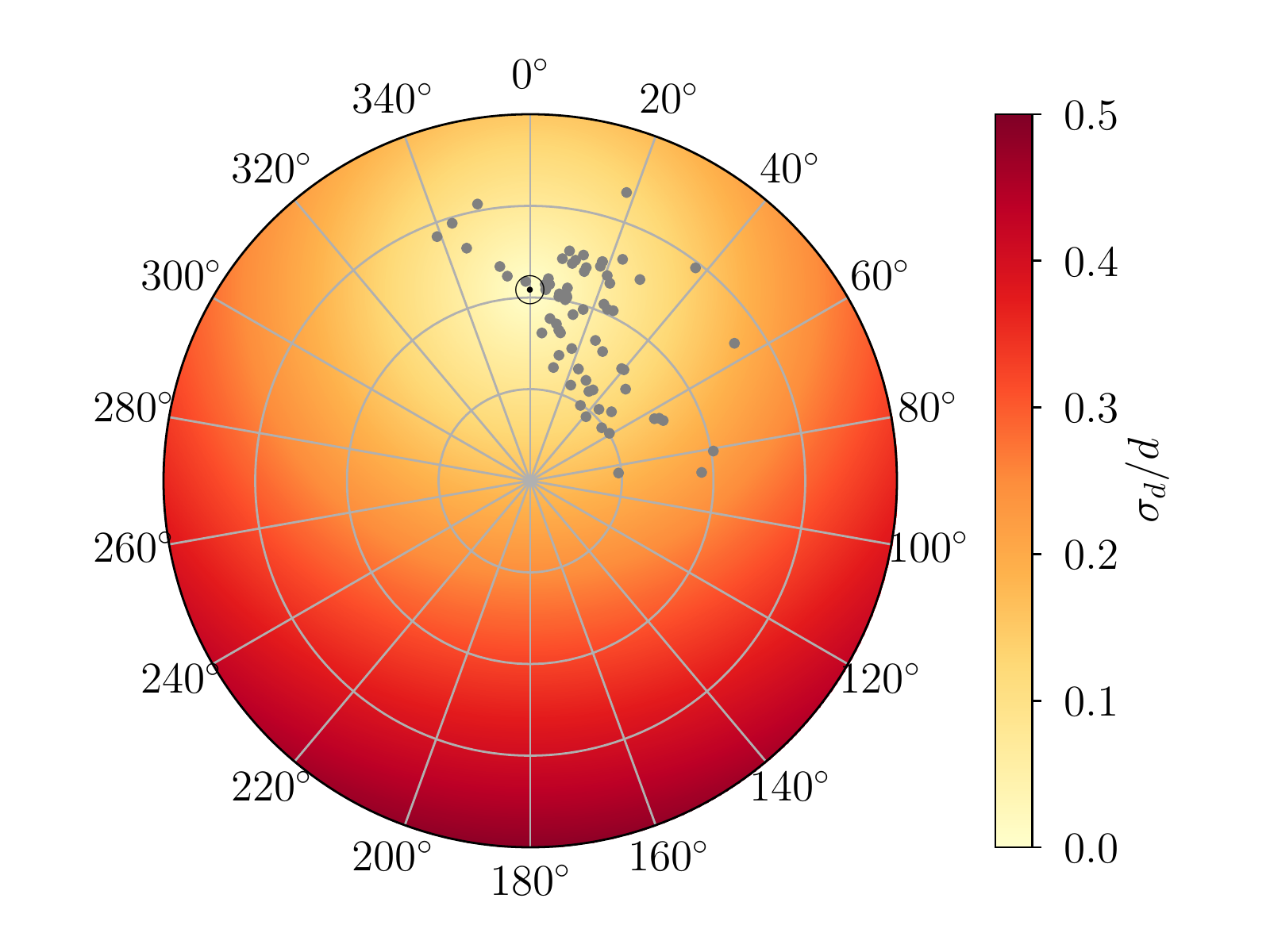}
  \caption{Face-on Galactic view of the parallax distance
    uncertainties assuming a typical parallax uncertainty of
    \(0.02\,\mu\text{as}\). The top panel is the absolute distance
    uncertainty and the bottom panel is the fractional distance
    uncertainty. The Galactic Center is located at the origin and the
    Sun is located 8.34 kpc in the direction \(\az=0^\circ\). The
    concentric circles are 4, 8, and 12 kpc in \(R\) and \(\az\) is
    given in degrees. The color represents the distance
    uncertainty. The black regions represent distance uncertainties
    greater than \(\sigma_d = 2\kpc\) (top) or \(\sigma_d/d = 0.5\)
    (bottom). The gray points are the HMSFRs in our sample.}
  \label{fig:parallax_dist_unc}
\end{figure}

\begin{figure}[ht]
  \centering
  \includegraphics[width=\linewidth]{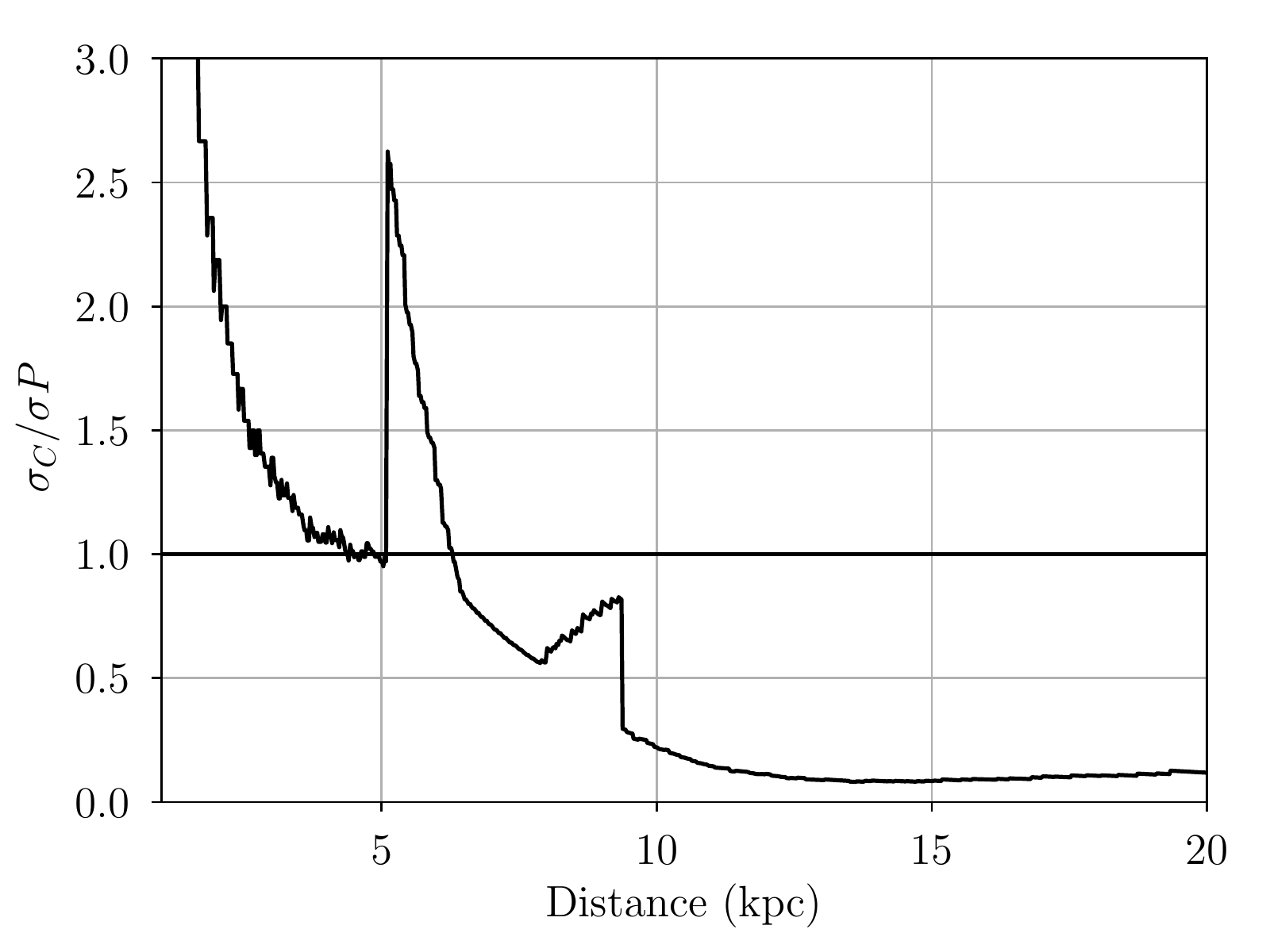}
  \caption{Ratio of the Method C Monte Carlo kinematic distance
    uncertainty to the typical parallax distance uncertainty as a
    function of distance in the direction \(\ell=30^\circ\). The solid
    vertical line indicates a ratio of one where \(\sigma_C =
    \sigma_P\). The spikes near \(5\kpc\) and \(9\kpc\) are at the
    boundaries of the ``tangent point region,'' defined where the LSR
    velocity is within \(20\kms\) of the tangent point velocity.}
  \label{fig:para_dist_err_ratio}
\end{figure}

Although kinematic distances are not as accurate as parallax distances
in the Solar neighborhood, their accuracy is much better in distant
regions of the Milky Way. To demonstrate this point, we generate a
face-on view of the typical parallax distance uncertainty in the
Galaxy (Figure~\ref{fig:parallax_dist_unc}). We assume a
characteristic parallax uncertainty of \(0.02\,\mu\text{as}\)
\citep{reid2014rev} which corresponds to a typical parallax distance
uncertainty of \(\sigma_d/\text{kpc} = 0.02(d/\text{kpc})^2\). This
figure uses the same color scale as the Method C Monte Carlo kinematic
distance uncertainty map in Figure~\ref{fig:pdf_uncertainty}.  By
comparing these figures we see that large regions of Galactic
quadrants I and IV (\(-90^\circ < \ell < 90^\circ\)) have Method C
kinematic distance uncertainties much smaller than the typical
parallax distance uncertainties. In
Figure~\ref{fig:para_dist_err_ratio} we show the ratio of the Method C
kinematic distance uncertainty to the typical parallax distance
uncertainty along \(\ell=30^\circ\). Beyond the tangent point at a
distance of \({\sim}8\kpc\), the Method C kinematic distance
uncertainties are smaller than the typical parallax distance
uncertainty. This ratio reaches a minimum at about \(14\kpc\) where
the Method C kinematic distance uncertainty is less than \(10\%\) of
the typical parallax distance uncertainty. The spikes near \(5\kpc\)
and \(9\kpc\) are, again, located at the boundaries of the ``tangent
point region,'' where the Method C kinematic distance uncertainties
are much larger.

\begin{figure}[ht]
  \centering
  \includegraphics[width=\linewidth]{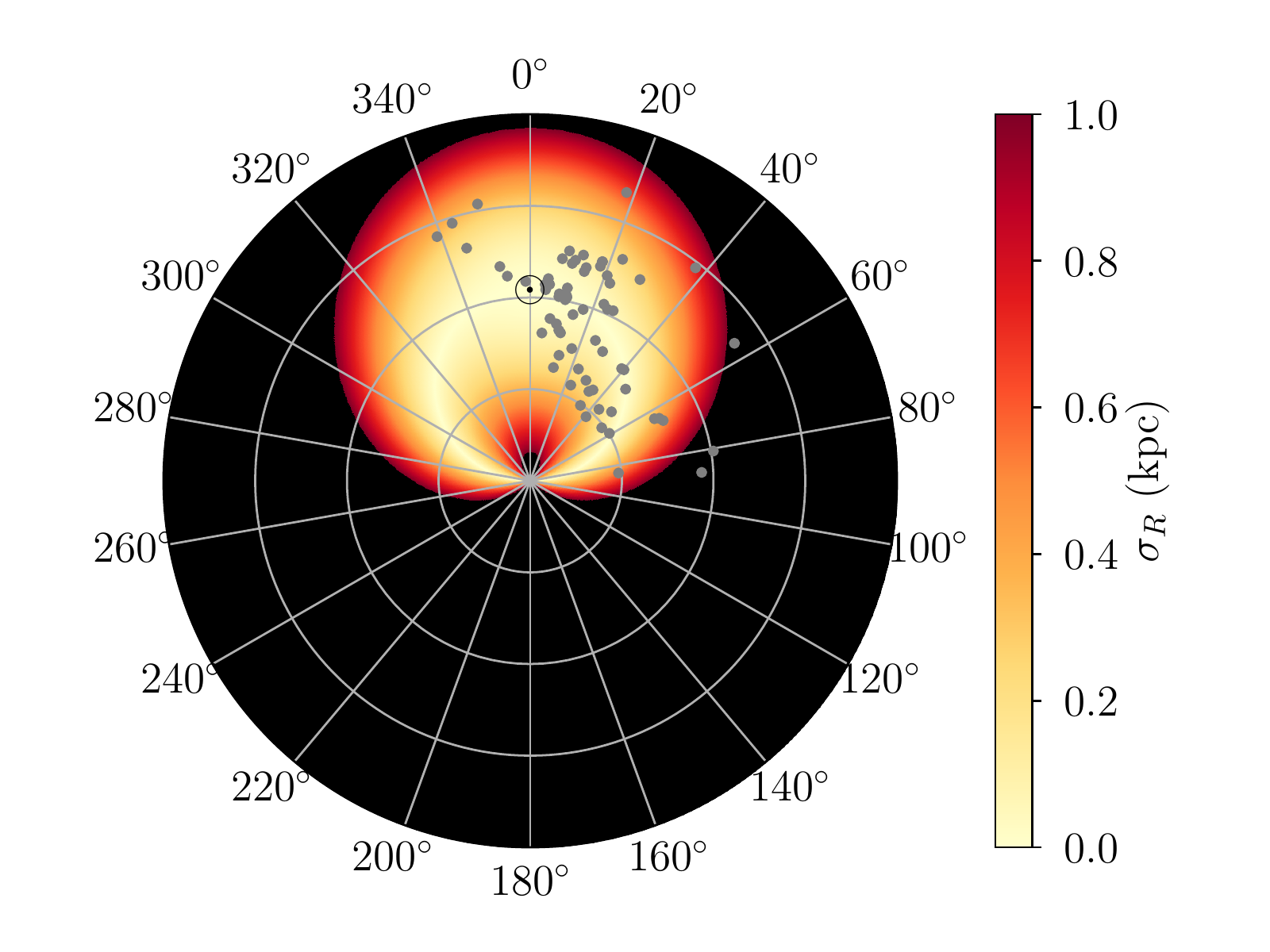} \\
  \includegraphics[width=\linewidth]{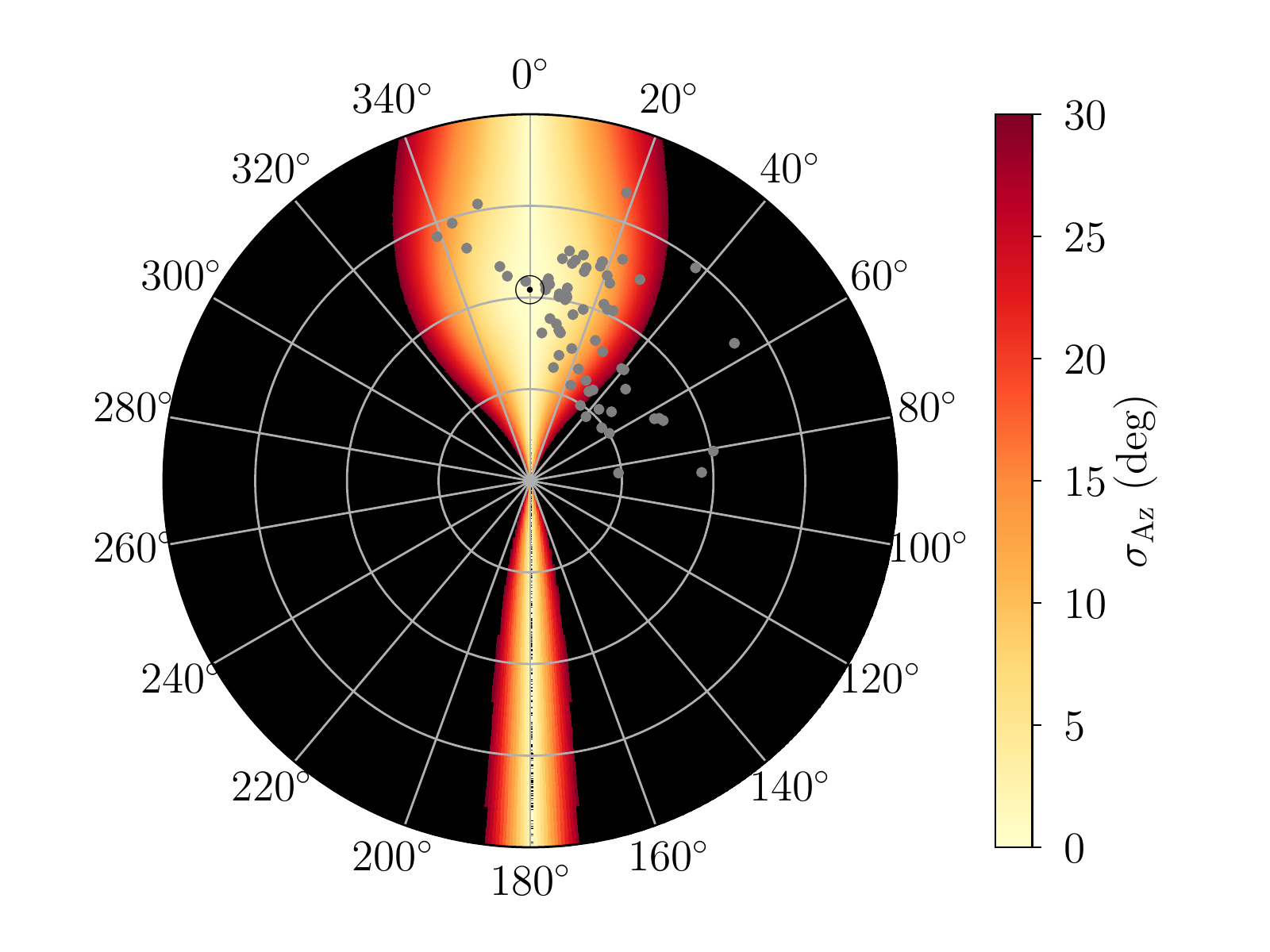}
  \caption{Face-on Galactic view of the typical parallax distance
    uncertainties converted to Galactocentric coordinates, \(R\) (top)
    and \(\az\) (bottom). The Galactic Center is located at the origin
    and the Sun is located 8.34 kpc in the direction
    \(\az=0^\circ\). The concentric circles are 4, 8, and 12 kpc in
    \(R\) and \(\az\) is given in degrees. The color represents the
    distance uncertainty. The black regions have uncertainties larger
    than the maximum value shown in the color scale. The gray points
    are the HMSFRs in our sample.}
  \label{fig:parallax_unc}
\end{figure}

\begin{figure*}[ht]
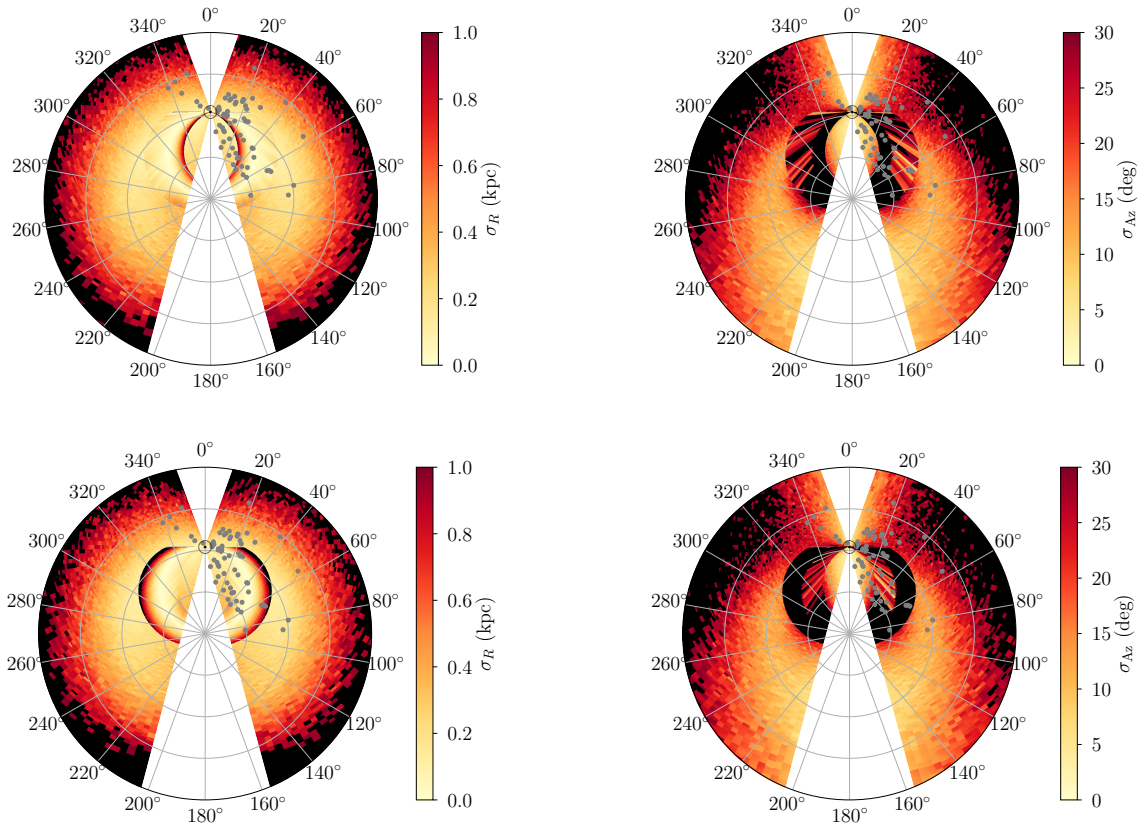

  \plottwo{{pdf_neg_Rgal}.pdf}{{pdf_neg_az}.pdf}
  \plottwo{{pdf_pos_Rgal}.pdf}{{pdf_pos_az}.pdf}
  \caption{Face-on Galactic view of the Monte Carlo kinematic distance
    uncertainties converted to Galactocentric coordinates, \(R\)
    (left) and \(\az\) (right). The top figures are the distance
    uncertainties in the negative direction while the bottom figures
    are the distance uncertainties in the positive direction. The
    Galactic Center is located at the origin and the Sun is located
    8.34 kpc in the direction \(\az=0^\circ\). The concentric circles
    are 4, 8, and 12 kpc in \(R\) and \(\az\) is given in degrees. The
    color represents the distance uncertainty. The black regions have
    uncertainties larger than the maximum value shown in the color
    scale. The regions \(-15^\circ < \gl < 15^\circ\) and \(160^\circ
    < \gl < 200^\circ\) are masked in white since kinematic distances
    are very inaccurate towards the Galactic Center and Galactic
    Anti-center. The gray points are the HMSFRs in our sample.}
  \label{fig:pdf_unc}
\end{figure*}

The accuracy of the Method C kinematic distances is especially
apparent when we consider that Galactic structure analyses are more
interested in the Galactocentric positions of structure tracers
(\(R,\az,z\)) than the heliocentric positions (\(\ell,b,d\)).  We
derive the relationship between the distance uncertainty and
uncertainties in \(R\) and \(\az\) in
Appendix~\ref{sec:unc_derivation}.  Figure~\ref{fig:parallax_unc}
shows the face-on uncertainties in Galactocentric position given these
uncertainties in parallax distance. The same analysis using the Monte
Carlo kinematic distance uncertainties is shown in
Figure~\ref{fig:pdf_unc}. In large regions of Galactic quadrants I and
IV (\(-90^\circ < \ell < 90^\circ\)), the Monte Carlo kinematic
distances have smaller uncertainties in both \(R\) and \(\az\) than
the parallax distances. Kinematic distances therefore determine not
only the distance of objects, but also the \textit{Galactocentric}
position of objects more accurately than parallax distances when the
object is far from the Solar neighborhood.

Streaming motions will have a \textit{systematic} effect on the
accuracy of kinematic distances rather than a \textit{random} effect
as we have assumed in this analysis. With a much larger catalog of
parallax observations of HMSFRs, we could compare kinematic and
parallax distances and uncover any systematic differences. We may then
be able to create a non-axisymmetric GRM that includes these
non-circular motions. Such a task requires parallax observations
uniformly across the entire Galactic disk.

\section{Conclusions}

We investigate the accuracy of kinematic distances by comparing the
kinematic and parallax distances of 75 Galactic HMSFRs. We derive the
kinematic distances using three different methods: the traditional
method using the \citet{brand1993} rotation curve and the IAU-defined
Solar motion parameters (Method A), the traditional method using the
\citet{reid2014} rotation curve and their revised Solar motion
parameters (Method B), and a new Monte Carlo method using the
\citet{reid2014} rotation curve and their revised Solar motion
parameters (Method C). The best agreement between the kinematic and
parallax distances is when we use Method C. In this case, the median
absolute difference between the kinematic distances and parallax
distances is \(0.71\kpc\) with a standard deviation of
\(0.83\kpc\). The Method C kinematic distance uncertainties are
smaller than those of Methods A and B for most of the Galaxy except
near the tangent point. Along the line-of-sight with
\(\ell=30^\circ\), for example, the Method C kinematic distance
uncertainty is \(50\%\) of the Method A and B uncertainties at a
distance of \(15\kpc\).  We test the accuracy of KDAR techniques using
the KDARs derived in the literature for 9 of our inner-Galaxy,
non-tangent point HMSFRs. The KDAR is incorrect in 3 cases when using
the \textit{WISE} catalog KDARs to compare the parallax distances to
our Monte Carlo kinematic distances, but each of these KDARs are
low-quality determinations.

We recommend a new prescription for deriving and applying kinematic
distances and their uncertainties: (1) correct the measured LSR
velocity using the \citet{reid2014} Solar motion parameters and
Equations~\ref{eq:helio} and \ref{eq:vlsr}; (2) use the corrected LSR
velocity and the Monte Carlo method (Method C) to derive the kinematic
distances and uncertainties; and (3) use only the highest quality
KDARs from the \textit{WISE} Catalog to resolve the kinematic distance
ambiguity. Based on the typical parallax distance uncertainties, we
show that, in a large region of Galactic quadrants I and IV
(\(-90^\circ < \ell < 90^\circ\)), both the distances and the
Galactocentric positions of HMSFRs are more accurately constrained by
the Method C kinematic distances than parallax distances. In the
direction \(\ell=30^\circ\), for example, the Method C kinematic
distance uncertainties are smaller than the parallax distance
uncertainties everywhere beyond the tangent point, reaching a minimum
of \(10\%\) of the parallax distance uncertainty at a distance of
\(14\kpc\). The code to derive the Method C Monte Carlo kinematic
distances and kinematic distance uncertainties is publicly available
and may be utilized through an on-line tool. In a future paper, we
will investigate the effects of using the Monte Carlo kinematic
distances on the interpretation of Galactic morphological and
metallicity structure.

\acknowledgments

TVW is supported by the NSF through the Grote Reber Fellowship Program
administered by Associated Universities, Inc./National Radio Astronomy
Observatory, the D.N. Batten Foundation Fellowship from the Jefferson
Scholars Foundation, the Mars Foundation Fellowship from the
Achievement Rewards for College Scientists Foundation, and the
Virginia Space Grant Consortium. LDA is supported by NSF grant
AST1516021. TMB is supported by NSF grant AST1714688. We thank the
anonymous referee for useful comments and suggestions that improved
the quality of this paper.

\nraoblurb

\software{Astropy \citep{astropy2013},
  KDUtils \citep{kdutils2017},
  Matplotlib \citep{matplotlib2007},
  NumPy \& SciPy \citep{numpyscipy2011},
  Pandas \citep{pandas2010},
  PyQt-Fit (\url{http://pyqt-fit.readthedocs.io/}),
  Python (\url{https://www.python.org/})}

\bibliography{distances}

\appendix
\section{Galactocentric Position Uncertainty Derivations \label{sec:unc_derivation}}

Here we derive the relationship between uncertainties in the distance
to an object, \(d\), and the uncertainties in its Galactocentric
position, (\(R,\az\)). For simplicity, we assume all
objects are in the Galactic plane (\(b=0^\circ\) and \(z = 0\)).

\begin{figure}[ht]
  \centering
  \includegraphics[width=0.6\linewidth]{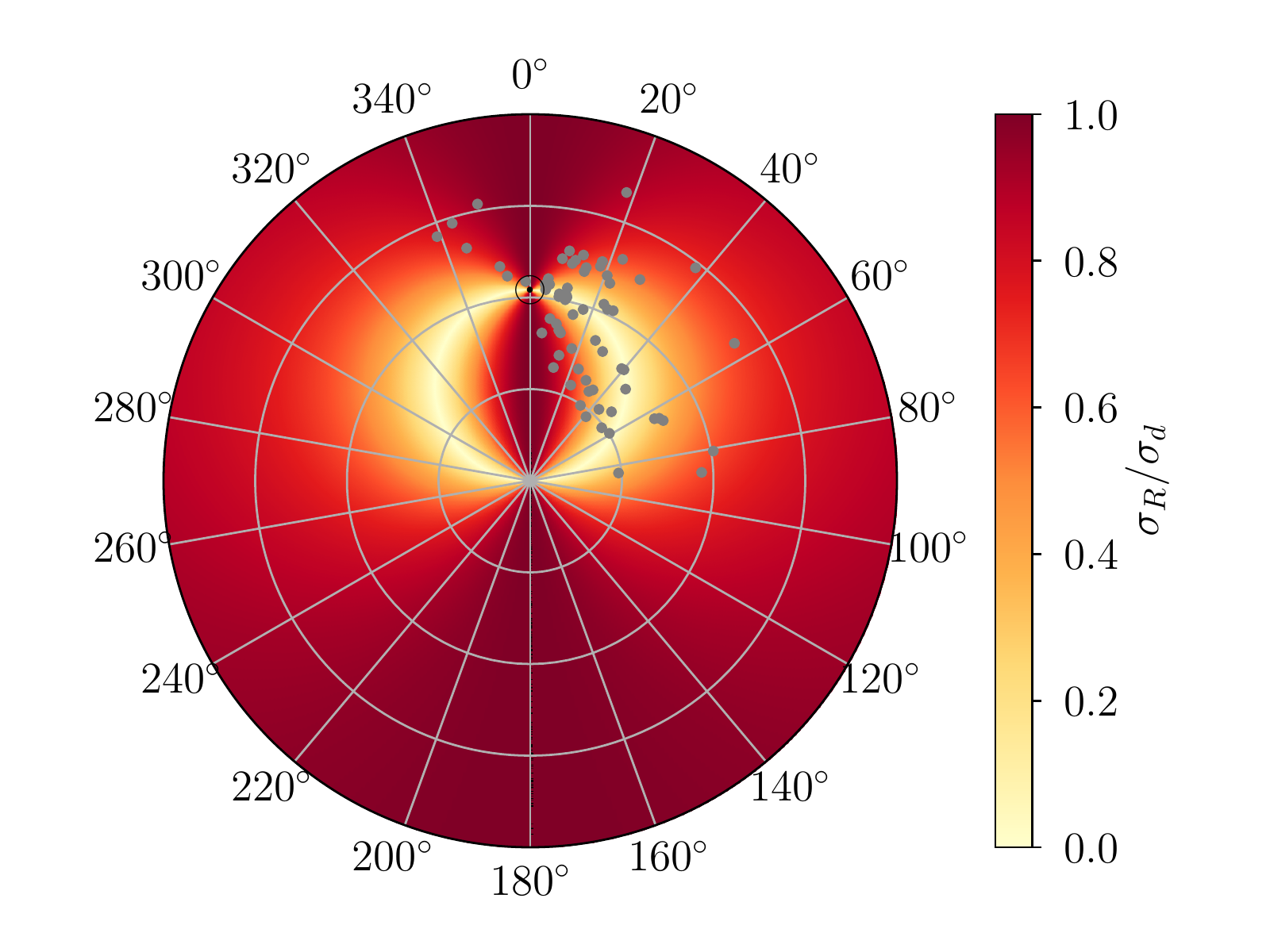}
  \includegraphics[width=0.6\linewidth]{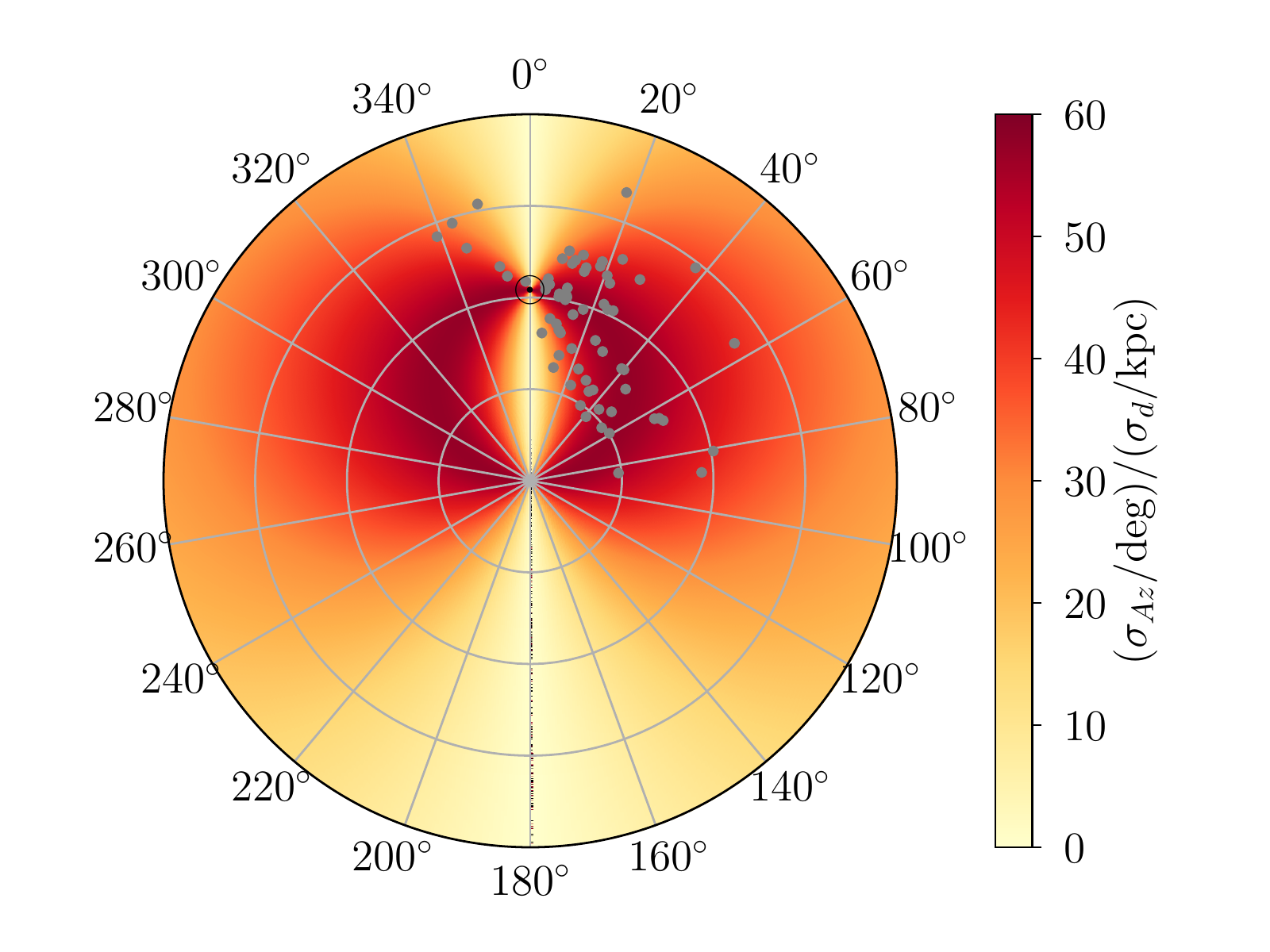}
  \caption{The relationship between the uncertainty in distance,
    \(\sigma_d\), and the uncertainty in Galactocentric radius
    (\(\sigma_R\), Equation~\ref{eq:sigma_R}, top) and Galactocentric
    azimuth (\(\azerr\), Equation~\ref{eq:sigma_Az}, bottom). The
    Galactic Center is located at the origin and the Sun is located
    8.34 kpc in the direction \(\az=0^\circ\). The concentric circles
    are 4, 8, and 12 kpc in \(R\) and \(\az\) is given in degrees. The
    color represents the uncertainty ratio. The gray points are the
    HMSFRs in our sample.}
  \label{fig:faceon_unc}
\end{figure}

An object's Galactocentric radius is given by
\begin{equation}
  R = \left(d^2 + R_0^2 - 2d R_0 \cos\gl\right)^{1/2} \label{eq:R}
\end{equation}
where \(R_0\) is the Galactocentric radius of the Solar orbit and
\(\gl\) is its Galactic longitude. The uncertainty in \(R\) is
\begin{equation}
  \sigma_R^2 = \sigma_d^2\left(\frac{\partial R}{\partial d}\right)^2 +  \sigma_{R_0}^2\left(\frac{\partial R}{\partial R_0}\right)^2 + \sigma_\gl^2\left(\frac{\partial R}{\partial \gl}\right)^2
\end{equation}
where \(\sigma_R\) is the uncertainty in \(R\), \(\sigma_{R_0}\) is
the uncertainty in \(R_0\), and \(\sigma_\gl\) is the uncertainty in
\(\gl\). For simplicity we ignore cross-terms and assume
\(\sigma_{R_0} = \sigma_\gl = 0\).  The above equation then
reduces to
\begin{equation}
  \sigma_R^2 = \sigma_d^2\left(\frac{\partial R}{\partial d}\right)^2.
\end{equation}
The partial derivative evaluates to
\begin{align}
  \frac{\partial R}{\partial d} & = \frac{1}{2}\left(d^2 + R_0^2 - 2dR_0 \cos\gl\right)^{-1/2}\left(2d - 2R_0\cos\gl\right) = \frac{d - R_0\cos\gl}{R}.
\end{align}
The uncertainty in the Galactocentric radius of an object is thus
related to the uncertainty in its distance from the Sun by
\begin{equation}
  \sigma_R = \frac{\sigma_d}{R}|d-R_0\cos\gl|. \label{eq:sigma_R}
\end{equation}
This relationship is shown in the top panel of
Figure~\ref{fig:faceon_unc}.

An object's Galactocentric azimuth is given by
\begin{equation}
  \cos\az = \frac{R^2 + R_0^2 - d^2}{2RR_0} = \frac{d^2+R_0^2-2dR_0\cos\gl+R_0^2-d^2}{2RR_0} = \frac{R_0 - d\cos\gl}{R}.
\end{equation}
Again ignoring the \(\sigma_{R_0}\) and \(\sigma_\gl\) terms, the
uncertainty in \(\az\) is
\begin{equation}
  \azerr^2 = \sigma_d^2\left(\frac{\partial\az}{\partial d}\right)^2.
\end{equation}
The partial derivative evaluates to
\begin{align}
  \frac{\partial \az}{\partial d} & = -\frac{1}{\sqrt{1 - \cos^2\az}}\frac{\partial \cos \az}{\partial d} = -\frac{1}{\sin\az}\frac{\partial \cos \az}{\partial d}. \nonumber
\end{align}
To evaluate the derivative of \(\cos \az\), we use our expression for
\(R\) (Equation~\ref{eq:R}) and define functions \(f(d)\) and \(g(d)\)
such that
\begin{equation}
  \cos\az = \frac{R^2 + R_0^2 - d^2}{2RR_0} = \frac{d^2+R_0^2-2dR_0\cos\gl+R_0^2-d^2}{2RR_0} = \frac{R_0 - d\cos\gl}{R} \equiv \frac{f(d)}{g(d)}
\end{equation}
where
\begin{align}
  \frac{\partial f(d)}{\partial d} & = \frac{\partial (R_0 - d\cos\gl)}{\partial d} = -\cos\gl \nonumber \\ 
  \frac{\partial g(d)}{\partial d} & = \frac{\partial R}{\partial d} = \frac{d - R_0\cos\gl}{R} \nonumber.
\end{align}
We find
\begin{align}
  \frac{\partial \cos \az}{\partial d} & = -\left[\frac{-R\cos\gl - \left(R_0 - d\cos\gl\right)\left(\frac{d - R_0\cos\gl}{R}\right)}{R^2\sin\az}\right] \nonumber \\ 
  & = \left[\frac{\cos\gl}{R\sin\az} + \frac{\left(R_0 - d\cos\gl\right)\left(d - R_0\cos\gl\right)}{R^3\sin\az}\right]. \nonumber
\end{align}
The uncertainty in the Galactic azimuth is thus
\begin{align}
  \azerr^2 & = \sigma_d^2\left(\frac{\cos\gl}{R\sin\az} + \frac{\left(R_0 - d\cos\gl\right)\left(d - R_0\cos\gl\right)}{R^3\sin\az}\right)^2 \nonumber \\ 
  \azerr & = \frac{\sigma_d}{R}\bigg|\csc\az\left[\frac{\cos\gl}{\sin\az} + \frac{\left(R_0 - d\cos\gl\right)\left(d - R_0\cos\gl\right)}{R^2\sin\az}\right]\bigg|.
\end{align}
Rearranging, we see that the uncertainty in the Galactocentric azimuth
of an object is related to the uncertainty in its distance from
the Sun by
\begin{equation}
  \azerr = \frac{\sigma_d}{R}\bigg|\frac{\cos\gl}{\sin\az} + \frac{\left(R_0 - d\cos\gl\right)\left(d - R_0\cos\gl\right)}{R^2\sin\az}\bigg|. \label{eq:sigma_Az}
\end{equation}
This relationship is shown in the bottom panel of
Figure~\ref{fig:faceon_unc}.

\end{document}